\documentclass[useAMS,usenatbib]{mnras}

\usepackage{amsmath,amssymb,mathrsfs,graphicx,float,latexsym,url,booktabs}
\usepackage{epstopdf,ragged2e}
\usepackage{natbib,url}
\usepackage{hyperref}
\usepackage{lmodern}

\usepackage[usenames,dvipsnames]{color}
\definecolor{darkred}{rgb}{0.5,0,0}
\definecolor{darkgreen}{rgb}{0,0.5,0}
\definecolor{darkblue}{rgb}{0,0,0.5}
\definecolor{prussian}{rgb}{0.0, 0.19, 0.33}
\definecolor{richelectricblue}{rgb}{0.03, 0.57, 0.82}
\definecolor{teal}{rgb}{0.0, 0.5, 0.5}
\definecolor{mediumseagreen}{rgb}{0.24, 0.7, 0.44}
\definecolor{lust}{rgb}{0.9, 0.13, 0.13}
\definecolor{ballblue}{rgb}{0.13, 0.67, 0.8}
\definecolor{darkcyan}{rgb}{0.0, 0.55, 0.55}
\definecolor{mountainmeadow}{rgb}{0.19, 0.73, 0.56}
\definecolor{palecarmine}{rgb}{0.69, 0.25, 0.21}
\definecolor{richcarmine}{rgb}{0.84, 0.0, 0.25}
\definecolor{tangelo}{rgb}{0.98, 0.3, 0.0}
\definecolor{venetian}{rgb}{0.784,0.031,0.082}
\definecolor{bdfrance}{rgb}{0.192,0.549,0.906}
\usepackage{mathtools}
\usepackage{amsbsy}
\usepackage{bm}
\usepackage{float}
\newcommand{\Rs}{R_{\star}}
\newcommand{\Rcc}{R_{\rm cc}}
\newcommand{\Ms}{M_{\star}}

\def\04a{{2004 a}}
\def\04b{{2004 b}}

\begin{document}

\title[Magnetoelastic avalanching in magnetars]
{Magnetoelastic avalanches in magnetar crusts with plastic islands}
\author[Arthur G. Suvorov]{Arthur G. Suvorov$^{1}$\thanks{a.suvorov@uni-tuebingen.de} \\
$^1$Theoretical Astrophysics, Eberhard Karls University of T{\"u}bingen, T{\"u}bingen, D-72076, Germany
}

\date{Accepted ?. Received ?; in original form ?}

\pagerange{\pageref{firstpage}--\pageref{lastpage}} \pubyear{?}

\maketitle
\label{firstpage}

\begin{abstract}
\noindent High-energy outbursts from magnetars may be associated with crustal failures or quakes. 
The magnetoelastic modes of the star can be excited to large amplitudes in such a framework, providing a natural explanation for the quasi-periodic oscillations observed in the X-ray tails of some events.
However, previous calculations of torsional modes have been carried out assuming an intact crust without accounting for the failure itself.
We develop a code in this paper that solves the axial sector of the linearised Einstein-Maxwell-Euler equations in a cavity that hosts inelastic regions or `plastic islands'.
As crustal elements are displaced, shear stresses are tracked as a function of time to allow cells to dynamically switch from elastic to inelastic states.
The presence of islands in the crust promotes phase mixing and, when reflected waves mix constructively, large sections fail collectively (`avalanche') if the initial kick is strong enough.
Transient spectral features are observed as a result of the progressive loss of elastic support, which may explain the temperamental nature of the oscillations observed following giant flares and magnetar storms.
\end{abstract}

\begin{keywords}
stars: magnetars, pulsations, magnetohydrodynamics, methods: numerical
\end{keywords}

%%%%%%%%%%%%%%%%%%%%%%%%%%%%%%%%%

%%%%%%%%%%%%%%%%%%%%%%%%%%%%%%%%%
\section{Introduction} 
\label{sec:intro}
%%%%%%%%%%%%%%%%%%%%%%%%%%%%%%%%%

%[mention $m=0$; others use broad kicks to excite ad hoc; ]

Magnetars are observationally classified by their propensity for bursting and flaring in the X- and gamma-ray bands.
The giant flares unleashed from these sources constitute the brightest events ever observed in the local universe in terms of peak luminosity \citep{maz79,palm05}. %the isotropic energy output from the 2004 giant flare from SGR 1806--20 was $\gtrsim 10^{46}$~erg .
Theoretical efforts to model the trigger of flares, including their rarity, rise times, and durations have led to significant shifts in our collective understanding of neutron-star phenomena \cite[see][for a review]{sotani2024}.
Given that quantum-electrodynamic, chromodynamic, and general-relativistic (GR) effects all interact non-trivially in magnetar environments \cite[e.g.,][]{suvpons25,wad26}, it is worth trying to understand how observed characteristics translate into a theoretical picture and vice versa. %to map data to fundamental physics.

A noteworthy feature associated with high-energy outbursts are quasi-periodic oscillations (QPOs).
Data from Rossi X-ray Timing Explorer (RXTE) for the August 1998 giant flare from SGR 1900+14, for instance, reveal a rich forest of spectral peaks throughout the decaying, pulsed tail spanning hundreds of seconds \citep{sw05}.
One interpretation of QPOs is that the quasinormal modes of the star were excited and subsequently imprinted on the signal \cite[e.g.,][]{stroh06}.
By matching observed frequencies to the perturbed spectra of model stars one can, in principle, approach the inverse problem of constraining nuclear and magnetic substructure \citep{sotani2008,gabler18}. 
As an example, \cite{col12} found that, for a $1.4 M_{\odot}$ star with a field strength of $\approx 4 \times 10^{15}$~G, the 28, 53 and 84 Hz QPOs observed from SGR 1900+14 could be identified as discrete Alfv{\'e}n modes for some equations of state (EOS) or global modes in a star of the same mass but different EOS.
Aside from this ambiguity, many subtle aspects complicate the identification problem; superfluidity \citep{gabler12,gabler2013} plus crust-core \citep{vh11,kj15}, crust-magnetosphere \citep{burn25}, and non-axisymmetric \citep{passl13} couplings all shape the spectrum and excitation amplitudes.

Mechanical stresses build up over time in a magnetar crust through the combined actions of Hall drift and Ohmic dissipation \citep{ponsv19}, which strain the lattice and inject twist into the magnetosphere \citep{belo09}.
In this respect, the two main models put forward to explain outbursts are either that (i) the magnetosphere becomes overtwisted and a reconnection is triggered, or (ii) those same mechanical stresses overwhelm the ionic lattice and instigate crustal failure \cite[see][for a review]{kaspbel17}.
Although the sub-ms rise times of giant flares hint towards a magnetospheric origin \citep{lyu15}, the quake picture finds some favour in that modes are naturally excited to large amplitudes.
Long-term magnetothermal evolutions allowing for quakes also reproduce observed outburst rates rather well \citep{pons11,perna11}.

However, previous quake-centered models for QPOs do not account for the fact that the crust must fail in the first place to excite the modes.
In this paper, we explore how crustal failure itself affects the development of magnetoelastic modes, purportedly tied to QPOs, by tracking stresses in a time-domain simulation that can dynamically switch a cell from an elastic to an inelastic state.
Generally speaking, the loss of elastic support in overstressed zones tends to decrease the frequency of torsional modes in a state-dependent manner that adjusts with the failure geometry.
Moreover, if the crust is already in a volatile, prestressed state when some section fails, the torrent of magnetoelastic waves that are launched will induce further stresses at mode crests as they propagate, potentially driving an avalanche \citep{lev16}.
The formalism developed here has the added benefit that the choice of initial data can be tied directly to the initial quake geometry and is thus (arguably) less \emph{ad hoc}.

We describe, and present the first results of, a new code that evolves the linearised conservation equations in an elastic crust with a prespecified failure geometry (`plastic island'). 
Stresses are tracked during the evolution and, if further cells exceed a nominal von Mises threshold, have their elastic traction removed before the simulation resumes.
We observe both prompt avalanching and beat-frequency phenomena such that, as more cells fail over time, the character of magnetoelastic modes drifts in time with energy cascading to smaller scales as fragmentary patterns emerge due to mode mixing \cite[see, e.g.,][]{HeyvaertsPriest83,gabler12}.
Although linear, a time-domain solution to the mode problem with a dynamical elastic response adds a step of realism towards the characterisation of QPO phenomena. 

\begin{figure}
\centering
 \includegraphics[width=0.48\textwidth]{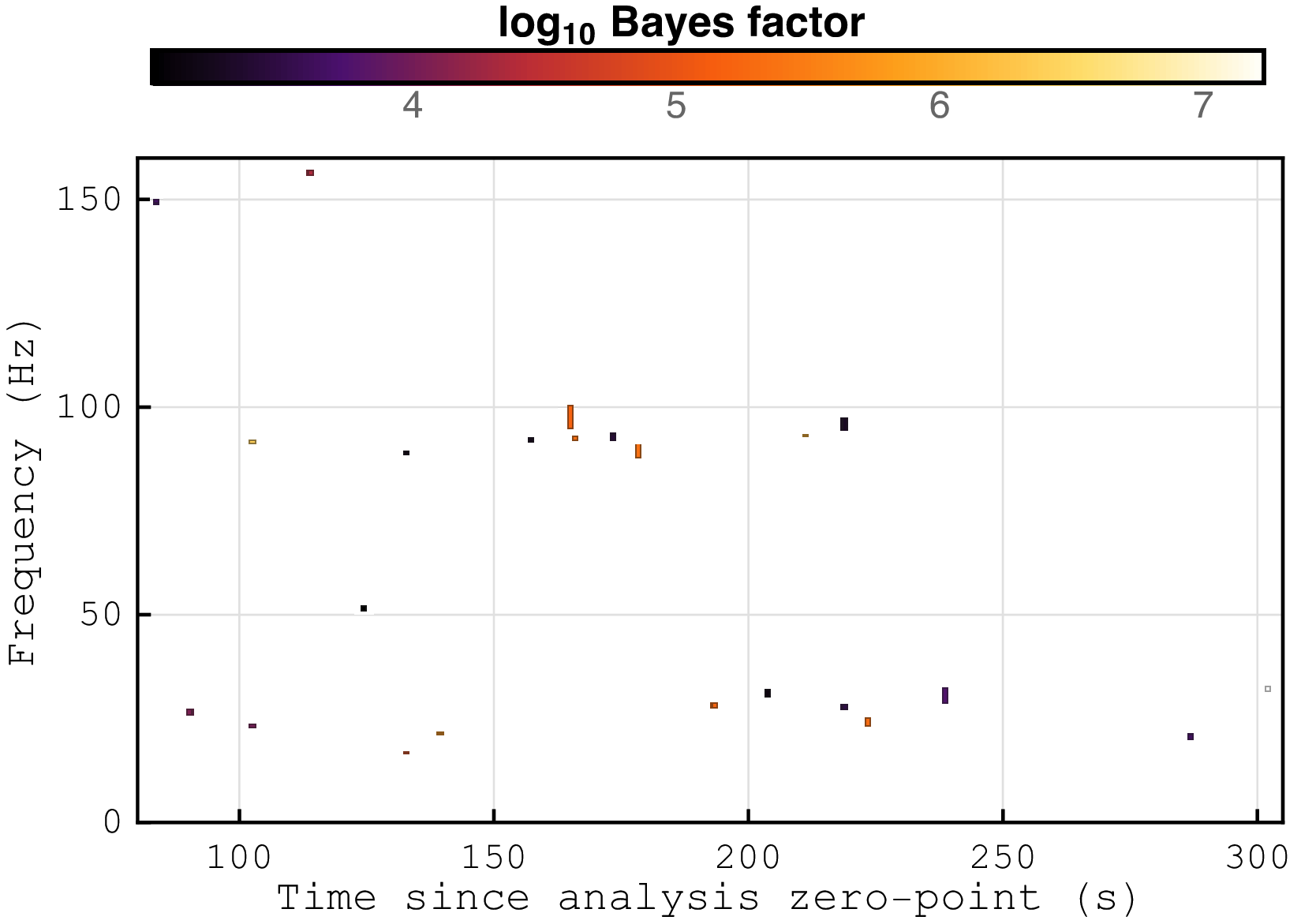}
 \caption{QPOs as a function of time following the giant flare from SGR 1806--20 in December 2004, colour-coded according to Bayes factor, as reported in Table 1 of \protect\cite{miller19qpos} (see also their Figure 3). 
 Brighter shades indicate greater significance.
 The vertical extent represents the reported frequency interval. 
 For ease of visibility, we impose a minimum width of $2$~s.}
  \label{fig:miller}
\end{figure}

Indeed, a key feature of the QPOs that has thus-far resisted adequate description is mode transience: dynamic spectra show oscillations apparently being excited and then decaying and/or re-emerging at later times with different frequencies, amplitudes, and longevities \citep{hupp14,pumpe17}.
For instance, we show in Figure~\ref{fig:miller} the spectrum determined by \cite{miller19qpos} following the December 2004 giant flare from SGR 1806--20.
By eye one can see a feature around $\sim 25$~Hz that seems to become excited and then fade away, only to sporadically reemerge with a slightly different frequency and amplitude.
Whether it is the `same mode' is unclear; nevertheless, transient features may hint towards nonlinear phenomena such as mode-mode seeding \citep{sot04,pl25}, damping due to various couplings \citep{gabler16}, or -- as we investigate here -- dynamical mode mixing promoted by the progressive loss of elasticity \cite[see also][]{koj24,cap2026,cap2026b}. 

This paper is organised as follows.
Section~\ref{sec:mathematicalmodel} introduces the mathematical formalism, including the perturbation equations and failure prescription (Sec.~\ref{sec:failureprescription}). 
The background ingredients needed to initialise a model (i.e., EOS, shear modulus, and magnetic field) are then detailed in Section~\ref{sec:hydrostatics}.
The numerical backbone, including how stresses are tracked and variables are evolved more generally, is described in Section~\ref{sec:numerical_methods}. 
Two representative runs are then presented, corresponding to global and local quakes and the associated phenomena, in Sections~\ref{sec:global} and \ref{sec:localised}, respectively.
Some discussion is provided in Section~\ref{sec:discussion}.
Various code tests and scaling relationships are explored in the Appendices.
Throughout this work, except where quantities are expressed in physical CGS units, we work in natural units where $G=c=1$. 

%%%%%% FIGURE %%%%%%
%%%%%%%%%%%%

%%%%%%%%%%%%%%%%%%%%%%%%%%%%%%%%%
\section{Mathematical model} \label{sec:mathematicalmodel}
%%%%%%%%%%%%%%%%%%%%%%%%%%%%%%%%%

We begin by describing the mathematical model used in terms of background quantities and their perturbations, following the treatment of \cite{sotani2007} for describing axisymmetric torsional oscillations.

We work with a static and spherically symmetric spacetime in Schwarzschild-like coordinates,
\begin{equation} \label{eq:metric}
    ds^2 = -e^{2\Phi(r)} dt^2 + e^{2\Lambda(r)}dr^2 + r^2 \left( d\theta^2+\sin^2\theta\,d\phi^2 \right),
\end{equation}
where the potentials $\Phi$ and $\Lambda$ are determined by solving the Tolman-Oppenheimer-Volkoff (TOV) equations in the standard manner. 
Details about the EOS choice to close the system are given in Sec.~\ref{sec:hydrostatics}.

%Equivalently, if $\epsilon$ denotes the energy density in conventional units, then $\epsilon=\epsilon/c^2$ when restoring factors of $c$.
In general, the fluid and magnetic parts of the energy-momentum tensor sum to give
\begin{equation} \label{eq:mhd}
\begin{aligned}
\hspace{-0.338cm}    T^{\mu\nu}_{\rm MHD} = \left( \epsilon+p+\frac{B^2}{4\pi} \right) u^\mu u^\nu\ + \left( p+\frac{B^2}{8\pi} \right)g^{\mu\nu} - \frac{B^\mu B^\nu}{4\pi},
\end{aligned}
\end{equation}
where $u^\mu u_\mu=-1$ for 4-velocity $u^{\alpha}$, the total matter energy-density is denoted by $\epsilon$ (not to be confused with the baryonic rest-mass density), $p$ is the pressure, and we have $u_\mu B^\mu =0$ where $B^2=B_\mu B^\mu$ is the magnetic intensity measured in the fluid frame.  
%We depart slightly from \cite{sotani2007} by using $B^\mu$ directly rather than introducing the rescaled magnetic variable $H^{\mu} =B^\mu/\sqrt{4\pi}$ to avoid confusion with enthalpy.
Elastic stresses are included through the linearised shear-stress tensor \citep{carter72}
\begin{equation}  \label{eq:elastic}
    \delta T^{\rm s}_{\mu\nu} = -2\mu_{\rm eff} \delta S_{\mu\nu},
\end{equation}
such that, for a displacement field $\xi^\mu$,
\begin{equation} \label{eq:shear_tensor}
\delta S_{\mu\nu} = h_\mu{}^\alpha h_\nu{}^\beta \left[\frac{1}{2}\left(\nabla_{\alpha}\xi_{\beta} + \nabla_{\beta} \xi_{\alpha} \right)- \frac{1}{3} h_{\alpha\beta} \nabla_\gamma\xi^\gamma \right], 
\end{equation}
where $h^\mu{}_\nu = \delta^\mu{}_\nu+u^\mu u_\nu$ denotes the fluid projection.
The quantity $\mu_{\rm eff}$ is just the shear modulus in elastic regions but is otherwise set to zero, viz.
\begin{equation} \label{eq:effective_mu}
    \mu_{\rm eff} = \begin{cases}
        \mu(r) & \text{elastic material}\\
        0     & \text{otherwise},
    \end{cases}
\end{equation}
and thus may be anisotropic in general (see Sec.~\ref{sec:failureprescription}).
%Details of the modulus and the failure mechanism we consider to shut off $\mu(r)$ are given in Secs.~\ref{sec:shear} and \ref{sec:failureprescription}, respectively.
Even in a magnetar, the ratio of magnetic-to-hydrostatic energy is expected to be very small and thus it is reasonable to treat the background field in the passive limit where we do not account for magnetic pressures at the metric level \eqref{eq:metric}. 
We further ignore crustal prestresses in the equilibrium force balance, meaning that we assume the initial currents are not themselves supported by elastic stresses \cite[cf.][for a critical discussion]{koj24}. 
The background construction is thus complete once the Grad-Shafranov equation is solved over a given TOV background to determine $B^{\alpha}$; see Sec.~\ref{sec:magneticbackground}.

In this paper, we consider purely torsional oscillations.
This is motivated by noting that the stratification in the crust, combined with the fact the hydrostatic pressure exceeds the Coulomb energy density by a couple orders of magnitude, limits vertical (i.e., $\xi_r \approx 0$) and compressible ($\nabla_{j} \xi^{j} \approx 0$) displacements \citep{bransgrove2018}.
Such a conclusion may change however if a strong toroidal field is present as magnetic buoyancy can help lift fluid elements.
For the purely-poloidal equilibrium fields considered here,
\begin{equation}
    B^\mu=(0,B^r,B^\theta,0),
\end{equation}
an azimuthal displacement generates only an azimuthal velocity and a twist ($\delta B^{\phi}$). 
To linear order, the Lorentz force is likewise azimuthal. 
The axial subsystem therefore closes without sourcing density, pressure, or meridional velocities,
\begin{equation}
    \delta\epsilon=\delta p = \delta u^r = \delta u^\theta=0 ,
\end{equation}
where we further impose the Cowling approximation ($\delta g_{\mu \nu} = 0$). 
As noted above, such a decoupling is limited to the case considered where we do not have a mixed poloidal-toroidal field, rotation, non-axisymmetry, or a non-spherical equilibrium any of which will, in general, blend the axial and polar sectors and lead to a considerably more complicated problem \cite[see, e.g.,][]{sotani2008}.

The dynamical equations solved in this paper are thus obtained directly from
\begin{equation} \label{eq:stress_conservation}
0 = \nabla_{\mu} \delta T^{\mu \nu} = \nabla_\mu \left( \delta T^{\mu\nu}_{\rm MHD} + \delta T^{\mu\nu}_{\rm s} \right),
\end{equation}
which is closed through the induction equation,
\begin{equation} \label{eq:induction}
0 =\nabla_\mu \left( \delta u^\mu B^\nu + u^\mu \delta B^\nu-\delta u^\nu B^\mu - u^\nu \delta B^\mu \right).
\end{equation}
The torsional displacement can be described by a single variable, $\xi^{\phi} = Y(t,r,\theta)$. 
The physical displacement as measured in a local orthonormal frame, henceforth denoted by overhead hats, is
\begin{equation} \label{eq:Y_definition}
\xi^{\hat\phi} = r\sin\theta \,Y .
\end{equation}
The associated perturbed velocity component is just $\delta u^\phi=e^{-\Phi}\dot{Y}$, where we use an overhead dot to denote (coordinate) time derivatives.
The trace term in equation \eqref{eq:shear_tensor} vanishes within the scheme of our approximations, and the two non-zero shear components are
\begin{equation} \label{eq:torsional_strain}
\delta S_{\hat r\hat\phi} = \frac{1}{2} e^{-\Lambda}r\sin\theta\,Y_{,r},
\end{equation}
and
\begin{equation}
\delta S_{\hat\theta\hat\phi} = \frac{1}{2} \sin\theta\,Y_{,\theta},
\end{equation}
while the magnetic perturbation follows from the $\phi$ component of the induction equation \eqref{eq:induction}. 

At this point it is convenient to introduce a new angular coordinate, $x=\cos\theta$, to circumvent polar singularities and write expressions in divergence form for later numerical use.
Since the background field has no azimuthal component we have
\begin{equation}
    \delta B^\phi  = \nabla_{\rm B} Y =  B^rY_{,r}+B^{x}Y_{,x},
    \label{eq:dBphi}
\end{equation}
where $\nabla_{\rm B} $ is a magnetic-aligned derivative.
The above just conveys that magnetic twisting is produced only by variations of the displacement \emph{along} the equilibrium field.
The three non-zero components entering equation \eqref{eq:stress_conservation} are easily found to be
\begin{equation} \label{eq:Ttphi}
\delta T^t{}_\phi = \left( \epsilon+p+\frac{B^2}{4\pi} \right) r^2(1-x^2) e^{-2\Phi}\dot{Y},
\end{equation}
\begin{equation} \label{eq:Trphi}
    \delta T^r{}_\phi = -r^2(1-x^2) \left[ \mu_{\rm eff}e^{-2\Lambda}Y_{,r} + \frac{B^r}{4\pi}\nabla_{\rm B} Y \right],
\end{equation}
and
\begin{equation} \label{eq:Tthetaphi}
    \delta T^x{}_\phi = -(1-x^2) \left[ (1-x^2) \mu_{\rm eff} Y_{,x} + r^2 \frac{B^x}{4\pi}\nabla_{\rm B} Y \right],
\end{equation}
allowing us to write down the final `master equation' of relevance,
\begin{equation} \label{eq:torsional_evolution} 
\begin{aligned}     
0  =& \left[ r^4 e^{\Phi} (1-x^2) \left( \mu_{\rm eff}e^{-\Lambda} Y_{,r} + \frac{e^{\Lambda} B^r}{4\pi} \nabla_{\rm B} Y \right) \right]_{,r} \\ 
&+ \left[r^2 e^{\Phi+\Lambda} (1-x^2) \left( \mu_{\rm eff}(1-x^2)Y_{,x} +\frac{r^2 B^x}{4\pi} \nabla_{\rm B} Y \right) \right]_{,x} \\
&-\left( \epsilon+p+\frac{B^2}{4\pi} \right) r^4e^{\Lambda-\Phi}(1-x^2)\ddot{Y}.
\end{aligned} 
\end{equation}

The aim of this work is to investigate how oscillations propagate in a magnetar crust that has regions that can dynamically switch from an elastic to an inelastic state.
It is most natural to carry out such a program in the time domain, as further stresses are applied throughout the crust as the displacement evolves in time (expression \ref{eq:shear_tensor}). 
Furthermore, if the shear modulus is anisotropic, as equation \eqref{eq:effective_mu} encapsulates, we generically expect that even an $\ell=2$ (for example) perturbation would excite a host of modes with $\ell=2,3,\ldots$, with amplitudes that depend on the initial kick. 
It is for these reasons we do not decompose equation \eqref{eq:torsional_evolution} into spherical harmonics like \cite{sotani2007} and others.
Tests of the island-free limit where $\mu_{\rm eff} = \mu(r)$ everywhere -- such that we expect to approximately recover standard results for torsional modes -- are given in Appendix~\ref{sec:noislands}.

%%%%%%%%%%%%%%%%%%%%%%%%%%%%%%%%%
\subsection{Failure prescription}  \label{sec:failureprescription}
%%%%%%%%%%%%%%%%%%%%%%%%%%%%%%%%%

We adopt the von Mises description as a failure criterion because of its simplicity.
That is, we say a local patch of the crust fails when it exceeds a critical strain threshold, 
\begin{equation} \label{eq:threshold}
    \varepsilon_{\rm max}=0.1,
\end{equation}
motivated by the molecular-dynamics calculations of \cite{hk09}.
The exact threshold is unimportant in our case, as what matters is only the relative difference between the imposed pre-stress and the elastic maximum.

For the torsional excitations considered here in the Cowling approximation, the only relevant components of the strain which combine to define\footnote{Strictly speaking, both stress measures are smaller by a factor $2\sqrt{3}$ than the conventional von Mises stress and its corresponding threshold \cite[see, e.g.,][]{Andersson19}. By reducing both together however the failure criterion is effectively unchanged, and our choices are merely convenient to avoid numerical prefactors.} the von Mises equivalent stress are $\delta T^{\rm s}_{\hat r\hat\phi}$ and $\delta T^{\rm s}_{\hat\theta\hat\phi}$, viz.
\begin{equation} \label{eq:von_mises}
\sigma_{\rm VM} = \frac{1}{2}\sqrt{\left(\delta T^{\rm s}_{\hat r\hat\phi}\right)^{2} + \left(\delta T^{\rm s}_{\hat\theta\hat\phi}\right)^{2}}.
\end{equation}
We translate the adopted breaking strain $\varepsilon_{\rm max}$ into the corresponding local yield stress through a Hookean approximation,
\begin{equation} \label{eq:yield_stress}
\sigma_{\rm yield} = \mu_{\rm eff}\varepsilon_{\rm max},
\end{equation}
and thus say a cell fails when
\begin{equation} \label{eq:failure_criterion}
\sigma_{\rm VM} \geq \sigma_{\rm yield}.
\end{equation}
Note that it is the orthonormal components of the stress tensor that are used to avoid spurious adjustments to the local yield criterion from coordinate-dependent scale factors.

The simulation begins with a pre-loaded strain profile and a plastic island of some spatial extent, the details of which are deferred to Sec.~\ref{sec:numerical_methods}.
Importantly, although we use the phrase `plastic' interchangeably with `inelastic' in this work some caution is warranted as we do not adopt a viscoplastic flow law, creep time, strain hardening, healing, or temperature-dependent backreactions.
Such effects are likely important in a realistic star \cite[see][]{lg19,cap2026,cap2026b}, though we effectively assume they operate on timescales longer than our simulations ($\sim$~seconds) and do not account for them.
For our purposes therefore, a failed cell simply corresponds to one in which $\mu =0$.
Extensions to a frequency-dependent response [i.e., $\mu_{\rm eff} = \mu(\boldsymbol{x},\omega)$] are deferred to future work.

%%%%%%%%%%%%%%%%%%%%%%%%%%%%%%%%%
\section{Equilibrium input} \label{sec:hydrostatics}
%%%%%%%%%%%%%%%%%%%%%%%%%%%%%%%%%

To solve equation \eqref{eq:torsional_evolution}, some background inputs are required.
The model choices are detailed in this section.

%%%%%%%%%%%%%%%%%%%%%%%%%%%%%%%%%
\subsection{Equation of state} \label{sec:equationofstate}
%%%%%%%%%%%%%%%%%%%%%%%%%%%%%%%%%

We adopt the BSk24 EOS\footnote{Available on the CompOSE database hosted at \url{https://compose.obspm.fr/eos/253} and the Ioffe repository at \url{http://www.ioffe.ru/astro/NSG/BSk}.} \citep{pearson2018} to construct the energy-pressure relation, $\epsilon(P)$, to close the system. 
This EOS was chosen because it is compatible with all existing multimessenger experiments and, conveniently, has crustal composition information available that is relevant for the construction of the shear modulus (see Sec.~\ref{sec:shearmodulus}).

%%%%%% FIGURE %%%%%%
\begin{figure}
\centering
 \includegraphics[width=0.48\textwidth]{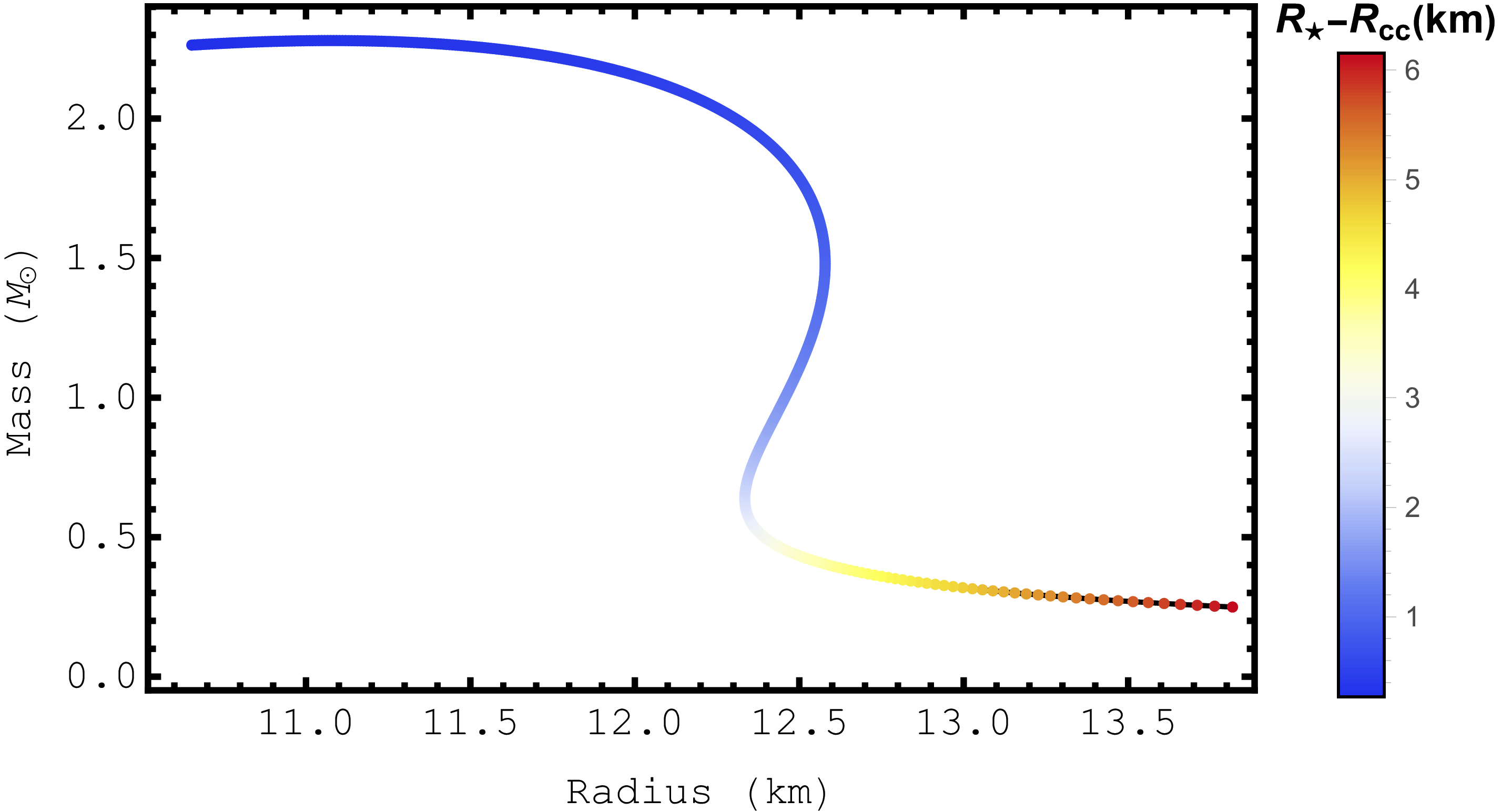}
 \caption{Mass-radius diagram for the BSk24 EOS used in this paper, colour-coded according to the thickness of the crust.}
  \label{fig:massrad}
\end{figure}
%%%%%%%%%%%%

For any given model star, the crust-core boundary ($r = R_{\rm cc}$) is identified through the transition density
\begin{equation} \label{eq:transdensity}
n_b(\Rcc)=n_{\rm cc}=0.081\ {\rm fm}^{-3},
\end{equation}
with $\Rcc$ found from the pressure profile using the tabulated BSk24 data, i.e., interpolating to locate $n_b[p(\Rcc)]$.
We ignore ocean layers in this work for simplicity, fixing the outer boundary where $p(r)$ drops to zero as the stellar surface ($r=\Rs$).
The obtained mass-radius diagram is depicted in Figure~\ref{fig:massrad}, colour-coded according to the size of the crustal cavity $(\Rs - R_{\rm cc})$.
Although the conventional statement is that the crust occupies the final $\sim 1$~km of the star, we see in fact that its thickness varies monotonically across the diagram because the central pressure decreases as we move to lower masses, which shifts the radius where the transition \eqref{eq:transdensity} occurs.
Although we fix $\Ms = 1.4M_{\odot}$ throughout in this work to avoid a large parameter-space survey, considering a magnetar of mass $\sim 1.0M_{\odot}$ would effectively double the thickness and shift the shear mode spectrum upwards by a factor $\sim 2$ \cite[see equation 18 in][]{sa07}.

%%%%%%%%%%%%%%%%%%%%%%%%%%%%%%%%%%%%%%%%%%%%%%%%%%%%%%%%%%%%%%%%
\subsection{Shear modulus} \label{sec:shearmodulus}
%%%%%%%%%%%%%%%%%%%%%%%%%%%%%%%%%%%%%%%%%%%%%%%%%%%%%%%%%%%%%%%%

Another ingredient we require at the background level is the shear modulus.
For our chosen BSk24 EOS, composition data is supplied by CompOSE: at a given baryon density $n_b$ in beta equilibrium, we have tables for the lepton fractions $Y_e$ and $Y_\mu$, the free-proton fraction $Y_p^{\rm free}$, and the charge number $Z$. 
Imposing charge neutrality then gives the cluster-number density as $n_N = n_b \left({Y_e+Y_\mu-Y_p^{\rm free}}\right)/{Z}$,
from which the Wigner-Seitz radius is found through $a=\left({3}/{4\pi n_N}\right)^{1/3}$.
The usual point-nucleus, Coulomb-crystal result from \cite{stroh91} can thus be computed directly through
\begin{equation} \label{eq:shearzero}
\mu_0 \approx \frac{3}{25} \frac{n_{N} Z^2 e^2}{a},
\end{equation}
where we ignore finite-temperature corrections \cite[which could be sizeable in a $\sim$~kyr-old magnetar;][]{ponsv19}.

%%%%%% FIGURE %%%%%%
\begin{figure}
\centering
 \includegraphics[width=0.48\textwidth]{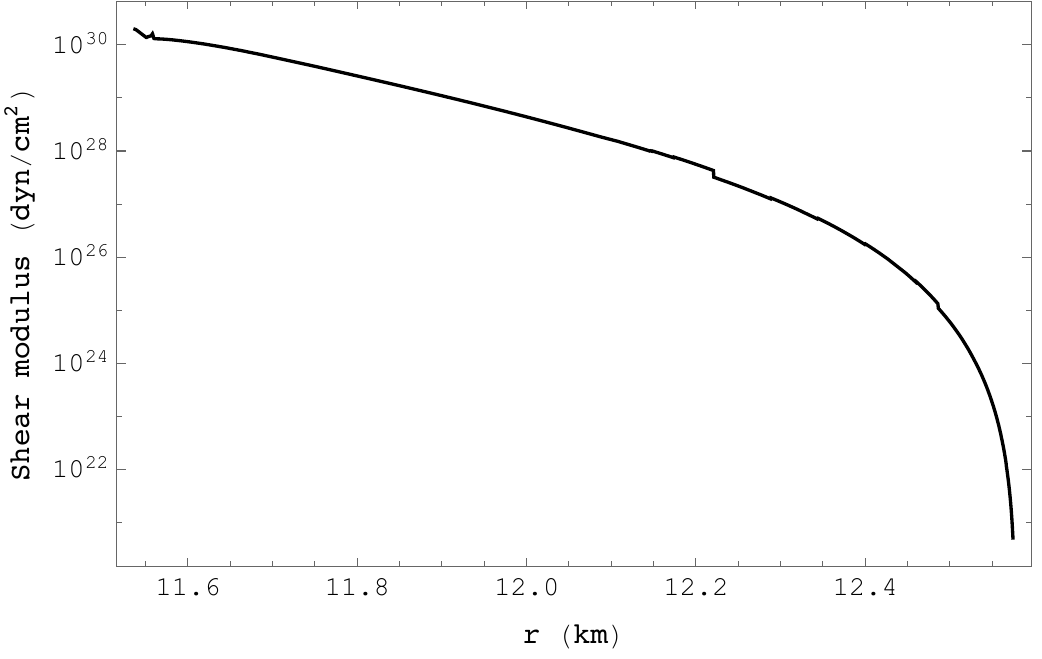}
 \caption{Shear modulus used in elastic portions of the crust accounting for corrections due to shape degrees of freedom through equation \eqref{eq:finalmodulus}, as detailed by \protect\cite{zem23}.}
  \label{fig:shearmod}
\end{figure}
%%%%%%%%%%%%

As described by \cite{zem23} however, expression~\eqref{eq:shearzero} overestimates the lattice rigidity by treating ions as point-like charges, neglecting the nuclear-shape degrees of freedom which are known to be important for magnetar evolution \citep{pons2013}.
Under applied strain, these finite-sized nuclei undergo quadrupolar deformations that relax electrostatic forces across the lattice, lowering the elastic energy cost and reducing the effective shear modulus.
While only of order $\sim 20\%$ in the inner crust, as it incurs no additional computational expense to use any given $\mu(r)$ over another we include such effects here.
To do so, we additionally require the cluster volume fraction $u$, which is not supplied directly by the table. 
We therefore used the fit (accurate to within $\sim 2\%$) for the equivalent nuclear-size ratio provided by equation (A9) in \cite{pot18},
\begin{equation}
x_{\rm nuc} = 0.1034 + \frac{2.005 \tilde{n}_b^{0.575}}{1+72.7 \tilde{n}_b^{2.45}} + 1.06 Z \tilde{n}_b ^{2.61},
\end{equation}
with $\tilde{n}_b = n_b/\text{fm}^{-3}$.
The shear modulus we implement is then \cite[see equation 9 in][]{zem23}
\begin{equation} \label{eq:finalmodulus}
\mu(r) = \mu_0 \left[ 1- \frac{u^{5/3}} {2-4u^{1/3}+3u} \right],
\end{equation}
where $u \approx x_{\rm nuc}^3$ if clusters are, on average, approximately spherical.
For a star of mass $1.4 M_{\odot}$, the profile is shown in Figure~\ref{fig:shearmod} for reference.
The sharp features are not interpolation artifacts but compositional jumps in the nuclear profile for the BSk24 crust. 
Some testing using expression \eqref{eq:shearzero} reveals that additional failures become possible in the inner crust because of the reductions for a given prestress loadout, but the total count changes are small ($\sim$~percent level).

%%%%%%%%%%%%%%%%%%%%%%%%%%%%%%%%%
\subsection{Poloidal fields} \label{sec:magneticbackground}
%%%%%%%%%%%%%%%%%%%%%%%%%%%%%%%%%

The background magnetic field is obtained from the GR Grad-Shafranov equation allowing, for simplicity, only a poloidal dipole.
Although the inclusion of higher-order multipoles is straightforward at the equilibrium level \cite[see, e.g.,][]{suvg23}, it makes for a more expensive evolution as smaller scales require finer timestepping to avoid the Courant–Friedrichs–Lewy (CFL) instability; see Sec.~\ref{sec:time_stepping}.

%%%%%% FIGURE %%%%%%
\begin{figure}
\centering
 \includegraphics[width=0.48\textwidth]{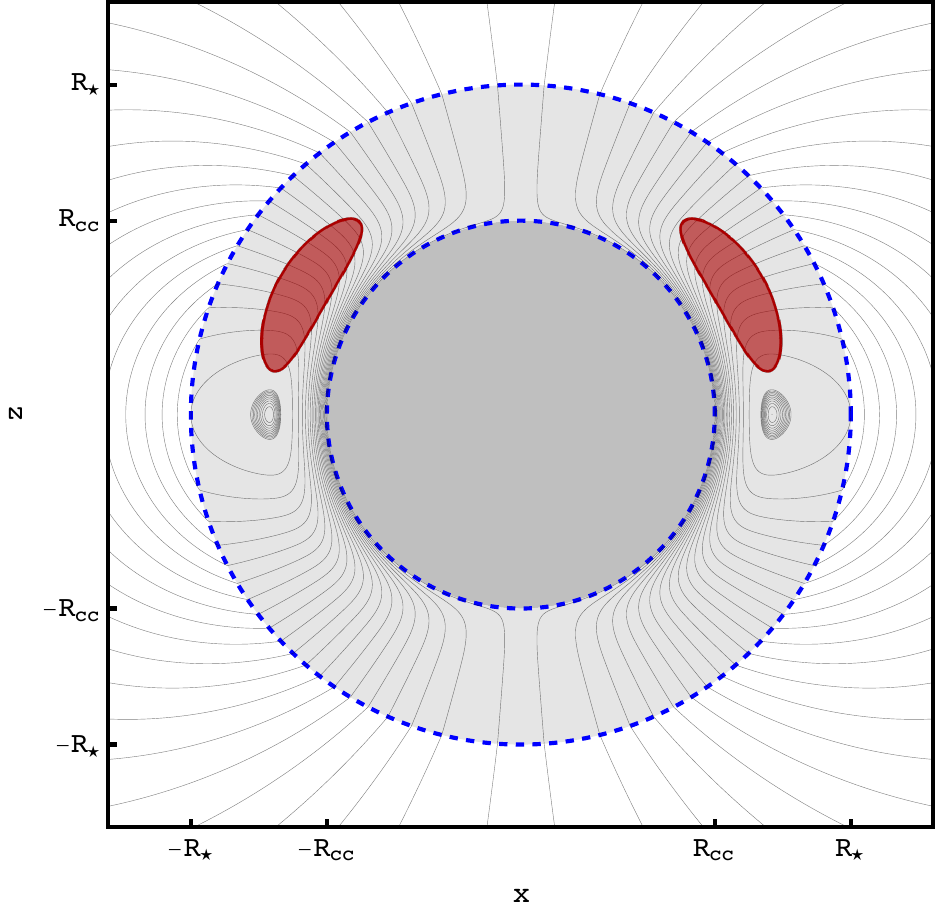}
 \caption{Crust-confined magnetic geometry, in a Cartesian-like projection, for a $M=1.4 M_{\odot}$ star used to supplement the TOV background. 
 The dashed, blue curves delimit the computational domain ($\Rcc \leq r \leq \Rs$) while the solid red region encased within represents an initial island that sources the magnetoelastic perturbations.
 The crust has been stretched by a factor 5 for visibility.
 }
  \label{fig:fieldlines}
\end{figure}
%%%%%%%%%%%%

For the dipolar stream function $a_1(r)$ defining the azimuthal component of the four-potential, $A_\phi=-a_1(r)\sin^2\theta$, we have \cite[see, e.g.,][]{sotani2007}
\begin{equation} \label{eq:GS}
0 = \frac{d^2a_1}{dr^2} + \left(\Phi'-\Lambda'\right) \frac{da_1}{dr} - \frac{2e^{2\Lambda}}{r^2}a_1 + 4\pi e^{2\Lambda}j(r).
\end{equation}
We adopt the simple \citet{konno99} current profile $j(r)\propto r^2(\epsilon+p)$, with the constant of proportionality fixed by the dipole moment as measured at infinity, $\mu_b$.
In any case, the local orthonormal frame vector components of the magnetic field are given by $B^{\hat r} = 2a_1 \cos\theta /r^2$ and $B^{\hat \theta} = -{e^{-\Lambda}a_1'(r)}\sin\theta/r$.
At the crust-core interface we impose
\begin{equation} \label{eq:a1inner}
    a_1(\Rcc)=0,
\end{equation}
so that $B^{\hat r}(\Rcc)=0$. 
We do not require $a_1'(\Rcc)=0$, however, meaning that the tangential field on the interface remains finite and supported by a current sheet. 
Such crust-confined configurations are commonly used as (idealised) initial conditions in magnetothermal/Hall evolutions by appealing to Meissner expulsions, thereby reducing magnetic lengthscales so that Joule heating is more efficient \cite[see, e.g.,][for discussions]{pv19,fuji14,lan25}.
Importantly, if such a field is realised in nature then the crust and core cannot couple through torsional modes as neither elastic nor magnetic traction extend past $R_{\rm cc}$.
The absence of a core in our model is therefore not a further approximation but actually required because of our crust-confined poloidal field \eqref{eq:a1inner} and symmetries. 
Whether this is realistic is, of course, debatable \cite[cf.][]{gabler12}.

At the stellar surface the equilibrium field is matched smoothly to the relativistic vacuum dipole,
\begin{equation} \label{eq:vacuum_dipole}
    a_1^{\rm ext} = -\frac{3\mu_b}{8M^3}r^2 \left[ \ln\left(1-\frac{2M}{r}\right) + \frac{2M}{r} + \frac{2M^2}{r^2} \right],
\end{equation}
which implies that the polar field strength reads
\begin{equation}  \label{eq:Bpole_mub}
    B_{p} = -\frac{3\mu_b}{4M^3} \left[ \ln\left(1-\frac{2M}{\Rs}\right) + \frac{2M}{\Rs} + \frac{2M^2}{\Rs^2} \right].
\end{equation}
A given configuration is thus fixed simply by a choice for $B_{p}$.
Although not directly relevant to our problem, expression \eqref{eq:Bpole_mub} conveys an important point through its GR weightings.
The polar-field strengths inferred for magnetars via timing data are typically quoted using a Newtonian expression ($M/\Rs \ll 1$), which can reduce by a factor $\lesssim 4$ when redshift, magnetospheric drag, and realistic EOS are all accounted for \cite[see equation 17 in][]{stef25}.
Thus, although we stipulate a `modest' polar dipole strength of $B_{p} = 10^{14}$~G in most of our simulations, such a value is probably closer to reality than $\sim 10^{15}$~G for objects like SGR 1900+14.
Although we emphasise again that our goal here is not to attempt an accurate and detailed matching with observed QPO spectra, it is worth highlighting that \cite{pumpe17} argue that the QPOs in SGR 1806--20 seem to favour relatively weak fields with $B \gtrsim 10^{14}$~G.

%We solve the relativistic dipole Grad–Shafranov equation as the sum of a homogeneous solution and a current-sourced particular solution, both integrated from the crust–core boundary, adjusting their relative amplitudes to match the  vacuum dipole at the surface

The resulting magnetic geometry described above, and solved for using a standard shooting routine, is shown in Figure~\ref{fig:fieldlines}.
Because of the crust-confined nature of the geometry, the field is locally stronger near the equator; we find $B_{\rm max}/B_{p} \approx 48$, meaning that if $B_{p} = 10^{14}$~G then the maximum is $\approx 5 \times 10^{15}$~G, which better matches heating requirements set by magnetar X-ray luminosities \citep{pv19}.
Overlaid in the figure for illustration is a large, initialised island where $\mu_{\rm eff}=0$. 
The initial kick for the magnetoelastic oscillations is provided by integrating the elastic energy stored there and distributing it; see Sec.~\ref{sec:initial_conditions}.

%%%%%%%%%%%%%%%%%%%%%%%%%%%%%%%%%%%%%%%%%%%%%%%%%%%%%%%%%%%%%%%%
\section{Numerical Methods}
\label{sec:numerical_methods}
%%%%%%%%%%%%%%%%%%%%%%%%%%%%%%%%%%%%%%%%%%%%%%%%%%%%%%%%%%%%%%%%

We solve equation \eqref{eq:torsional_evolution} on a uniform, cell-centred grid in $r$ and $x$ with $N_r\times N_x$ cells of spacing
\begin{equation} \label{eq:res}
    \Delta r=\frac{\Rs-\Rcc}{N_r}, \quad \Delta x=\frac{2}{N_x},
\end{equation}
such that
\begin{equation} \label{eq:grid}
    r_i=\Rcc+\left(i-\frac12\right)\Delta r, \quad x_j=-1+\left(j-\frac12\right)\Delta x .
\end{equation}
Where no confusion can arise with spatial indices, we use the subscripts $i$ and $j$ to denote the value of a quantity in the appropriate cell. 
As mentioned above, the use of a uniform $x$ grid avoids placing cell centres on the polar singularities.
It has the secondary benefit of giving finer resolution towards the equator, where the background dipole is strongest (see Fig.~\ref{fig:fieldlines}).
%The spatial discretisation follows directly from the divergence form of .  
Transport terms in equation \eqref{eq:torsional_evolution} are evaluated on the appropriate \emph{cell faces}, and the contributions from the faces surrounding a cell determine the local restoring force.  
In this way the traction leaving one cell is associated with the same face as that entering its neighbour.
Importantly, although we only evolve the displacement at cell centres, some quantities are evaluated on shared faces because that is where shearing occurs.
The strategy is thus a finite-difference scheme but with a subroutine that evaluates certain quantities along faces to shutoff traction if a failure is detected.

Spatial derivatives normal to a face are obtained from the two cells that share it. 
For example,
\begin{equation} \label{eq:face_differences}
\left.Y_{,r}\right|_{i+1/2,j} = \frac{Y_{i+1,j}-Y_{i,j}}{\Delta r},
\end{equation}
with the analogous construction on angular faces. 
The background geometric and magnetic components appearing in equation~\eqref{eq:torsional_evolution} are evaluated at the corresponding face positions.
For the radial elastic coupling, the shear modulus at a common face is represented by the harmonic mean,
\begin{equation} \label{eq:harmonic_mean}
\mu_{i+1/2} = \frac{2\mu_i\mu_{i+1}}{\mu_i+\mu_{i+1}}.
\end{equation}
Expression \eqref{eq:harmonic_mean} encodes an appropriate `series average' for adjacent layers that ensures that elastic traction is not transmitted across faces adjoining a failed cell.  
Since the background shear modulus depends only on radius, angular faces require no additional interpolation of $\mu$, although their elastic coupling is likewise removed if a neighbouring cell has failed.

%%%%%% FIGURE %%%%%%
\begin{figure}
\centering
 \includegraphics[width=0.35\textwidth]{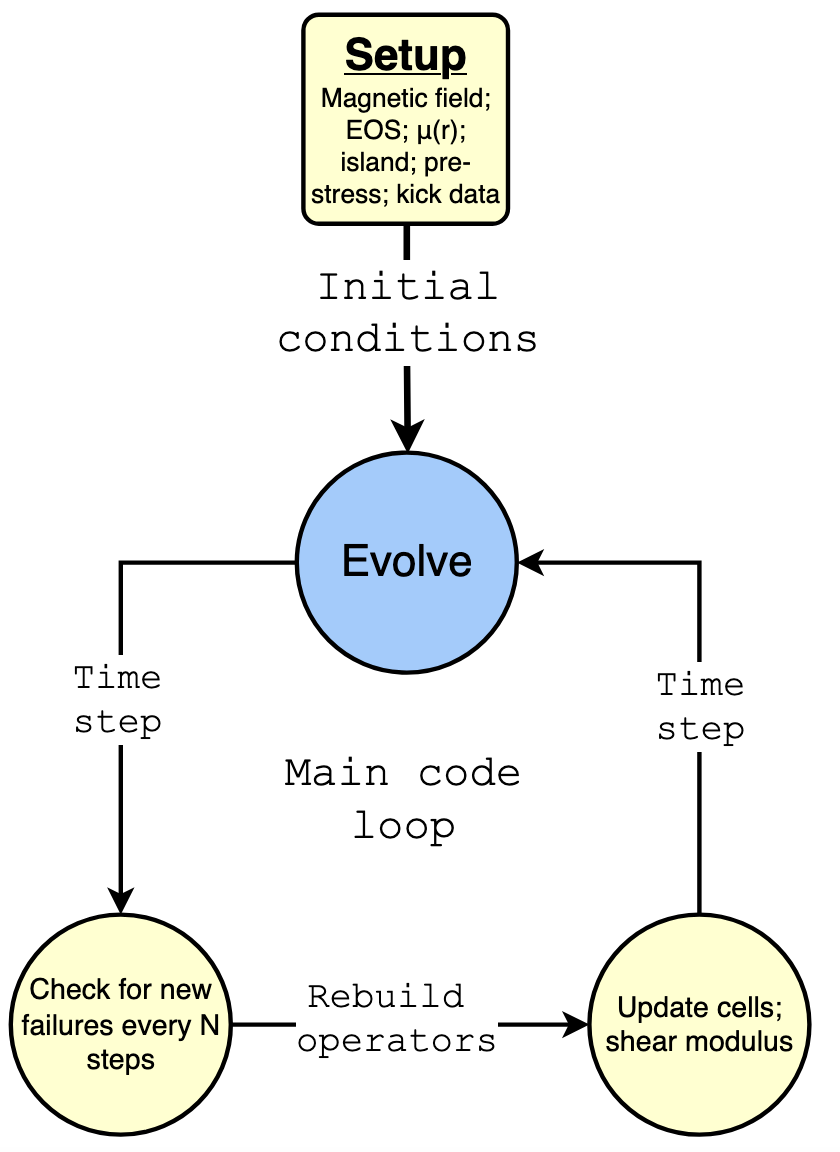}
 \caption{Schematic depicting the main features of the core code loop used throughout this study.}
  \label{fig:schematic}
\end{figure}
%%%%%%%%%%%%

Because the magnetic restoring force contains both radial and angular fluxes, the field-line derivative $\nabla_{\rm B}Y$ defining $\delta B^\phi$ in equation~\eqref{eq:dBphi} is required on two different sets of faces. 
On a radial face, the radial derivative is obtained directly from the two cells separated by that face, while the angular derivative is evaluated at the neighbouring cell centres and averaged to the same location.  
On an angular face the roles are reversed: $Y_{,x}$ is obtained across the face and the neighbouring values of $Y_{,r}$ are averaged onto it.  
This reversal is necessary because the derivative normal to a given face is defined by the two cells that share it, whereas the horizontal derivative must be brought to that same location.
The two face-based evaluations then supply the magnetic contributions in the discretised version of equation \eqref{eq:torsional_evolution}.  
They are combined so that the magnetic coupling between neighbouring cells is reciprocal and no preferred direction is introduced.
The transverse cell-centred derivatives use second-order centred differences in the interior and second-order one-sided differences
in numerical corners.  
The cell-centred restoring terms are then passed to the time stepper detailed below.

%%%%%%%%%%%%%%%%%%%%%%%%%%%%%%%%%%%%%%%%%%%%%%%%%%%%%%%%%%%%%%%%
\subsection{Time stepping and failure updating}
\label{sec:time_stepping}
%%%%%%%%%%%%%%%%%%%%%%%%%%%%%%%%%%%%%%%%%%%%%%%%%%%%%%%%%%%%%%%%

After spatial discretisation, equation~\eqref{eq:torsional_evolution} essentially reads
\begin{equation} \label{eq:semidiscrete}
    \mathcal I_{ij}\,\ddot Y_{ij} = \mathcal S_{ij}(Y,\partial Y),
\end{equation}
for an inertia coefficient $\mathcal{I}_{ij}$ and operator $\mathcal{S}_{ij}$ depending on $Y$ and its derivatives.
%At a given time $t_n=n\Delta t$, the complete numerical state is the set $Y^n_{ij}=Y(t_n,r_i,x_j)$ and its derivatives.
To step forward a unit in time, $t_{n+1} - t_{n} = \Delta t$, we use a centred leapfrog scheme
\begin{equation} \label{eq:leapfrog}
    Y_{ij}^{n+1} = 2Y_{ij}^{n} - Y_{ij}^{n-1} + \frac{\Delta t^2}{\mathcal I_{ij}} \mathcal S_{ij}(Y,\partial Y),
\end{equation}
where we use the shorthand $Y^n_{ij}=Y(t_n,r_i,x_j)$; the velocity is reconstructed via
\begin{equation} \label{eq:centered_velocity}
\dot Y_{ij}^{n} = \frac{Y_{ij}^{n+1}-Y_{ij}^{n-1}}{2\Delta t}.
\end{equation}
Given some initial data $Y^0$ and $\dot Y^0$, the `ghost step' is set through $Y_{ij}^{-1} = Y_{ij}^{0} - \Delta t \dot Y_{ij}^{0} + {\Delta t^2 \mathcal S_{ij}(Y^0,\partial Y^{0})}/{2\mathcal I_{ij}}$.
Although some numerical dispersion is unavoidable, the leapfrog method is useful here because it is both non-dissipative and inexpensive as elastic connectivity changes during a run.
At each step the displacement is first advanced using the elastic state from the preceding time level according to \eqref{eq:leapfrog}.
If that step happens to coincide with a failure check (see below), the stresses are then evaluated from the newly obtained displacement, freshly failed cells are identified, and the elastic face couplings are rebuilt \emph{before} the next leapfrog.  

To choose a suitable timestep, we evaluate the largest local magnetoelastic wave speed on the grid,
\begin{equation} \label{eq:vme}
v_{\mu B} = \left[ \frac{ \mu+B^2/(4\pi)}{\epsilon+p+B^2/(4\pi)} \right]^{1/2},
\end{equation}
and define the GR-weighted maximum coordinate crossing rates ($\Gamma$) throughout the crust via
\begin{equation}
    \Gamma_r = \max_{i,j} \left[ e^{\Phi(r_i)-\Lambda(r_i)} \frac{v_{\mu B}(r_i,x_j)}{\Delta r} \right],
\end{equation}
and
\begin{equation}
\Gamma_x = \max_{i,j} \left[ e^{\Phi(r_i)} \frac{v_{\mu B}(r_i,x_j)\sqrt{1-x_j^2}}{r_i\Delta x} \right].
\end{equation}
We then pick
\begin{equation} \label{eq:cfl_condition}
    \Delta t = \frac{C_{\rm CFL}} {\sqrt{\Gamma_r^2+\Gamma_x^2}},
\end{equation}
for CFL parameter $C_{\rm CFL}$. 
Based on some routine tests of stability (see also Appendix~\ref{sec:resolutiontests}), we find $C_{\rm CFL} = 0.25$ is sufficiently small to not affect the results in any meaningful way. 
For $B_{p} \gtrsim 10^{14}$~G and a $96 \times 448$ resolution, we have $\Delta t \sim \mathcal{O}(30~\text{ns})$.
Importantly, we will see later that fault lines develop as cells fail and have their traction diverted horizontally and it is important therefore that the timestep is small enough to have wavefronts not jump through multiple cells per step.

Any given run proceeds by evolving in time and recomputing stresses by tracking the relevant components of $\delta T^{\rm s}_{\mu \nu}$ in equation \eqref{eq:von_mises}.
It is prohibitively expensive to check these every timestep, and thus we do this only every $N$ steps. 
For all cases shown in this paper we use $N=20$, as reducing this interval to five or ten timesteps produces no significant change in the macroscopic plastic fraction or avalanche/creep morphology.
At each failure check, the dynamical shear stresses are reconstructed from the same face-centred gradients used in the elastic evolution (using expression \ref{eq:harmonic_mean} for the shear modulus).
These updates are added to the hand-set prestress (see Sec.~\ref{sec:initial_conditions}), and the von Mises criterion is checked to determine whether a cell should remain elastic or fail.
A schematic depicting the code flow is shown in Figure~\ref{fig:schematic}.

Once a cell fails, its elastic support is removed for the remainder of the simulation.
That is, we do not consider the possibility of healing or viscoelastic flow here \cite[cf.][]{lg19}.
Neither $Y$ nor $\dot Y$ is reset though, and the cell continues to carry hydromagnetic inertia.
The elastic energy lost when rigidity is removed is therefore treated as having dissipated into unresolved channels (e.g., heat) rather than sourcing more perturbations.
While new kicks could be easily incorporated after each failure, we avoid such complications here to present a scenario where plastic accumulation is only driven by the initial pulse itself.

%%%%%%%%%%%%%%%%%%%%%%%%%%%%%%%%%%%%%%%%%%%%%%%%%%%%%%%%%%%%%%%%
\subsection{Boundary conditions} \label{sec:boundary_conditions}
%%%%%%%%%%%%%%%%%%%%%%%%%%%%%%%%%%%%%%%%%%%%%%%%%%%%%%%%%%%%%%%%

The physical boundary conditions we impose are that of zero traction at the crust-core interface and a torque-free condition at the surface following previous works \cite[see, e.g.,][]{schut83}.
That is, the discrete counterpart of
\begin{equation} \label{eq:stressbc}
    \left.\delta T^{\rm s}_{\hat r\hat\phi}\right|_{r=\Rcc} = \left.\delta T^{\rm s}_{\hat r\hat\phi}\right|_{r=\Rs} =0,
\end{equation}
evaluated using expression \eqref{eq:Trphi}.
Note in particular that these conditions apply when evaluating the radial elastic contribution to the master equation.  
For an interior cell, the stress is measured from the difference between elastic fluxes through the two radial interfaces.  
For the innermost cell, the flux through $\Rcc$ is taken to be zero, while the interior flux is calculated from $Y_{2j}-Y_{1j}$ using the shear-modulus and geometric factors. 
The flux through $\Rs$ is similarly zero, while the interior contribution is calculated from $Y_{N_r,j}-Y_{N_r-1,j}$.  
The \emph{dynamical} boundary fluxes remain zero throughout the evolution to machine precision where we emphasise that \eqref{eq:stressbc} applies to additional stresses and not the prestress profile described in Sec.~\ref{sec:initial_conditions}.
No condition is required at the angular boundaries because the geometrical factors in the fluxes force the angular traction to vanish at the symmetry axis.

The magnetic perturbation is likewise evolved only within the crustal cavity, but is not required to vanish at the surface.
The edge cells remain free to twist the background poloidal field since, in general, $\delta B^{\phi}$ does not vanish there even though we do not involve any magnetospheric degrees of freedom. 

%The magnetic perturbation is likewise evolved only within the crustal domain, but its boundary treatment is not identical to the elastic one since, 

%%%%%%%%%%%%%%%%%%%%%%%%%%%%%%%%%%%%%%%%%%%%%%%%%%%%%%%%%%%%
\subsection{Prestress and initial perturbation}
\label{sec:initial_conditions}
%%%%%%%%%%%%%%%%%%%%%%%%%%%%%%%%%%%%%%%%%%%%%%%%%%%%%%%%%%%%

%%%%%% FIGURE %%%%%%
\begin{figure}
\centering
 \includegraphics[width=0.48\textwidth]{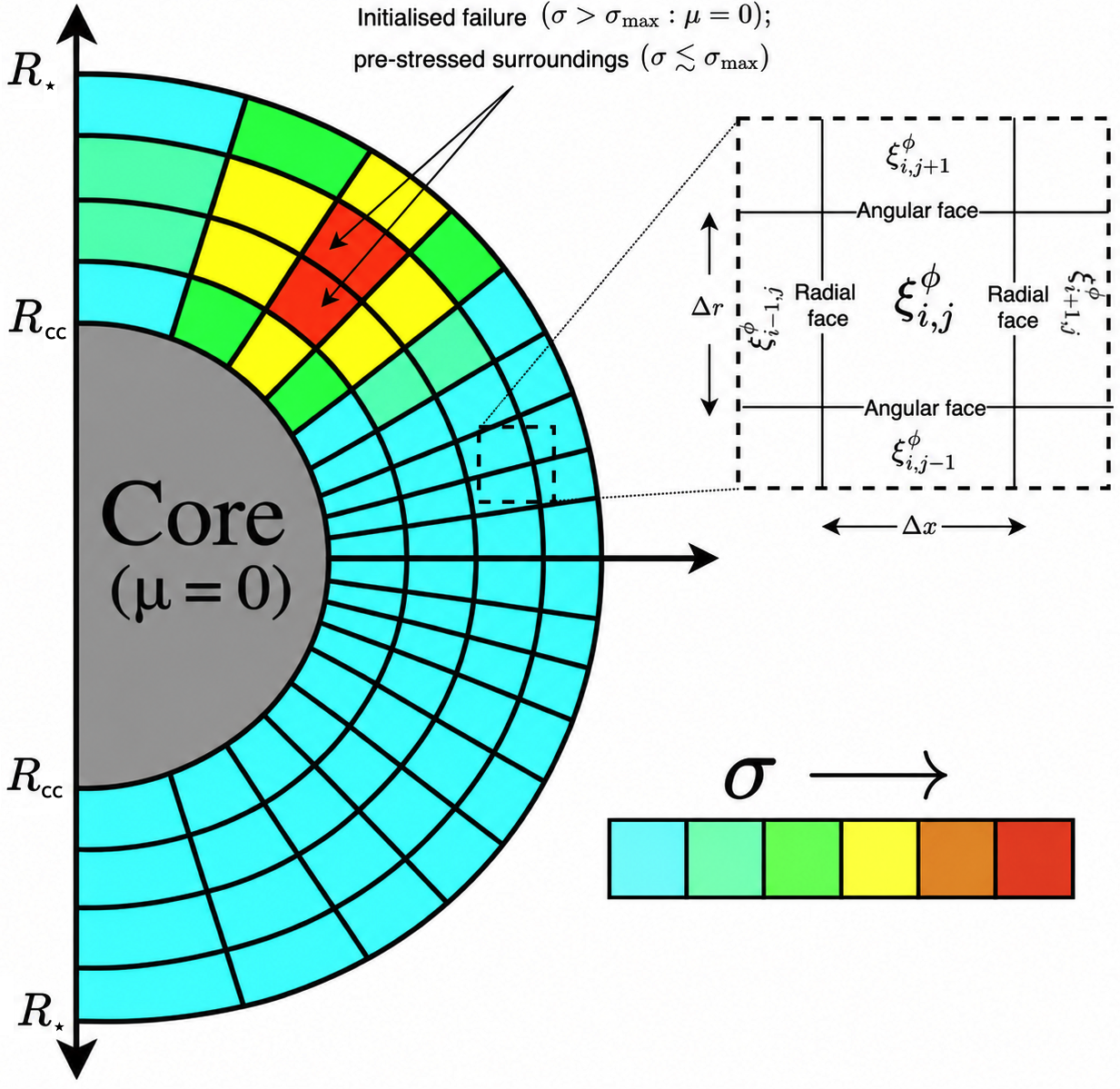}
 \caption{Cartoon depiction of the simulation domain ($\Rcc \leq r \leq R_{\star}, -1 < x < 1$), prestressing, and how the torsional displacement is defined with respect to the grid. 
 Darker shades for grid cells indicate greater prestress: those that are red/brown indicate the initial failure [where $\mu =0$ in the elliptical region \eqref{eq:island_coordinate}], while the surroundings become progressively lighter with distance in accord with the profile \eqref{eq:prestress_profile}. 
 The depicted angular spacing between the cells is intentionally reduced towards the equator, reflecting the nature of the uniform in $x=\cos \theta$ grid.}
  \label{fig:schematic2}
\end{figure}
%%%%%%%%%%%%

The initial event is specified by two distinct ingredients:
(i) a localised velocity perturbation, representing the recoil produced when an initial region loses elastic support, and 
(ii) a prescribed static prestress, determining how close the rest of the crust is to failure.

We fix the initially failed region via an ellipse profile,
\begin{equation} \label{eq:island_coordinate}
s^2(r,\theta) = \left( \frac{r-r_c}{\Delta r_c} \right)^2 + \left( \frac{\theta-\theta_c}{\Delta\theta_c} \right)^2,
\end{equation}
with $s\leq1$ meaning initially failed and $r_{c}, \Delta r_c, \theta_c, \Delta \theta_c$ are parameters associated to the island (e.g., a local or global quake at some position).
The surrounding crust is assigned a prestress that is conveniently expressed as a fraction, $f(s)$, of the local breaking threshold.
We take a Gaussian-like profile for this in the form
\begin{equation} \label{eq:prestress_profile}
f(s>1) = f_{\rm collar} \exp \left[-\frac{(s-1)^2}{2h^2}\right],
\end{equation}
with the prestress placed initially in the $\hat r \hat\phi$-shear component with $T^{\rm s,pre}_{\hat r\hat\phi} = -2\mu(r)\varepsilon_{\rm max}f(s)$.
This gives $\sigma_{\rm VM}^{\rm pre}/\sigma_{\rm yield}=f(s)$, meaning physically that the failed island begins above the adopted breaking threshold while the immediately surrounding `collar' is such that passing waves can, in principle, trigger further yielding if carrying enough energy.
The value of $f_{\rm collar}$ determines how close the surroundings are to failure while $h$ defines an effective lengthscale (see below).
Other choices are possible of course, though expression \eqref{eq:prestress_profile} captures an intuitive case where stresses have built up around a region \cite[e.g., tracing a large-scale Hall wave;][]{brans25} rather than being scattered throughout the crust.
A thorough exploration lies beyond the scope of this paper and will be considered elsewhere.

At the start of any given run we set
\begin{equation}
    Y(0,r,\theta)=0 ,
\end{equation}
but $\dot{Y} \neq 0$ so that the lattice elements are kicked rather than displaced.
This serves as a simple surrogate for the recoil associated with stress release that can excite oscillations.
%The pulse covers both the failed region and its intact surroundings, extending to $s=1.5$, with additional weight around the interface over $0.5<s<1.5$, such that the inertia-weighted angular momenta cancel.
On the other hand, it is easy to see that $Y_{\rm rigid}(t,r,\theta)=\Omega_0 t$ is a solution to equation \eqref{eq:torsional_evolution} which represents a kind of zero mode for rigid rotation.
We want to avoid exciting this component since it contains no oscillatory content \cite[though cf. plastic flow discussions;][]{lg19,coop24}.
%We do this by fixing the relative amplitudes of the island and collar components such that that their contributions to the initial angular momentum cancel.
Numerically this amounts to
\begin{equation} \label{eq:no_rigid_rotation}
    \sum_{ij}\mathcal I_{ij}\dot Y^0_{ij}=0.
\end{equation}
The spatial form of the recoil is chosen through $b(s) = b_{\rm broad}(s) + b_{\rm col}(s)$ with $b_{\rm broad}(s) = \left(1-4s^2/9\right)^3$ for $0<s<1.5$ and $b_{\rm col}(s) = 3\left[1-4(s-1)^2\right]^3$ for $0.5<s<1.5$.
As the naming suggests the first term provides a broad recoil across the failed region and its neighbourhood while the second excites waves around the island interface.  
The cubic form makes the profile and its first two derivatives vanish at the edge, avoiding an acceleration discontinuity.
The initial displacement velocity is then constructed through
\begin{equation} \label{eq:initial_velocity_profile}
    \dot Y(0,r,\theta) = Y_{\rm kick} \frac{e^{\Phi(r)}}{r\sin\theta_c} \left[b^2(r,\theta)-\tilde{C} b(r,\theta) \right],
\end{equation}
where $\tilde{C}$ is a constant chosen such that the sum \eqref{eq:no_rigid_rotation} vanishes.
Again we emphasise that such a profile \eqref{eq:initial_velocity_profile} is somewhat arbitrary, but reasonably captures the physical picture of a quake that imparts a localised recoil both within the failed region and its surroundings without inducing rigid rotation; some testing with different exponents and parameters suggests that the exact choice is not too important.
It is important, however, that the collar perturbation is issued otherwise Alfv{\'e}n modes remain localised to the island and do not couple much to the elastic surroundings (see Sec.~\ref{sec:creep}).

We set the overall amplitude of the kick, $Y_{\rm kick}$, through the elastic energy released from the sourcing quake.
A fraction, $\eta\leq1$, of the elastic energy stored in the failed island is `converted' into kinetic energy through
\begin{equation} \label{eq:kick_efficiency}
    E_{\rm kin,0} = \eta E_{\rm el}.
\end{equation}
Hence, the velocity amplitude is not a free scale but follows from the island geometry and a choice of $\eta$, the value of which is uncertain but could be of order a few percent \cite[see, e.g.,][]{brans20}.
In the context of earthquakes, the amount of elastic energy transferred into seismic waves is thought to be of order $\sim 6\%$ \cite[see chapter 14 in][]{stacey08}.

In the following result sections, we fix a spatial resolution of $N_{r} \times N_{x} = 96 \times 448$ (for a total of 43008 cells) based on a convergence study of failure morphology; see Appendix~\ref{sec:resolutiontests}.
We further fix $\eta = 0.05$, though some tests are provided in Appendix~\ref{sec:fieldstrength} demonstrating that, for instance, as $\eta \to 0$ we monotonically approach a scenario where no further failures occur.
Each run extends up to $t_{\rm max}=2.5$~s, at which time the simulation has approximately saturated.
This time is long enough to capture both avalanching and creep phenomena, while being shorter than characteristic damping timescales associated with viscosity or gravitational radiation-reaction; see Sec.~\ref{sec:fourier}.
A schematic of the computational domain and prestressing, including how the displacement is constructed on cell faces, is depicted in Figure~\ref{fig:schematic2}.

%%%%%%%%%%%%%%%%%%%%%%%%%%%%%%%%%
\section{Results: global quakes} \label{sec:global}
%%%%%%%%%%%%%%%%%%%%%%%%%%%%%%%%%

%%%%%% FIGURE %%%%%%
\begin{figure}
\centering
 \includegraphics[width=0.48\textwidth]{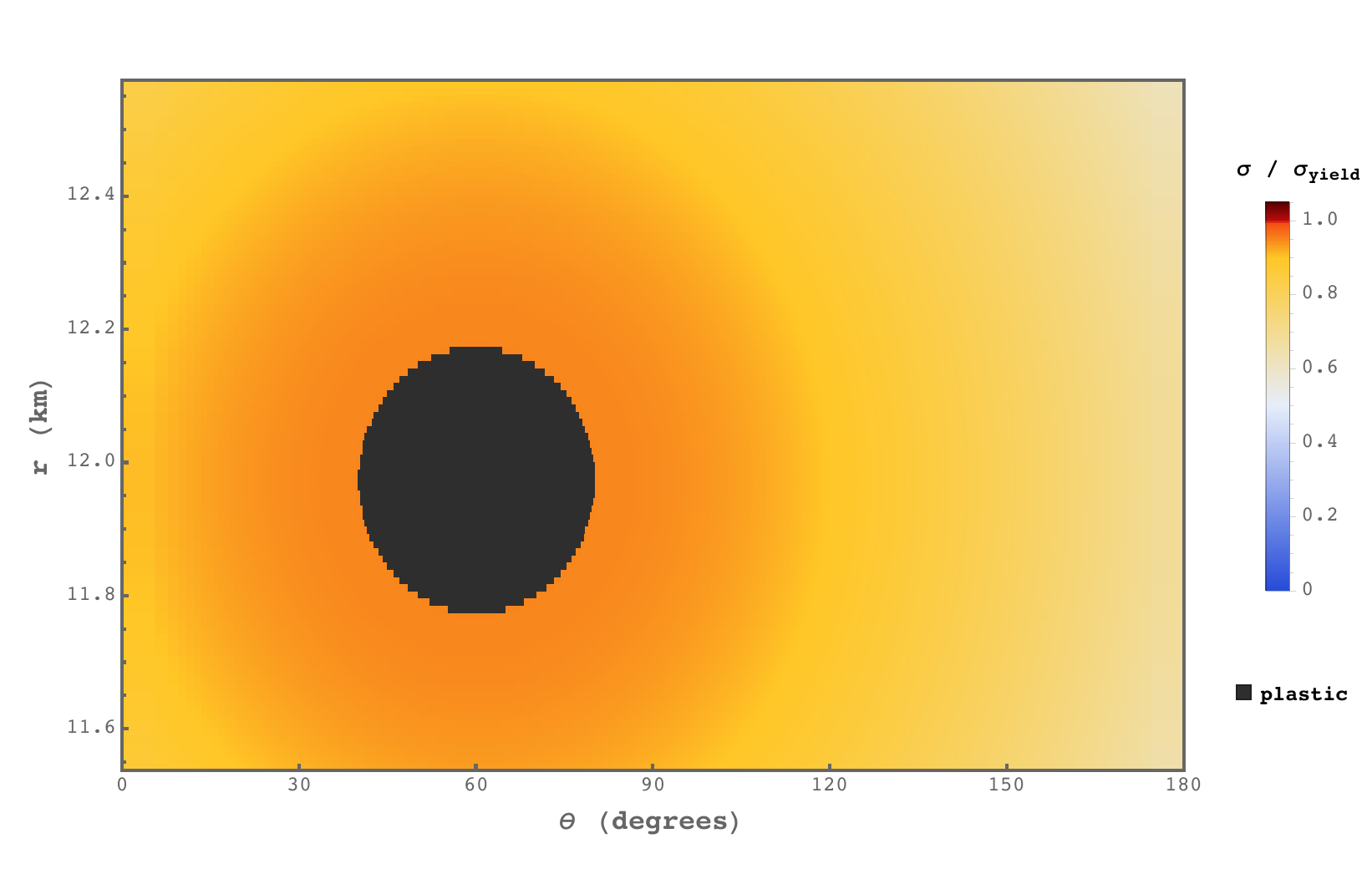}
 \caption{A depiction of the initial failure geometry in polar coordinates and pre-stressing for the global quake.
 Cells which are completely opaque carry no elastic traction, while redder shades correspond to prestress levels approaching the von Mises limit \eqref{eq:failure_criterion}.}
  \label{fig:initfail}
\end{figure}
%%%%%%%%%%%%

In this section, we set up a big initial island to mimic a global event, where a significant fraction ($\lesssim 10\%$) of the crust is initially overstressed and a sizeable kick issues forth to drive oscillations.
For this model, we choose $r_c=\Rs-0.60$~km and $\theta_c=60^\circ$ with respective thicknesses of $\Delta r_c=0.20$~km and $\Delta\theta_c=20^\circ$.
The adopted prestress profiles use $f_{\rm collar}=0.95$ and $h=6$. 
A depiction of the initial island and the pre-stressed surroundings is shown in Figure~\ref{fig:initfail}.

%%%%%%%%%%%%%%%%%%%%%%%%%%%%%%%%%%%%%%%%%%%%%%%%%%%%%%%%%%%%%%%%%%%%%%%%
\subsection{Failure progression}
%%%%%%%%%%%%%%%%%%%%%%%%%%%%%%%%%%%%%%%%%%%%%%%%%%%%%%%%%%%%%%%%%%%%%%%%

%[change to tau50 and note its not exactly that.]

\begin{figure}
\centering
 \includegraphics[width=0.48\textwidth]{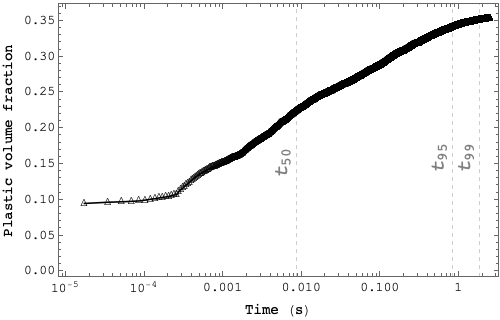}
 \caption{Plastic volume fraction, as a function of time, $f_{\rm fail}$ from expression \eqref{eq:plastic_fraction} , for the global quake model detailed in text.
 Overlaid by vertical lines (from left to right) are values of $t_{50}$, $t_{95}$, and $t_{99}$, with the respective subscript denoting  the time by which the total fraction of failures by simulation end have been achieved.
 }
  \label{fig:globalacc}
\end{figure}

%[use the proper ffail. note its volume not cell counts.]
Initially, $\approx 9\%$ of the total volume is set to failed in accord with the choices made above (3879 out of 43008 cells).
As the simulation progresses, further failures are accumulated as magnetoelastic modes exert stress.
The failed volume is measured using the proper spatial volume of the grid,
\begin{equation}
    V_{\rm fail} = 2\pi \sum_{i,j\in{\rm fail}} e^{\Lambda_i} r_i^2 \Delta r \Delta x ,
\end{equation}
so that we what we call the failure fraction is given by
\begin{equation} \label{eq:plastic_fraction}
    f_{\rm fail} = \frac{V_{\rm fail}}{V_{\rm crust}} ,
\end{equation}
with $V_{\rm crust}$ defined in the obvious way.
The value of $f_{\rm fail}$, as a function of time, is depicted in Figure~\ref{fig:globalacc}. 
Noting the log scale of the horizontal axis, two distinct phases can be seen to occur.
The first involves a prompt avalanche: within $\approx 8$~ms over half the total failure count by simulation end has occurred (denoted by $t_{50}$ and depicted by a dashed, vertical line). 
More precisely, we define $t_{50}$ implicitly as $f_{\rm fail}(t_{50})=f_{\rm fail}(0)+\tfrac{1}{2}[f_{\rm fail}(t_{\rm max})-f_{\rm fail}(0)]$.
By the end of the run, we have $f_{\rm fail}(2.5~\text{s}) = 35.4\%$, indicating a factor $\sim 4$ increase over the initial plastic volume.

%%%%%% FIGURE %%%%%%
%%%%%%%%%%%%

%%%%%%%%%%%%%%%%%%%%%%%%%%%%%%%%%%%%%%%%%%%%%%%%%%%%%%%%%%%%%%%%%%%%%%%%
\subsubsection{Prompt avalanching} \label{sec:avalancheglobal}
%%%%%%%%%%%%%%%%%%%%%%%%%%%%%%%%%%%%%%%%%%%%%%%%%%%%%%%%%%%%%%%%%%%%%%%%

The fact that the failure fraction increases by a factor $\gtrsim 2$ within $\sim 10$~ms is due to the large kick amplitude and high $f_{\rm collar}$.
Picking $\eta = 0.05$ results in a significant amount of energy being distributed to both collar and island via the initial velocity \eqref{eq:initial_velocity_profile}.
Integrating the elastic energy of the island we find $E_{\rm island} = 3.06 \times 10^{44}$~erg, meaning that $E_{\rm kin,0} = 1.53 \times 10^{43}$~erg is distributed approximately equally between the kick components.
The island energy choice is motivated by the isotropic energy output from the giant flares from SGRs 0526--66 and 1900+14, estimated at $\sim 5 \times 10^{44}$~erg \citep{maz79} and $\sim 10^{44}$~erg \citep{hurley99}, respectively.
Given that the surroundings are heavily prestressed with $f_{\rm collar} = 0.95$, such a kinetic energy is sufficient to overwhelm neighbouring parts of the crust.

\begin{figure}
\centering
 \includegraphics[width=0.48\textwidth]{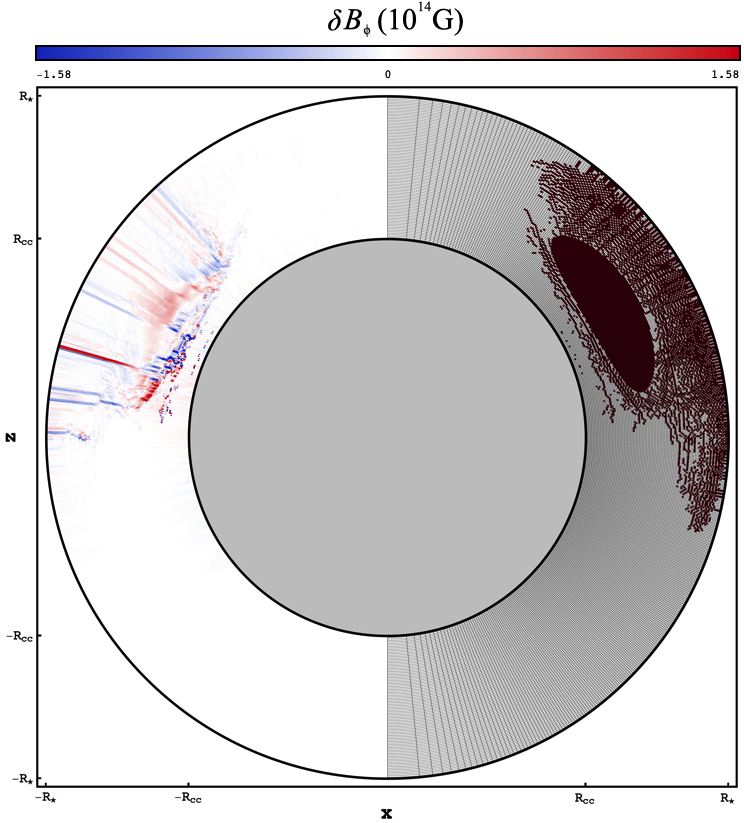}
 \caption{Snapshot depicting the strength and structure of magnetic perturbations (left hemispheres, colour-bar) and in-cell failure geometry (right hemispheres) at time $t=t_{50}$, as marked in Fig.~\ref{fig:globalacc}.
 The crust is stretched for visibility, though notches along the axes mark the relevant radii.
 }
  \label{fig:globalsnap50}
\end{figure}

Interestingly, however, failures do not simply encase the island and fragmentary patterns emerge.
To better understand the failure morphology itself, a snapshot at $t=t_{50}$ for both the profile of $\delta B_{\phi}$ (left hemisphere) and failed cell structure (right) is shown in Figure~\ref{fig:globalsnap50}.
A non-trivial, anisotropic failure pattern can be seen, primarily concentrated near the surface as the shear modulus is weaker there, with a network of cells remaining elastically intact.
Aside from the plastic web, some connected fault lines are also visible beneath the island demonstrating that `knock-on' effects can also occur with the loss of a neighbour's elasticity promoting further failure.
The fact that the failure geometry is itself not symmetric about the island is not surprising given the angular weighting of operators within equation \eqref{eq:torsional_evolution}, though the filamentary progression can be traced back to how perturbations develop.

%%%%%% FIGURE %%%%%%
%%%%%%%%%%%%

The narrow, angular structure in $\delta B_{\hat{\phi}}$ visible in Fig.~\ref{fig:globalsnap50} can be understood through how the initial disturbance travels through the crust.
The kick launches torsional waves both radially and across latitude, which subsequently reflect and traverse regions with different restoring forces. 
Since the effective inertia, $p + \epsilon + B^2/4\pi$, varies throughout the crust, different parts of the disturbance acquire different propagation speeds and oscillation phases. 
Regions that initially twist together can therefore become progressively out of phase, eventually producing alternating signs of the displacement over increasingly small angular separations. 
Such a phase-mixing mechanism was also observed by \cite{gabler12,gabler16}, who showed that spatial gradients of shear/Alfv{\'e}n waves tend to increase through accumulated timing differences \cite[see also][]{HeyvaertsPriest83,lev15}.
In our case with islands this phase mixing is even more pronounced because elastic shear is not transmitted across failed zones, producing alternating bands of $\delta B$ over increasingly small angular separations as disturbance scatter (which is why a large $N_{x}$ is needed; see Appendix~\ref{sec:resolutiontests}).
Reflections additionally allow outgoing and returning disturbances to interfere constructively, increasing stress at crests.
In a nutshell, faster returning waves from the inelastic regions mix with slower outgoing ones to decohere the perturbations and thus fragment the crust as narrow wavefronts overstress cells.

Equatorial migration (see also Sec.~\ref{sec:creep}) likely follows from the same loss of synchronisation.
As the initial disturbance travels and reflects within the crust, neighbouring twists decohere.
This implies that the location of perturbation crests shifts between neighbouring latitudes over time, appearing as a horizontally-moving band even though the motion is azimuthal \citep[see also][]{lev16}.
\cite{Kaneko2015} describe a similar phenomenon in solar-plasma models, where MHD waves with spatially varying periods produced apparently moving fronts without any actual propagation of energy across magnetic surfaces.
If successive maxima drift closer to the equator (pole), the perturbation belt appears to drift towards the equator (pole).
In fact, both forms of drift are visible but the equatorial one appears stronger, likely due to a combination of the more intense background field there (see Fig.~\ref{fig:fieldlines}), the $1-x^2$ weightings in equation \eqref{eq:torsional_evolution}, and island position ($\theta_c = \pi/3$).

%%%%%%%%%%%%%%%%%%%%%%%%%%%%%%%%%%%%%%%%%%%%%%%%%%%%%%%%%%%%%%%%%%%%%%%%
\subsubsection{Creeping failures} \label{sec:creep}
%%%%%%%%%%%%%%%%%%%%%%%%%%%%%%%%%%%%%%%%%%%%%%%%%%%%%%%%%%%%%%%%%%%%%%%%

\begin{figure*}
\centering
 \includegraphics[width=\textwidth]{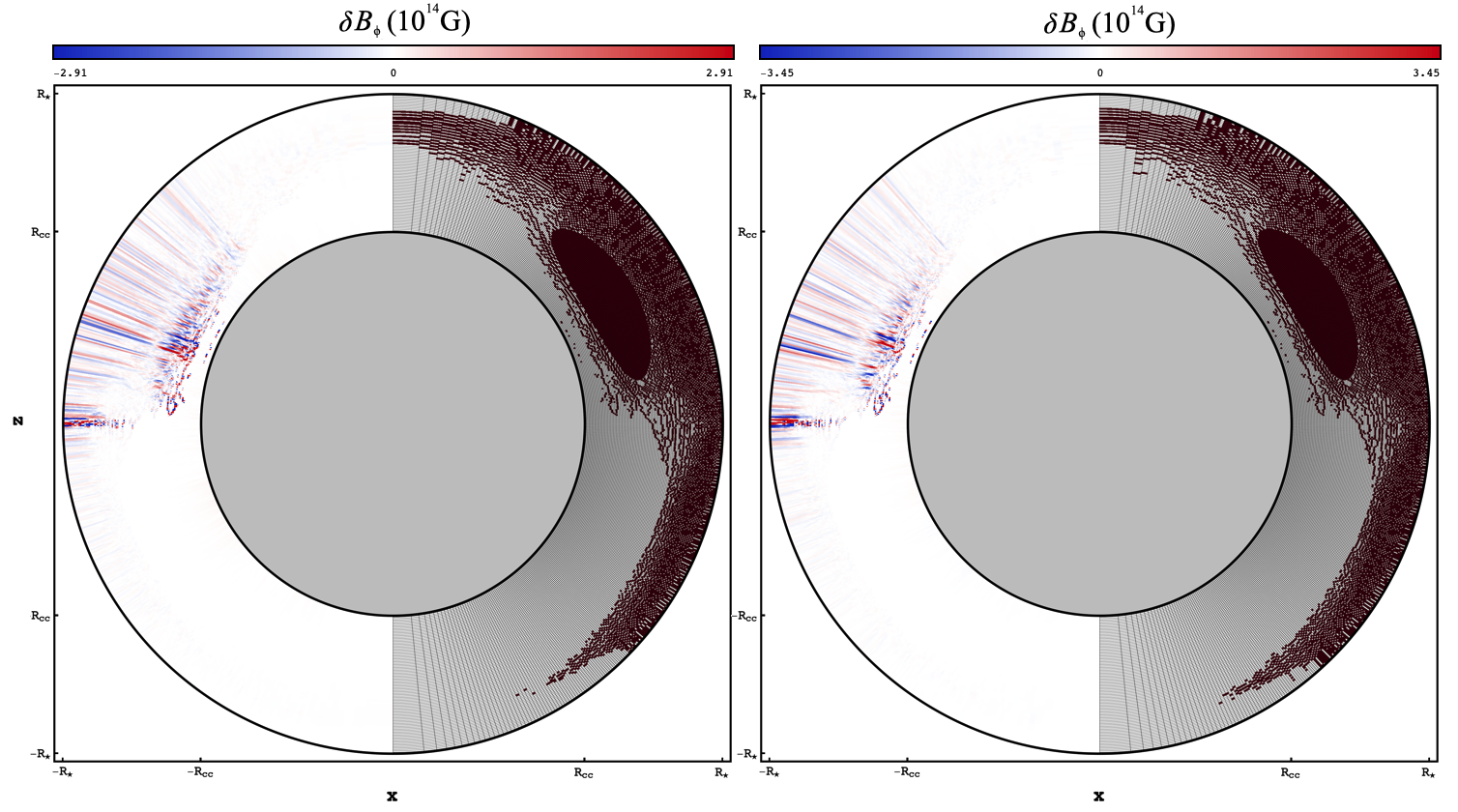}
 \caption{Similar to Fig.~\ref{fig:globalsnap50}, but instead for $t=t_{95}$ (left) and $t=t_{99}$ (right).}
  \label{fig:globalsnaps95}
\end{figure*}

After the prompt phase, failures continue to accumulate slowly over the next $\gtrsim 2$~seconds.
Such increases are steady though eventually saturate, as visible in Figure~\ref{fig:globalacc} where we see that $95\%$ of the failures occur by $\sim 1$~s and $99\%$ by $\sim 1.8$~s out of the 2.5~s run.
Snapshots of the perturbation and failure geometry, similar to Fig.~\ref{fig:globalsnap50} but for $t=t_{95}$ (left) and $t=t_{99}$ (right), are shown in Figure~\ref{fig:globalsnaps95}.
  
%%%%%% FIGURE %%%%%%
%%%%%%%%%%%%

%$t=t_{95}$ (middle), and $t=t_{99}$ (right) 

Delayed failure and further drifting can arise from constructive interference between long-lived modes. 
To see this, consider schematically two waves contributing (at fixed position) to the shear stress through
\begin{equation}
    T^{\rm s,dyn}_{\hat r\hat\phi}(t) = A_1\cos(\omega_1 t+\varphi_1) + A_2\cos(\omega_2 t+\varphi_2),
\end{equation}
where the $A$, $\omega$, and $\varphi$ denote amplitudes, angular frequencies, and phases, respectively.
For approximately constant $A$ and $\omega$, the amplitude in the interference `envelope' is $\approx \left[ A_1^2 + A_2^2 + 2A_1A_2\cos\Delta\varphi(t) \right]^{1/2}$ with $\Delta\varphi(t)=(\omega_1 - \omega_2)t+\varphi_1-\varphi_2$. 
The envelope achieves a maximum of $A_1+A_2$ at alignment and a minimum $|A_1-A_2|$ at destructive maximum.
Constructive alignments therefore occur on the beat timescale, $T_{\rm beat} \sim |\omega_1-\omega_2|^{-1}$, which can greatly exceed the respective oscillation periods (e.g., if $\omega_{1} \sim \omega_{2} \pm 2\pi$~Hz then $T_{\rm beat} \sim 1$~s). 
%It describes successive maxima of the envelope, not necessarily exact simultaneous positive crests of both stresses.  
Where the modes have overlapping shear patterns, these intervals of reinforcement can raise the total stress above the failure threshold when either alone would be futile. 
Such intermittent threshold crossings may explain the gradual failing that we refer to as `creep', which can also develop dynamically as the various $A$ and $\omega$ adjust as elastic support wanes and magnetic perturbations grow.

Indeed, the secular increase in $|\delta B_{\phi}|$ between the panels (note the colour ranges are different) is likely due to how energy is partitioned.  
The initial pulse is kinetic, and magnetic twisting develops as the crust is displaced. 
Kinetic energy subsequently exchanges with elastic and magnetic energy: when material fails, the removal of elastic support naturally shifts this partition and can funnel energy into the magnetic channel despite the initial energy being fixed by $E_{\rm kin,0}$.
We have validated that the total energy, $E_{\rm pert}=E_{\rm kin}+E_{\rm el}+E_{\delta B}$, is conserved \emph{between} failures but notably \emph{decreases} when a new cell fails because $E_{\rm el}$ drops and we do not redistribute that energy.
A growing magnetic energy is not attributable to failures leaking energy therefore.
Although this appears to be a genuine feature of the model, it has the unfortunate consequence of allowing $|\delta B/B|$ to no longer be much less than unity at late times: the maximum $|\delta B_{\hat \phi}|$ is a factor $\gtrsim 2$ larger by $t_{99}$ relative to $t_{50}$ (compare colourbars on Figs.~\ref{fig:globalsnap50} and \ref{fig:globalsnaps95}).
Although we still have $E_{\delta B} \lesssim E_{B}$ by simulation end, quadratic perturbation theory \citep{pl25} or full GRMHD may therefore be required to model creep for $t \gtrsim 1$~s and our model should be considered tentative.

\begin{figure}
\centering
 \includegraphics[width=0.48\textwidth]{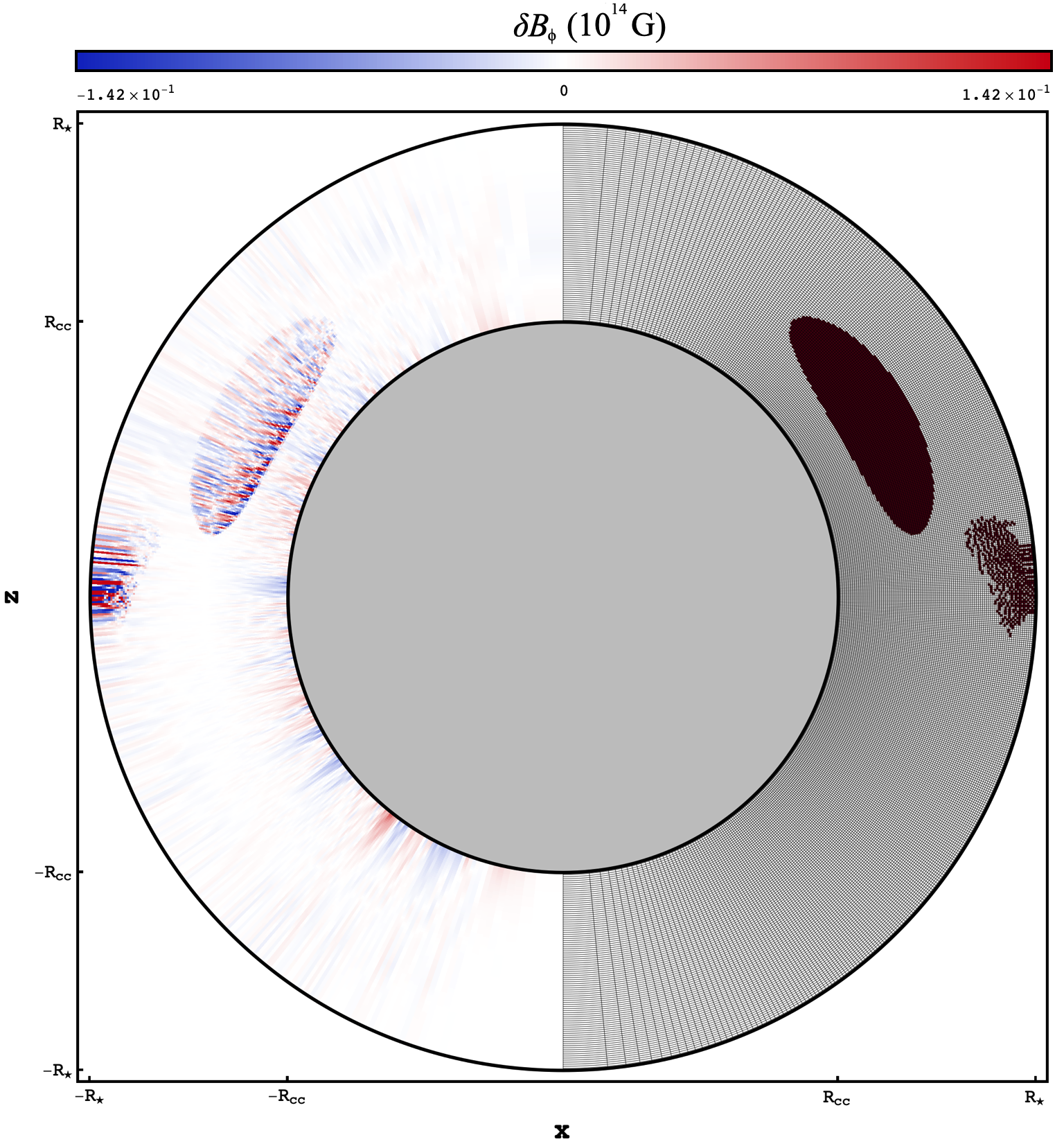}
 \caption{Similar to Fig.~\ref{fig:globalsnap50} but at $t=0.3$~s for a star with a stronger field, $B_{p} = 4 \times 10^{14}$~G, and kicked with an efficiency \eqref{eq:kick_efficiency} lower by a factor 100.
 }
  \label{fig:weakerkick}
\end{figure}

More generally, the representative examples presented in this paper, using an efficiency of $\eta = 0.05$, are on the extreme end. 
The elastic energy imparted to the kick may be smaller in reality, which would limit the degree of failure development.
If the field is much stronger than $10^{14}$~G the preferential excitation of Alfv{\'e}n modes also reduces the overall elastic stress.
We show one case here with $\eta = 5 \times 10^{-4}$ (i.e., a factor 100 reduction) and $B_{p} = 4 \times 10^{14}$~G but otherwise identical macroscopics (further examples are provided in Appendix~\ref{sec:fieldstrength}).
In this case, the energy imparted to the kick is a modest $E_{\rm kin,0} = 1.5 \times 10^{41}$~erg.
A snapshot, similar to Fig.~\ref{fig:globalsnap50} but at $t = 0.3$~s and for this case, is shown in Figure~\ref{fig:weakerkick}.
We find that $f_{\rm fail}(0.3~\text{s}) = 0.105$, a factor $\sim 2$ smaller than the large-kick case (Fig.~\ref{fig:globalacc}).
Moreover, the bulk of the magnetic perturbations are confined to the initial island as we have $\langle B^2 / 4 \pi \mu \rangle \gg 1$ and the magnetic tension limits differential twisting and coupling \cite[see][for discussions]{gabler12,lev15}.
Nevertheless, streaks at all latitudes can be seen showing that some waves also extend throughout the domain even though most failures occur near the surface around the equator (where $B \approx 2\times10^{16}$~G).

%%%%%% FIGURE %%%%%%
%%%%%%%%%%%%

%%%%%%%%%%%%%%%%%%%%%%%%%%%%%%%%%%%%%%%%%%%%%%%%%%%%%%%%%%%%%%%%%%%%%%%%
\subsection{Global spectra} \label{sec:fourier}
%%%%%%%%%%%%%%%%%%%%%%%%%%%%%%%%%%%%%%%%%%%%%%%%%%%%%%%%%%%%%%%%%%%%%%%%

There is a general issue of `mode' identification in the problem at hand.
As the presence of the initial plastic island already leads to mixing and dispersion, it becomes difficult to pose questions such as \emph{how much did a given oscillation drift in frequency due to island creep}?
This is compounded by the fact that the avalanche takes place within $\sim 10$~ms, which is faster than the oscillation period of low-node torsional modes ($\lesssim 0.1$~s). 
To that degree, if one wishes to interpret QPO frequency drifts via plastic development one must remain agnostic to mode labels and rely only on the amplitudes and frequencies that cells experience as waves pass.
In this section, we present Fourier spectra computed over the entire simulation. 
Windowing is introduced in Sec.~\ref{sec:dynamic} to try and track drifts.

To characterise oscillations without storing the full displacement field at every timestep, we record $\xi^{\hat \phi}$ and its derivatives at a set of fixed `observer' locations in intervals of $\Delta t_{\rm s} = 500 \Delta t$ throughout the crust.
The radial positions are specified by some depths together with a set of angles, for a total of $N_{\rm obs} = 44$ evenly-spaced observers throughout both hemispheres, where state values are stored during a run.
Sampling many locations is important because the amplitude of a `mode' can vary strongly throughout the domain.
For example, a weak signal at one observer may simply indicate that the observer lies close to a node -- or that waves have yet to reach them -- and inaccurately suggest an absence of oscillations.
Numerical data at each observer are analysed separately before their spectral power is combined with other observers to further avoid cancellation between displacements of opposite phase. 

For an interval containing $J=t_{\rm max}/\Delta t_{\rm s}$ recorded samples, we subtract the best-fitting straight line from the recorded displacement over the analysed interval.
This is again related to the problem of the zero mode ($Y_{a} \sim \Omega_0 t$) being excited, which requires detrending to capture oscillatory content. 
We combine the observers by averaging their powers rather than their displacement signals,
\begin{equation} \label{eq:mean_observer_power}
    \langle {P}(f) \rangle = \frac{1}{N_{\rm obs}} \sum_{a=1}^{N_{\rm obs}} P_a(f),
\end{equation}
to produce a global one-sided power spectral density (PSD). 
We apply a Hann taper (see below) to combat spectral leakage introduced by the recording interval (i.e., put less weight near the window endpoints where oscillations may be prematurely truncated).

%%%%%% FIGURE %%%%%%
\begin{figure}
\centering
 \includegraphics[width=0.48\textwidth]{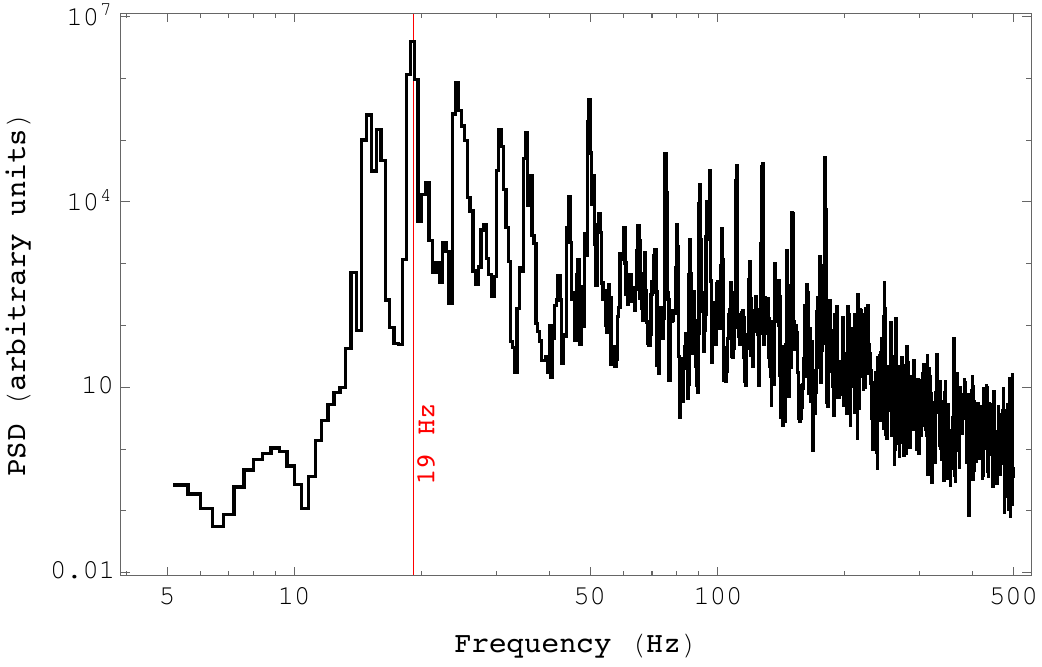}
 \caption{Global one-sided PSD for the entire signal produced for the global quake run.
 The vertical line highlights the dominant frequency at $\approx 19$~Hz.}
  \label{fig:gloablpsd}
\end{figure}
%%%%%%%%%%%%

The results for the global quake model are shown in Figure~\ref{fig:gloablpsd}.
The 10 strongest spectral peaks we find, in order of decreasing mean power, are $18.79$, $24.01$, $49.52$, $14.72$, $30.48$, $15.63$, $35.11$, $75.19$, $179.52$, and $127.41$~Hz.
In any case, it is clear that a forest of oscillations can be excited and mixing phenomena takes place given the initially large-scale nature of the island+collar perturbation.
Moreover, the peak frequencies are distinct from the usual magnetoelastic spectrum one gets in the island-free limit \cite[][see also Appendix~\ref{sec:noislands}]{sotani2007} and the distribution of power is highly non-uniform.
As mentioned previously, we cannot reliably attempt to match these to the QPO spectra associated with any given X-ray tail (e.g., those shown in Figure~\ref{fig:miller}) because they are, in some appropriate sense, averaged over the entire simulation.
At late times once the creep has saturated however, one may expect such a global spectrum to reflect the oscillatory content observed in X-ray tails.

On the other hand, viscosity and gravitational-wave emission provide damping channels with timescales that can be estimated from $\tau \approx E/\langle \dot{E} \rangle$ using an oscillation average \citep{cutl87}.
The relevant microscopic viscosity in the crust is (typically) assumed to be electron-ion scattering and, at temperatures of $\sim 10^8$~K, one expects a timescale of $\tau_{\nu} \sim 10^{10} (L/\text{1 km})^2$~s for lengthscale $L$ \citep{chug05} \cite[though see also][]{ofeng15}.
Gravitational radiation is likewise inefficient for predominantly axial motions with long damping times, $\tau_{\rm GW} \sim 10^4$~yr \citep{schut83}.
Both estimates are far longer than our seconds-long model except for `modes' whose gradients vary over lengthscales of $L \lesssim 10$~cm (i.e., smaller than our cell size).
The question therefore remains as to why QPO spectra continue to evolve over $\gtrsim 100$~s that cannot be addressed in this study.
Viscoelastic damping is likely to be much more relevant \citep{cap2026,cap2026b}, though its inclusion is deferred to future work as are nonlinear couplings, which become important when $|\delta B/B| \gtrsim 1$, and parametric mode-mode seedings \citep{sotani2024,pl25}.
%Thus, even an initially large-scale perturbation assocaited with the island+collar can lead to fragmenting and 

%Each spectrum is assigned the mean time of its contributing samples, and only complete windows are retained.
%The output is therefore a table of window-centre time, frequency, and mean spectral density, here restricted to $5$--$150\,{\rm Hz}$.
%%%%%%%%%%%%%%%%%%%%%%%%%%%%%%%%%%%%%%%%%%%%%%%%%%%%%%%%%%%%%%%%%%%%%%%%
\subsection{Dynamic spectra} \label{sec:dynamic}
%%%%%%%%%%%%%%%%%%%%%%%%%%%%%%%%%%%%%%%%%%%%%%%%%%%%%%%%%%%%%%%%%%%%%%%%

%%%%%% FIGURE %%%%%%
\begin{figure}
\centering
 \includegraphics[width=0.48\textwidth]{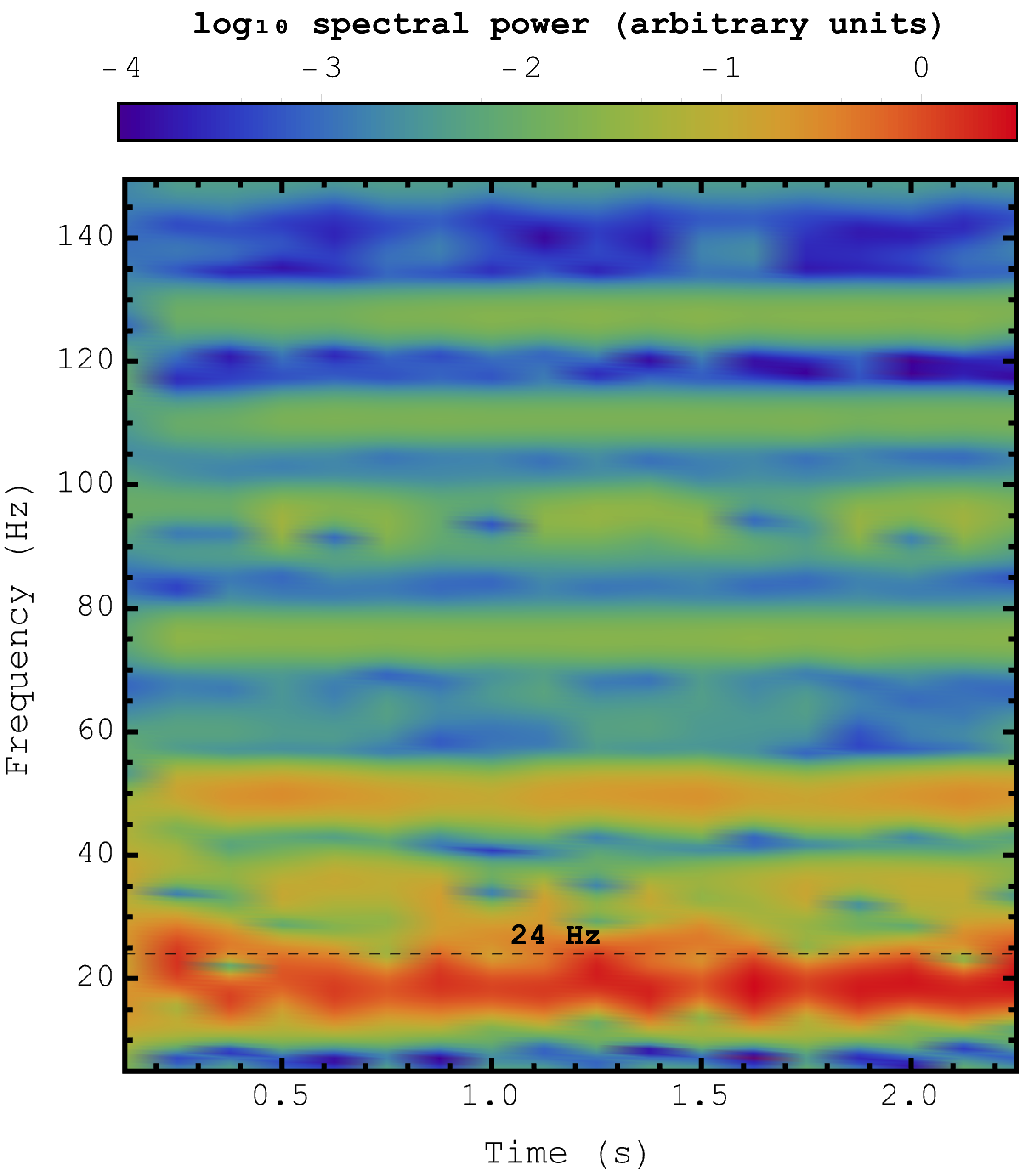}
 \caption{Dynamic spectrum associated with the development of oscillations.
 Overlaid by the dashed line is a frequency of $24$~Hz, corresponding roughly to the fundamental mode in the island-free limit.}
  \label{fig:globalspec}
\end{figure}
%%%%%%%%%%%%

To examine spectral content \emph{during} the failure cascade, we repeat the spectral analysis but in $0.25$~s segments with $50\%$ overlap, fitting and subtracting linear trends independently within each segment.
Such segments are long enough to capture the low-frequency content without being so long as to completely blur the effects of cascading; some tradeoff is inevitable however. 
For a segment centred at $t_c$, if $\mathcal Y_{a}(t_c,f)$ denotes the discrete Fourier transform of the detrended displacement at observer $a$, the one-sided PSD of relevance reads
\begin{equation} \label{eq:powersegment}
    \langle P(t_c,f)\rangle = \frac{2\Delta t_{\rm s}} {N_{\rm obs}\sum_{m=0}^{J-1}w_m^2} \sum_{a=1}^{N_{\rm obs}} \left|\mathcal{Y}_{a}(t_c,f)\right|^2,
\end{equation}
where $2w_m=1-\cos[2\pi m/(J-1)]$ are the Hann weights multiplying each time series so that their endpoints independently taper to zero.
This taper broadens spectral peaks, while the sum over $w_m^2$ accounts for the normalisation of the weights.
The window duration sets a Fourier scale of $\approx 4$~Hz, with additional broadening from the Hann taper. 

%%%%%% FIGURE %%%%%%
\begin{figure}
\centering
 \includegraphics[width=0.48\textwidth]{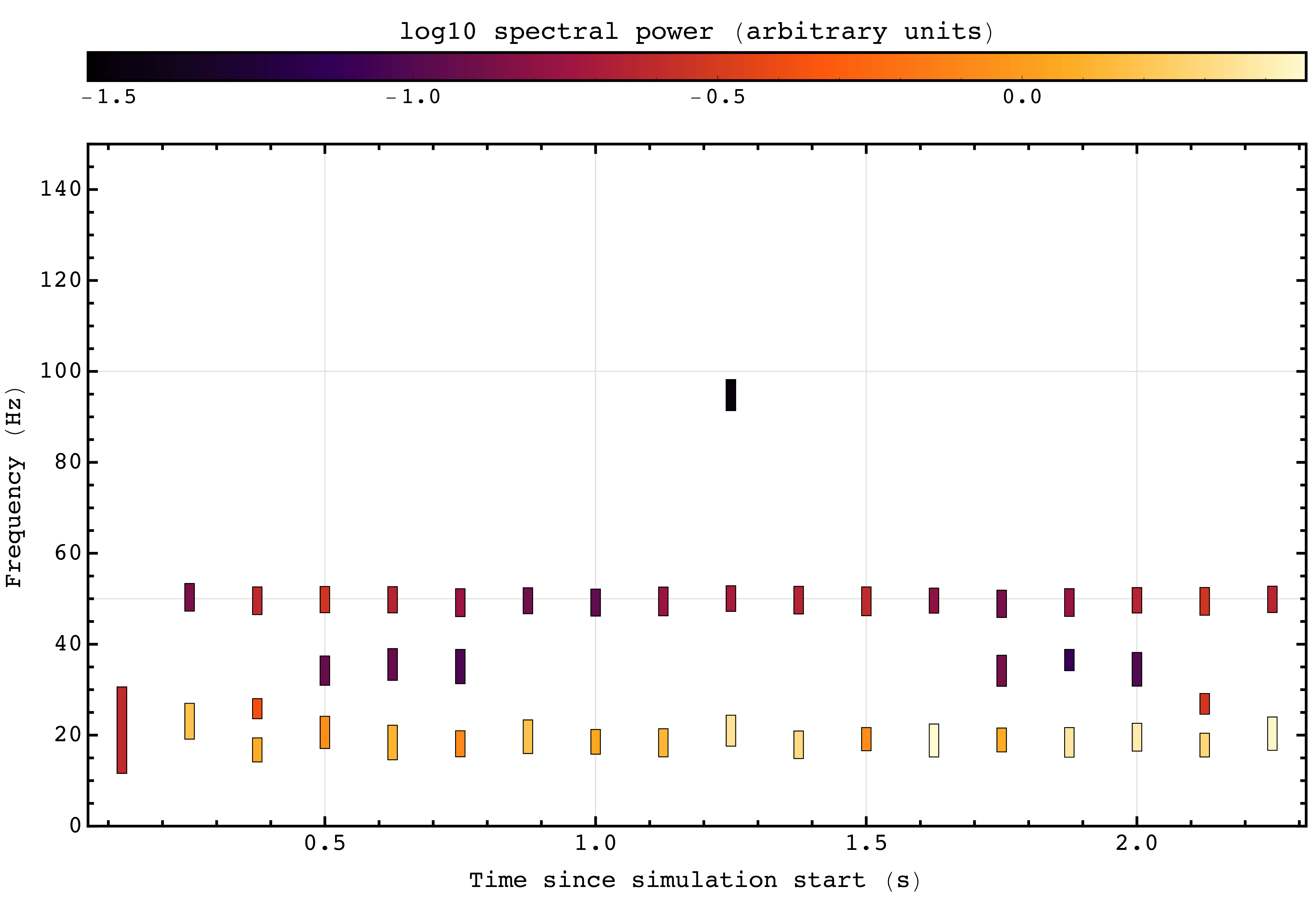}
 \caption{Low-frequency peaks in the observer-averaged dynamic spectrum \eqref{eq:powersegment}, with brighter shades indicating more spectral power.
 Each rectangle represents a $0.25$~s window but is drawn only $0.02$~s wide for clarity. 
 Missing markers at a given frequency reflect the selection criteria, not necessarily a cessation of oscillations.}
  \label{fig:globalspec2}
\end{figure}

The resulting `dynamic spectrum', computed using equation \eqref{eq:powersegment}, is shown in Figure~\ref{fig:globalspec}.
We see that there is initially large spectral power in the neighbourhood of $\approx 24$~Hz, which decreases to $\approx 19$~Hz in the following segments. 
As seen in Fig.~\ref{fig:gloablpsd}, the dominant peak appears at $\approx 19$~Hz and it is tempting therefore to interpret this feature as the drifted frequency associated with an $\ell=2$ torsional mode, which in the island-free limit has a frequency of $\approx 23.7$~Hz (see Appendix~\ref{sec:noislands}).
It is clear though that some power is redistributed throughout the course of the simulation with some bins gaining power and others losing; a transient feature can be seen around $\approx 90$~Hz, for example, appearing and subsequently disappearing around $\approx 1.3$ seconds in.

Another way to visualise the changing spectral content is shown in Figure~\ref{fig:globalspec2}.
To highlight distinct features, we analyse successive $0.25$~s windows with centres approximately $0.125$~s apart.
We keep up to three peaks per window, requiring the power be at least $1\%$ of the maximum within the interval and at least $3$ times the surrounding levels while maintaining separation of $\geq 8$~Hz from other peaks.
For example, we see a gap of $\approx 0.1$~s between the first two features because the markers are centred at $t \approx 0.125$ and $0.25$~s, despite the underlying windows overlapping. 
These choices were made to roughly mimic QPO detection strategies \cite[see, e.g.,][]{hupp14b}.
We see that the power concentrated in bins surrounding $\approx 20$~Hz varies significantly throughout the run and a single, high-frequency ($\approx 90$~Hz) component enters the significance window after only $\approx 1.25$~s.
This may explain QPO transience, in that (i) the signal may drop out of the visibility range of the detector because the `mode' donated power to other oscillations or (ii) changed character due to plastic development \cite[see also][]{pl25}.
While a thorough comparison lies beyond the scope of this work where we have considered only one representative star, such transience is reminiscent of the data reported by \cite{miller19qpos} for the QPOs in SGR 1806--20 (see Fig.~\ref{fig:miller}) and other sources, considered below.

%%%%%%%%%%%%

%%%%%%%%%%%%%%%%%%%%%%%%%%%%%%%%%
\section{Results: Localised quakes}  \label{sec:localised}
%%%%%%%%%%%%%%%%%%%%%%%%%%%%%%%%%

In this section we consider instead a more localised event with the same island position but thicknesses that are smaller by a factor 4, viz. $\Delta r_c=0.05$~km and $\Delta\theta_c=5^\circ$.
In this case, the elastic energy stored in the island is $E_{\rm el} = 1.41 \times 10^{43}$~erg and, with $\eta = 0.05$, the initial pulse therefore contains an energy content of $E_{\rm kin,0} = 7.07 \times 10^{41}$~erg -- a factor $\approx 22$ smaller than that of the global quake.
Such an island energy is more than sufficient to power the X-ray storm that occurred in 2009 from 1E 1547.0--5408, for example, where QPOs at $\approx 91, 93, 127$~Hz were detected \citep{hupp14b}. 
This object has an estimated polar-dipole field of $\approx 1.6 \times 10^{14}$~G when GR and magnetospheric corrections are accounted for \citep{stef25}. 
Thus -- although noting again we are not attempting any detailed QPO matching in this paper -- the local model of this section may be thought of as roughly corresponding to that source.

For the sake of completeness, we compile in Table 1 a list of QPOs reported in the literature for short/intermediate bursts from Galactic magnetars (i.e., not giant flares).
Although some of the lower-energy bursts exhibit higher-frequency oscillations and the sample size is small, it is worth noting that there is no clear correlation between the isotropic burst energy, $E_{\rm iso}$, and QPO frequencies.

%%%%%%%%%%%%%%%%%%%%%%%%%%%%%%%%%%%%%%%%%%%%%%%%%%%%%%%%%%%%%%%%%%%%%%%%
%%%%%%%%%%%%%%%%%%%%%%%%%%%%%%%%%%%%%%%%%%%%%%%%%%%%%%%%%%%%%%%%%%%%%%%%
\begin{table} \label{tab:qpos}
\centering
\caption{
QPOs observed in various short bursts from magnetars, as reported in the literature.
Here, $d_X$ denotes a distance with $X$ in kpc.
For sources with multiple events, the energy range covers individual bursts or storms, not their sum.
The SGR 1806--20 signal is associated with a storm \citep{hupp14c}.
For 1E 1547.0--5408 (aka SGR J1550--5418), the $\sim 91, 93, 127$~Hz QPOs appear in three separate bursts, while the $\sim 15.7, 28, 260$~Hz signals come from individual bursts \citep{hupp14b,coll15,dem25}.
The (tentative) 4U 0142+61 QPOs are quoted from \citet{chen23}.
For SGR J1935+2154, the three events are reported by \citet{li22} (FRB associated), \citet{rob23}, and \citet{tan26}, respectively.}
\begin{tabular}{@{}lcl@{}}
\hline
\hline
Object & $E_{\rm iso}$ ($10^{39}$ erg) & QPOs (Hz) \\
\hline
SGR 1806$-$20 (1996 Nov)
& $\sim 0.36 d_{8.7}^{2}$
& $\sim57$ \\

1E 1547.0$-$5408 (2009 Jan)
& $\sim (1-19) d_{4.5}^{2}$
& $\sim91, 93, 127$ \\
& & $\sim 15.7, 28, 260$ \\

4U 0142+61 (2015 Feb)
& $\sim 0.45 d_{3.6}^{2}$
& $57.5^{+1.2}_{-1.2}$, $112.2^{+1}_{-1}$ \\

SGR J1935+2154
& $\sim(4-39)d_{9}^{2}$
& $39.14^{+1.88}_{-1.94}$ \\
2020 Apr; 2022 Jan, Nov
& & $\sim 42$, $17.97^{+1.56}_{-1.22}$ \\
\hline
\hline
\end{tabular}
\end{table}
%%%%%%%%%%%%%%%%%%%%%%%%%%%%%%%%%%%%%%%%%%%%%%%%%%%%%%%%%%%%%%%%%%%%%%%%
%%%%%%%%%%%%%%%%%%%%%%%%%%%%%%%%%%%%%%%%%%%%%%%%%%%%%%%%%%%%%%%%%%%%%%%%

%%%%%% FIGURE %%%%%%
\begin{figure}
\centering
 \includegraphics[width=0.48\textwidth]{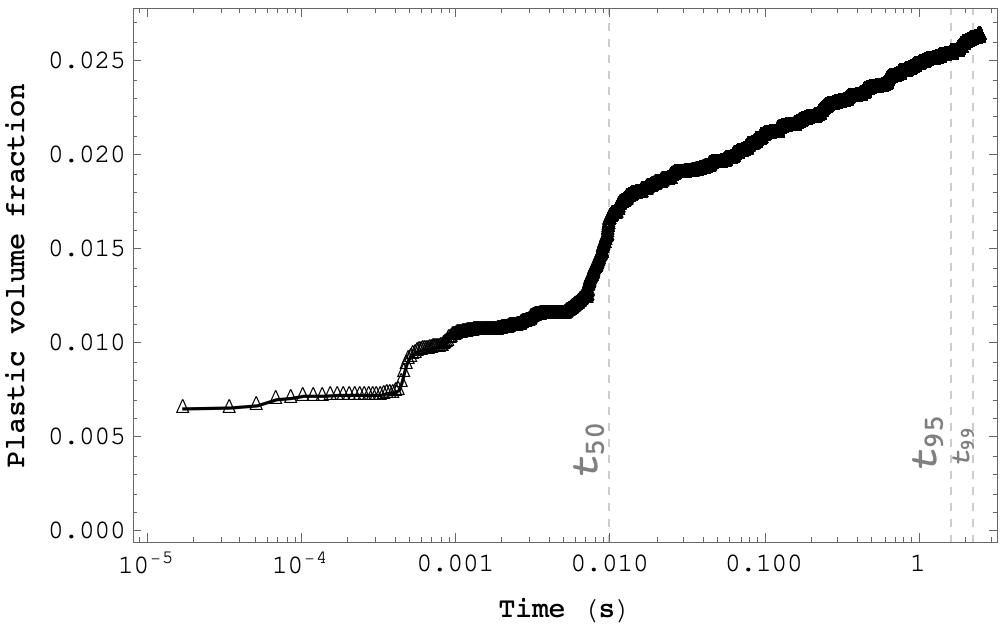}
 \caption{Similar to Fig.~\ref{fig:globalacc} but for the local quake model described at the beginning of Sec.~\ref{sec:localised}.}
  \label{fig:globalaccS}
\end{figure}
%%%%%%%%%%%%

Repeating the diagnostic steps described in the previous section, we show in Figure~\ref{fig:globalaccS} the accumulation of the plastic volume for the local quake.
Initially we have $f_{\rm fail} \approx 0.6\%$ (245 cells out of 43008), though this climbs to $\approx 2.6\%$ by simulation end (1099 cells). 
We thus have a factor $\sim 4$ increase in the total count, as in the global case (Fig.~\ref{fig:globalacc}).
The implication is thus that it is the efficiency, $\eta$, that primarily controls the extent of further failures rather than the initial island geometry itself (validated in Appendix~\ref{sec:fieldstrength}).
The morphology of the failure development is different, however, as there are two avalanches, one around $\approx 0.5$~ms and the second at $\approx 8$~ms. 
After this second phase, slow creep continues and failures develop in a gradual manner (though approximate saturation at simulation end is more questionable in this case).

%%%%%% FIGURE %%%%%%
\begin{figure*}
\centering
 \includegraphics[width=\textwidth]{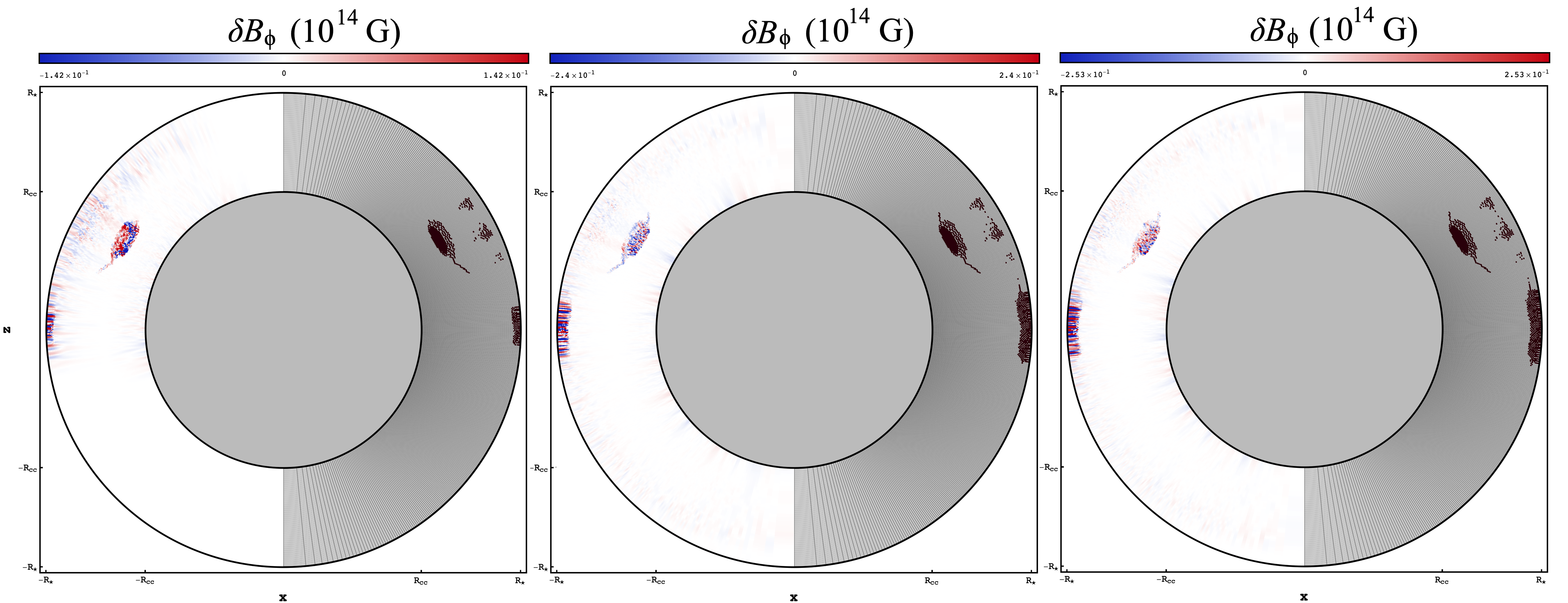}
 \caption{Similar to Fig.~\ref{fig:globalsnap50} but for the local quake model at times $t=t_{50}$ (left), $t=t_{95}$ (middle), and $t=t_{99}$ (right).}
  \label{fig:globalsnapsS}
\end{figure*}
%%%%%%%%%%%%

Snapshots of the cell phase and magnetic perturbation are shown in Fig.~\ref{fig:globalsnapsS}. 
The first avalanche takes less time to develop relative to the global case, likely because the smallness of the island promotes overlap with high-frequency oscillations, though has essentially the same explanation.
Note that the ellipsoidal prestress \eqref{eq:prestress_profile} is not circular in physical distance, and the characteristic half-widths (arclengths) are
\begin{equation} \label{eq:lengthscales}
L_r \sim \Delta r_c, \quad L_\theta \sim r_c\Delta\theta_c.
\end{equation}
For $\Delta r_c=0.05$~km, $\Delta\theta_c=5^\circ$, and $r_c \approx 12$~km, we have $L_{\theta}/L_{r} \approx 22$.
This explains the narrow features in the angular direction (noting also the radial stretching of the crust applied for visualisation) and the more rapid development: high-$\ell$ waves are preferentially excited from the outset and mix earlier.
The second avalanche, by contrast, is associated with failures taking place near the surface ($r \sim \Rs$) and equator ($\theta \sim \pi/2$).
This development takes longer as waves launched from the collar take time to propagate through the crust and mix constructively there.
As before, waves reach that boundary and then crash back into slightly slower ones and produce a cascade there.
A notable fault line can also be seen to protrude from the island as the initial pulse causes nearby failures and promotes mixing there as waves divert around the inelastic `wall'.
%such that a second avalanche takes place only after a delay.

\begin{figure}
\centering
 \includegraphics[width=0.48\textwidth]{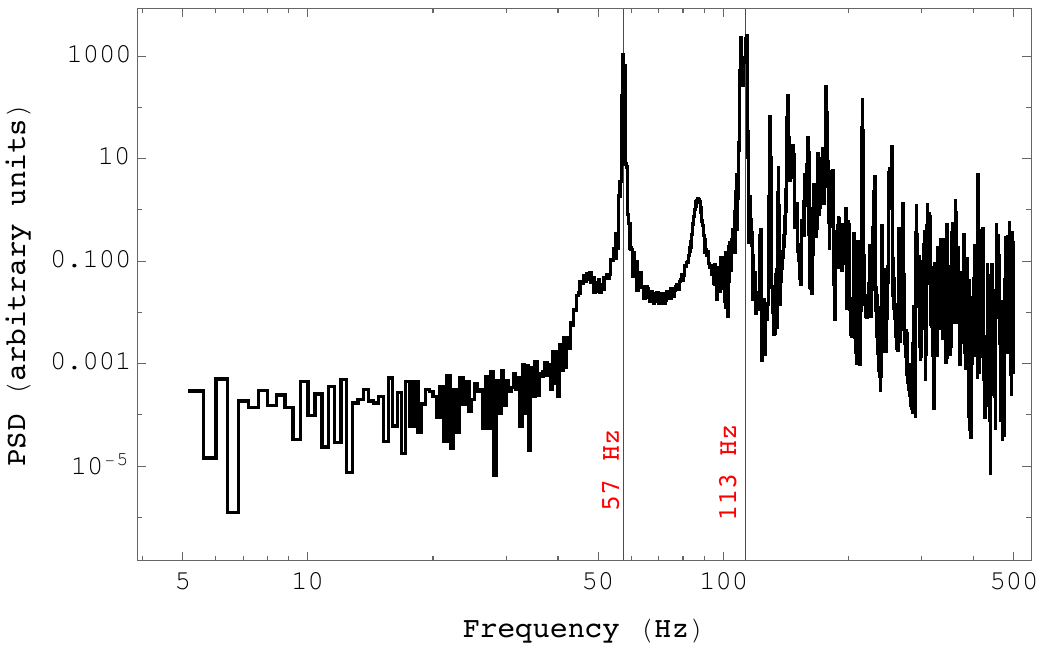}
 \caption{Similar to Fig.~\ref{fig:gloablpsd} but for the local quake case.}
  \label{fig:gloablpsdS}
\end{figure}

The global spectrum in this case is shown in Figure~\ref{fig:gloablpsdS}.
We see a dominant $\approx 57$~Hz feature, though notably the bulk of the power resides at higher frequencies.
The 10 most significant frequencies we find, in order of mean power, are $113.45$, $110.24$, $57.32$, $112.47$, $176.58$, $142.94$, $216.39$, $129.51$, $160.08$, and $108.95$ Hz.
For the prominent $113.5$~Hz feature, the cleanest interpretation is that the smaller initial failure injects substantially more high-$\ell$ structure than the broad kick of the global case.
In the idealised, crust-confined picture for the $n=0$ torsional family, the frequencies scale as \cite[see equation 33 in][]{kozh20}
\begin{equation} \label{eq:torsionalscaling}
f_{\ell,n=0}\propto \sqrt{(\ell-1)(\ell+2)}.
\end{equation}
If we take $24$~Hz for the fundamental mode (see Appendix~\ref{sec:noislands}), then the ratio $113/24$ indicates that an $\ell \sim 10$ mode would have the requisite frequency.
An $\ell \sim 10$ pattern varies over an angular scale of order $\Delta\theta_{10} \sim {\pi}/{10}\approx 18^\circ$, which can be compared with the angular extent of the kick, $3\Delta\theta_c=15^\circ$.
A localised kick, especially one with enhanced structure around the island boundary/collar, therefore has a strong harmonic projection onto angular wavelengths of this order and the few degrees discrepancy can be attributed to the influence of the magnetic field and loss of elasticity.
Moreover, the two features at $113.45$ and $112.47$~Hz, separated by just $\approx 1$~Hz, may be related by the magnetic field through a Zeeman-like splitting.

Nevertheless, high-frequency activity is expected in the localised case due to our initial data, and the preferential selection of high-frequency oscillations may be responsible for driving faster avalanches.
Overall, the model spectrum is in rough agreement with the QPOs observed in the X-ray storm from 1E 1547.0--5408 \citep{hupp14b}, and a different crustal thickness or magnetic field strength could shift spectral content to $\sim 90$~Hz with additional features at higher frequencies.
The frequencies we observe are consistent with QPOs in other sources too.
Notably, we find a prominent $57.3$~Hz feature which agrees well with the $\sim 57$~Hz QPOs detected in the 1996 November storm data from SGR 1806--20 \citep{hupp14c}.

%%%%%% FIGURE %%%%%%
%%%%%%%%%%%%

%%%%%%%%%%%%%%%%%%%%%%%%%%%%%%%%%%%%%%%%%%%%%%%%%%%%%%%%%%%%%%%%%%
\subsection{Equatorially-symmetric quakes and interpulses}
%%%%%%%%%%%%%%%%%%%%%%%%%%%%%%%%%%%%%%%%%%%%%%%%%%%%%%%%%%%%%%%%%%

Had we instead considered an island centred on the equator, we mathematically expect to develop a north-south symmetric failure pattern provided the kick, prestress, and background retain their symmetries\footnote{Adding even-order multipoles to the poloidal background could change the story though, not only because the magnetic geometry is known to reorganise magnetoelastic spectra \citep{gabler2013} and quadrupolar patterns \citep{mast15b}, but because equatorial symmetry would be broken.}.  
In particular, the function $s$ in expression \eqref{eq:island_coordinate} satisfies $s(r,\pi-\theta)=s(r,\theta)$ and the same for $\dot{Y}^{0}$ in expression \eqref{eq:initial_velocity_profile}.
Despite the reversal of $B^r$ between hemispheres, the dipolar magnetic restoring terms preserve this symmetry, meaning that the stress is also mirror symmetric, so mirror cells should satisfy the failure criterion simultaneously.
Although it is interesting for its own sake to consider different island initial positions, we place an equatorially-centred quake in this section as an independent test of the code to ensure symmetries are respected.

\begin{figure}
\centering
 \includegraphics[width=0.48\textwidth]{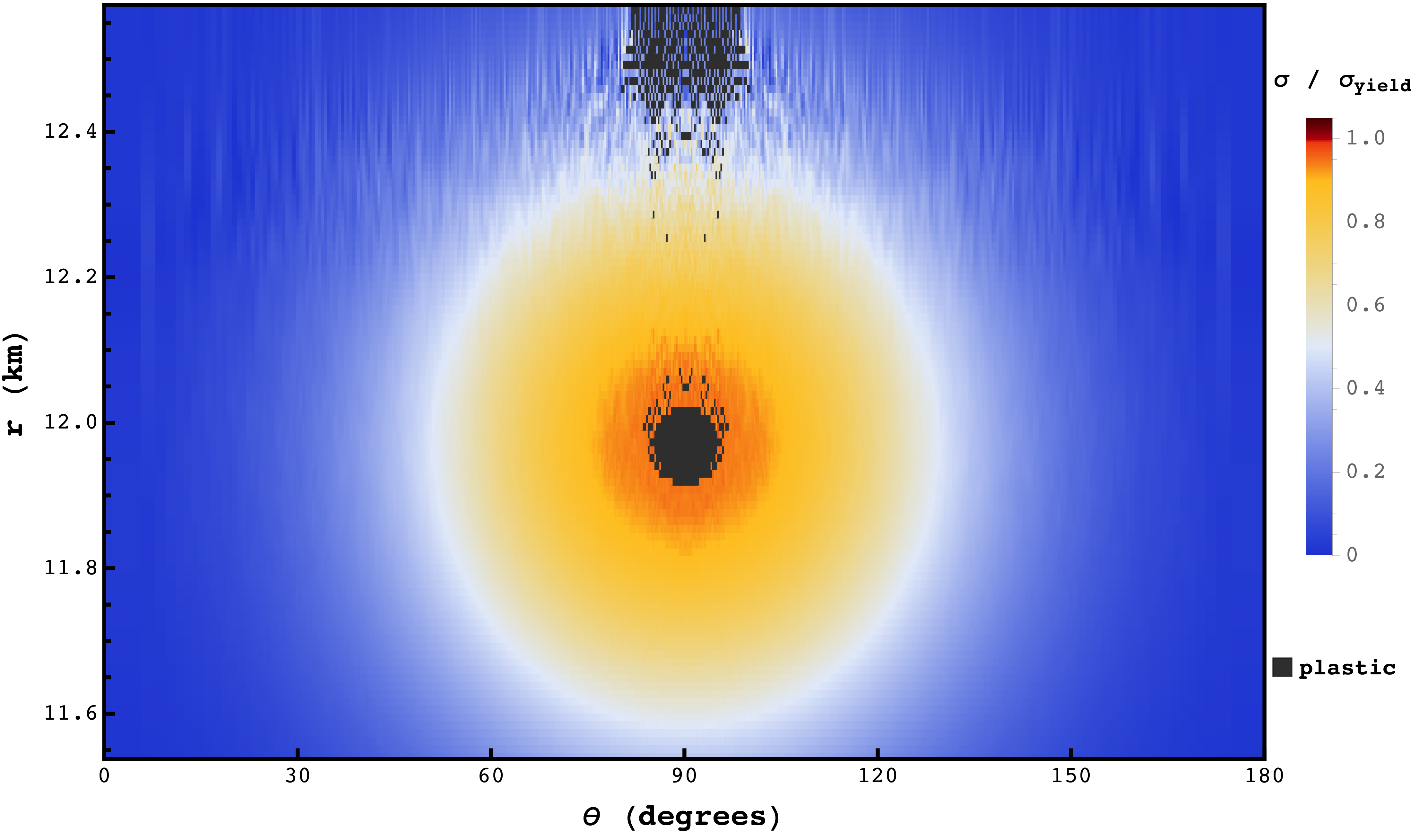}
 \caption{Depiction of the stress distribution at time $t=0.3$~s, including prestress, for a local quake centred about the equator.}
  \label{fig:equator}
\end{figure}

Figure~\ref{fig:equator} displays a snapshot of the stress profile and failure pattern at $t=0.3$~s for such a case. 
We see an equatorially-symmetric development, as expected, and further that the plastic development has extended outward from the equator by $\approx 10^{\circ}$ on either side of the island.
The failure pattern is fragmented as in previous examples, though is strongly concentrated near the surface going down to depths of $\sim 200$~m.
By $0.3$~s we have that 1008 out of 43008 cells have failed, and that $f_{\rm fail} \approx 0.024$, similar to that of the lopsided island case (see Fig.~\ref{fig:globalaccS}).

One motivation for considering such a scenario concerns interpulse phenomena observed recently from two long-period transients (LPTs), namely ASKAP J183950.5--075635.0 \citep{lee25} and LPT J1555--5631 \citep{hyman26}. 
What exactly these sources are constitutes an active area of research, with some studies advocating for neutron stars \cite[e.g.,][]{suvm23,coop24} and others for binaries involving white dwarfs \cite[e.g.,][]{qu25,skk26}. %see \cite{rea26} for a review.
Interestingly though, the LPTs named above show two pulses per cycle separated by phases of $\sim 180^\circ$ which, in a neutron-star scenario, indicates two mirrored emission regions as in interpulse models for orthogonal rotators.
As described by \cite{coop24}, radio activity could be sourced by discharges from antipodal islands with $u^{\phi}_{\rm plastic} \gg u^{\phi}_{\rm rigid}$ there, allowing the object to rise out of pulsar graveyard while the plastic flow is active. 
Our model predicts antipodal emission sites for an equatorial quake because the equivalent stress is mirror symmetric.

To determine whether our configuration could theoretically produce two pulses that are phase-separated by $\sim 180^\circ$, we appeal to the familiar orthogonal-rotator picture where two pulses are emitted half a rotation apart \citep{lor04}.
Although our hypothetical discharge sites are located away from the poles and are instead associated with failure development, a similar effect is possible if the magnetic axis is tilted by an angle, $\alpha$, relative to the rotation axis. 
Using standard spherical trigonometry, the instantaneous magnetic colatitude measured by an observer, $\zeta$, varies with rotational phase according to
\begin{equation}
\cos\zeta = \cos\alpha\cos\iota+\sin\alpha\sin\iota\cos\phi,
\end{equation}
where $\iota$ is the viewing angle. 
For an orthogonal rotator, $\alpha=\pi/2$, this implies $\zeta(\phi+\pi)=\pi-\zeta(\phi)$ such that emission from two mirrored patches could be observed at phases separated by exactly $\pi$. 
Although a more careful analysis will be presented in a future paper, it is plausible therefore that interpulses in LPTs -- if indeed they are magnetars -- could be tied to quake activity around the equator without requiring a simultaneous failure at diametrically-opposite positions.

%%%%%%%%%%%%%%%%%%%%%%%%%%%%%%%%%
\section{Discussion}  \label{sec:discussion}
%%%%%%%%%%%%%%%%%%%%%%%%%%%%%%%%%

In this paper we develop a code that solves the magnetoelastic master equation \eqref{eq:torsional_evolution} in cases where the shear modulus can be set to zero in failed regions of the crust (expression~\ref{eq:effective_mu}).
We allow for, and track, shear stresses that develop as a result of mode propagation and, if a critical threshold is met in accord with the von Mises criterion \eqref{eq:von_mises}, dynamically switch the local modulus in a cell to zero.
This study is motivated by the fact that if quake activity is invoked to explain magnetar outbursts and their QPOs, the loss of elastic support to fuel the event should be accounted for.
The loss of elastic traction across the interfaces of failed cells generally promotes mode mixing which can trigger either rapid failure cascading (an `avalanche') or the growth of a fault line. 
Constructive interference occurring on a beat-frequency timescale can also gradually extend the reach of the failure web (`creep').
This work is primarily of a `methods' nature, in that we have not attempted a thorough parameter survey, but aim to demonstrate the main effects through some representative examples. 
Key results for representative global quakes are described in Sec.~\ref{sec:global} and local quakes in Sec.~\ref{sec:localised}.

%%%%%% FIGURE %%%%%%
%%%%%%%%%%%%

One example of a parameter we have not explored is the island depth.
Changing the value of $r_{c}$ in expression \eqref{eq:island_coordinate} changes both the prestress and, under our normalisation, the energy of the initial event for a given $\eta$.
A comparable island placed deeper in more rigid regions of the crust would therefore supply a larger initial kinetic energy at fixed efficiency, but unleashed waves would also encounter greater hydromagnetic inertia.
A shallower quake would have less energy available but act on lighter material. 
The failure pattern may consequently change in both extent and spatial distribution, without any obvious, monotonic scaling(s) with depth.
The development of small-scale structure that we call `creep' would likely be sensitive to this, and would be worth investigating for models of post-burst cooling as more localised currents are more susceptible to Ohmic decay. 
One case we ran (though not shown) with $r_{c} = \Rs-0.2$~km (i.e., near the surface) but otherwise identical to the global case of Sec.~\ref{sec:global} had a failure fraction $f_{\rm fail}(10~\text{ms}) = 0.16$, which is $\approx 30\%$ lower than that accrued for a deeper quake (Fig.~\ref{fig:globalacc}).
%and could change with pre-stress specifics.

Additionally, core couplings are almost certainly important in the general QPO problem \citep{2007MNRAS.377..159L,vh11}. 
Allowing magnetic coupling to the core would change the boundary physics rather than merely add another damping coefficient. 
In general though, this requires an equilibrium field crossing the crust-core interface, which we have not used.
If the field extended there, the core could accept wave energy from the crust and return stresses through global oscillations.
We expect that early-time failures near the island will remain similar before the disturbance reaches the core but resonant transfer will quench repeated restressings.
For the dipolar models of \citet{gabler12}, crustal shear-mode damping at a field strength of $\sim 10^{14}$~G occurred within $0.1$~s, well inside our integration times.
Our conclusions about creep especially require caution therefore for core-threading fields.

The final point required to interpret our findings is what is, and is not, represented by a `failed' cell.
Failure removes the local elastic shear support, but the material itself is not removed and continues to carry magnetic twist.
The irreversible prescription naturally generates speckled patterns when a large region of the crust begins close to threshold; the use of a viscoplastic relation to change the redistribution of elastic energy could play a big role.
Allowing subsequent failures to receive velocity kicks would also change our conclusions.
Indeed, although the shutting off of elastic traction is deliberate in that the initial event is considered a trigger, donating more kinetic energy to waves could change cascade morphology.
Moreover, if heat is generated following a failure then thermally-activated creep can substantially lower the effective breaking stress through a kind of lattice fatigue \citep{chugunov10}.
Quantum-electrodynamic force balance may also influence failure development, especially near the equator where $B/B_{\rm QED} \gtrsim 100$ in the models considered here \cite[see][for an outline of how to include such effects]{suvpons25}.
In general, because the magnetic tension is decreased in Born-Infeld models (for instance), the restoring forces for transverse displacements are weaker and could lead to a significant, position-dependent reduction in crustal shear mode-mixing efficiencies compared to standard Maxwell electrodynamics.

All the above discussion aims to demonstrate how complicated the problem is in reality.
Still, the presented code and the representative results we obtained may contribute to our collective understanding of magnetar phenomena (e.g., QPO transience; see Figs.~\ref{fig:globalspec} and \ref{fig:globalspec2}).
Other directions may be worth exploring also, such as whether gravitational waves produced by avalanche activity could be detected for nearby magnetars with next-generation instruments.

%%%%%%%%%%%%%%%%%%%%%%%%%%%%%%%%%%%%%%%%%%%
\section*{Acknowledgements}
I would like to thank the attendees of the \texttt{BLADES} Workshop 2026 held in Alicante for feedback.
Some sections of code were both optimised and scrutinised by iterative interactions with a GPT 5.6 interface following initial, and eventually final, implementations by the author (who takes full responsibility for the contents of this manuscript).
I am grateful for the European Union's Horizon MSCA-2022 research and innovation programme `EinsteinWaves' under grant agreement No. 101131233, the Deutsche Forschungsgemeinschaft individual research grant 570901071, and the High Performance and Cloud Computing Group at the Zentrum f{\"u}r Datenverarbeitung of the University of T{\"u}bingen (Project `AnqaGW') for support.
%I acknowledge support from the Open Access Publishing Fund of the University of T{\"u}bingen.
%%%%%%%%%%%%%%%%%%%%%%%%%%%%%%%%%%%%%%%%%%%

%%%%%%%%%%%%%%%%%%%%%%%%%%%%%%%%%%%%%%%%%%%
\section*{Data availability statement}
Observational data used in this paper are quoted from the cited works, and only numerical data were produced. Computational routines are available on reasonable request and will eventually be made public. Some video files showing simulation development can also be made available on request.
%%%%%%%%%%%%%%%%%%%%%%%%%%%%%%%%%%%%%%%%%%%

\bibliographystyle{mn2e}
\bibliography{magnetarbib}

@ARTICLE{sot04,
       author = {{Sotani}, Hajime and {Suvorov}, Arthur G. and {Kokkotas}, Kostas D.},
        title = "{Couplings of torsional and shear oscillations in a neutron star crust}",
      journal = {\prd},
         year = 2024,
        month = jul,
       volume = {110},
       number = {2},
          eid = {023035},
        pages = {023035},
          doi = {10.1103/PhysRevD.110.023035},
archivePrefix = {arXiv},
       eprint = {2406.17195},
 primaryClass = {astro-ph.HE},
       adsurl = {https://ui.adsabs.harvard.edu/abs/2024PhRvD.110b3035S}
}

@ARTICLE{hurley99,
       author = {{Hurley}, K. and {Cline}, T. and {Mazets}, E. and {Barthelmy}, S. and {Butterworth}, P. and {Marshall}, F. and {Palmer}, D. and {Aptekar}, R. and {Golenetskii}, S. and {Il'Inskii}, V. and {Frederiks}, D. and {McTiernan}, J. and {Gold}, R. and {Trombka}, J.},
        title = "{A giant periodic flare from the soft {\ensuremath{\gamma}}-ray repeater SGR1900+14}",
      journal = {\nat},
         year = 1999,
        month = jan,
       volume = {397},
       number = {6714},
        pages = {41-43},
          doi = {10.1038/16199},
archivePrefix = {arXiv},
       eprint = {astro-ph/9811443},
 primaryClass = {astro-ph},
       adsurl = {https://ui.adsabs.harvard.edu/abs/1999Natur.397...41H}
}

@ARTICLE{kaspbel17,
       author = {{Kaspi}, Victoria M. and {Beloborodov}, Andrei M.},
        title = "{Magnetars}",
      journal = {\araa},
         year = 2017,
        month = aug,
       volume = {55},
       number = {1},
        pages = {261-301},
          doi = {10.1146/annurev-astro-081915-023329},
archivePrefix = {arXiv},
       eprint = {1703.00068},
 primaryClass = {astro-ph.HE},
       adsurl = {https://ui.adsabs.harvard.edu/abs/2017ARA&A..55..261K}
}

@ARTICLE{sw05,
       author = {{Strohmayer}, Tod E. and {Watts}, Anna L.},
        title = "{Discovery of Fast X-Ray Oscillations during the 1998 Giant Flare from SGR 1900+14}",
      journal = {\apjl},
         year = 2005,
        month = oct,
       volume = {632},
       number = {2},
        pages = {L111-L114},
          doi = {10.1086/497911},
archivePrefix = {arXiv},
       eprint = {astro-ph/0508206},
 primaryClass = {astro-ph},
       adsurl = {https://ui.adsabs.harvard.edu/abs/2005ApJ...632L.111S}
}

@ARTICLE{skk26,
       author = {{Suvorov}, Arthur G. and {Karnesis}, Nikolaos and {Korol}, Valeriya},
        title = "{Population synthesis and detection prospects for Galactic long-period transients with LISA}",
      journal = {\aap},
         year = 2026,
        month = oct,
       volume = {714},
          eid = {A172},
        pages = {A172},
          doi = {10.1051/0004-6361/202660574},
       adsurl = {https://ui.adsabs.harvard.edu/abs/2026A&A...714A.172S}
}

@ARTICLE{maz79,
       author = {{Mazets}, E.~P. and {Golentskii}, S.~V. and {Ilinskii}, V.~N. and {Aptekar}, R.~L. and {Guryan}, Iu. A.},
        title = "{Observations of a flaring X-ray pulsar in Dorado}",
      journal = {\nat},
         year = 1979,
        month = dec,
       volume = {282},
       number = {5739},
        pages = {587-589},
          doi = {10.1038/282587a0},
       adsurl = {https://ui.adsabs.harvard.edu/abs/1979Natur.282..587M}
}

@ARTICLE{sa07,
       author = {{Samuelsson}, Lars and {Andersson}, Nils},
        title = "{Neutron star asteroseismology. Axial crust oscillations in the Cowling approximation}",
      journal = {\mnras},
         year = 2007,
        month = jan,
       volume = {374},
       number = {1},
        pages = {256-268},
          doi = {10.1111/j.1365-2966.2006.11147.x},
archivePrefix = {arXiv},
       eprint = {astro-ph/0609265},
 primaryClass = {astro-ph},
       adsurl = {https://ui.adsabs.harvard.edu/abs/2007MNRAS.374..256S}
}

@ARTICLE{kozh20,
       author = {{Kozhberov}, A.~A. and {Yakovlev}, D.~G.},
        title = "{Deformed crystals and torsional oscillations of neutron star crust}",
      journal = {\mnras},
         year = 2020,
        month = nov,
       volume = {498},
       number = {4},
        pages = {5149-5158},
          doi = {10.1093/mnras/staa2715},
archivePrefix = {arXiv},
       eprint = {2009.04952},
 primaryClass = {astro-ph.HE},
       adsurl = {https://ui.adsabs.harvard.edu/abs/2020MNRAS.498.5149K}
}

@ARTICLE{chug05,
       author = {{Chugunov}, A.~I. and {Yakovlev}, D.~G.},
        title = "{Shear Viscosity and Oscillations of Neutron Star Crust}",
      journal = {Astronomy Reports},
         year = 2005,
        month = sep,
       volume = {49},
       number = {9},
        pages = {724-738},
          doi = {10.1134/1.2045323},
archivePrefix = {arXiv},
       eprint = {astro-ph/0511300},
 primaryClass = {astro-ph},
       adsurl = {https://ui.adsabs.harvard.edu/abs/2005ARep...49..724C}
}

@ARTICLE{pumpe17,
       author = {{Pumpe}, Daniel and {Gabler}, Michael and {Steininger}, Theo and {En{\ss}lin}, Torsten A.},
        title = "{Search for quasi-periodic signals in magnetar giant flares. Bayesian inspection of SGR 1806-20 and SGR 1900+14}",
      journal = {\aap},
         year = 2018,
        month = feb,
       volume = {610},
          eid = {A61},
        pages = {A61},
          doi = {10.1051/0004-6361/201731800},
archivePrefix = {arXiv},
       eprint = {1708.05702},
 primaryClass = {astro-ph.HE},
       adsurl = {https://ui.adsabs.harvard.edu/abs/2018A&A...610A..61P}
}

@ARTICLE{hupp14c,
       author = {{Huppenkothen}, D. and {Heil}, L.~M. and {Watts}, A.~L. and {G{\"o}{\u{g}}{\"u}{\textcommabelow s}}, E.},
        title = "{Quasi-periodic Oscillations in Short Recurring Bursts of Magnetars SGR 1806-20 and SGR 1900+14 Observed with RXTE}",
      journal = {\apj},
         year = 2014,
        month = nov,
       volume = {795},
       number = {2},
          eid = {114},
        pages = {114},
          doi = {10.1088/0004-637X/795/2/114},
archivePrefix = {arXiv},
       eprint = {1409.7642},
 primaryClass = {astro-ph.HE},
       adsurl = {https://ui.adsabs.harvard.edu/abs/2014ApJ...795..114H}
}

@ARTICLE{coll15,
       author = {{Collazzi}, A.~C. and {Kouveliotou}, C. and {van der Horst}, A.~J. and {Younes}, G.~A. and {Kaneko}, Y. and {G{\"o}{\u{g}}{\"u}{\textcommabelow s}}, E. and {Lin}, L. and {Granot}, J. and {Finger}, M.~H. and {Chaplin}, V.~L. and {Huppenkothen}, D. and {Watts}, A.~L. and {von Kienlin}, A. and {Baring}, M.~G. and {Gruber}, D. and {Bhat}, P.~N. and {Gibby}, M.~H. and {Gehrels}, N. and {McEnery}, J. and {van der Klis}, M. and {Wijers}, R.~A.~M.~J.},
        title = "{The Five Year Fermi/GBM Magnetar Burst Catalog}",
      journal = {\apjs},
         year = 2015,
        month = may,
       volume = {218},
       number = {1},
          eid = {11},
        pages = {11},
          doi = {10.1088/0067-0049/218/1/11},
archivePrefix = {arXiv},
       eprint = {1503.04152},
 primaryClass = {astro-ph.HE},
       adsurl = {https://ui.adsabs.harvard.edu/abs/2015ApJS..218...11C}
}

@ARTICLE{li22,
       author = {{Li}, X. and {Ge}, M. and {Lin}, L. and {Zhang}, S.-N. and {Song}, L. and {Cao}, X. and {Zhang}, B. and {Lu}, F. and {Xu}, Y. and {Xiong}, S. and {Tuo}, Y. and {Tan}, Y. and {Jiang}, W. and {Qu}, J. and {Zhang}, S. and {Wang}, L. and {Wang}, J. and {Zhang}, B. and {Zhang}, P. and {Li}, C. and {Liu}, C. and {Li}, T. and {Bu}, Q. and {Cai}, C. and {Chen}, Y. and {Chen}, Y. and {Chang}, Z. and {Chen}, L. and {Chen}, T. and {Chen}, Y. and {Cui}, W. and {Du}, Y. and {Gao}, G. and {Gao}, H. and {Gu}, Y. and {Guan}, J. and {Guo}, C. and {Han}, D. and {Huang}, Y. and {Huo}, J. and {Jia}, S. and {Jin}, J. and {Kong}, L. and {Li}, B. and {Li}, G. and {Li}, W. and {Li}, X. and {Li}, X. and {Li}, Z. and {Liang}, X. and {Liao}, J. and {Liu}, H. and {Liu}, H. and {Liu}, X. and {Lu}, X. and {Luo}, Q. and {Luo}, T. and {Ma}, B. and {Ma}, R. and {Ma}, X. and {Meng}, B. and {Nang}, Y. and {Nie}, J. and {Ou}, G. and {Ren}, X. and {Sai}, N. and {Song}, X. and {Sun}, L. and {Tao}, L. and {Wang}, C. and {Wang}, P. and {Wang}, W. and {Wang}, Y. and {Wen}, X. and {Wu}, B. and {Wu}, B. and {Wu}, M. and {Xiao}, S. and {Yang}, S. and {Yang}, Y. and {Yi}, Q. and {Yin}, Q. and {You}, Y. and {Yu}, W. and {Zhang}, F. and {Zhang}, H. and {Zhang}, J. and {Zhang}, W. and {Zhang}, W. and {Zhang}, Y. and {Zhang}, Y. and {Zhao}, H. and {Zhao}, X. and {Zheng}, S. and {Zhou}, D.},
        title = "{Quasi-periodic Oscillations of the X-Ray Burst from the Magnetar SGR J1935-2154 and Associated with the Fast Radio Burst FRB 200428}",
      journal = {\apj},
         year = 2022,
        month = may,
       volume = {931},
       number = {1},
          eid = {56},
        pages = {56},
          doi = {10.3847/1538-4357/ac6587},
archivePrefix = {arXiv},
       eprint = {2204.03253},
 primaryClass = {astro-ph.HE},
       adsurl = {https://ui.adsabs.harvard.edu/abs/2022ApJ...931...56L}
}

@ARTICLE{chen23,
       author = {{Chen}, R.-C. and {Deng}, C.-M. and {Wang}, X.-G. and {Zhou}, Z.-M. and {Yang}, X. and {Lin}, D.-B. and {Wang}, Q. and {Liang}, E.-W.},
        title = "{Detection of Quasi-periodic Oscillations in SGR 150228213}",
      journal = {Research in Astronomy and Astrophysics},
         year = 2023,
        month = aug,
       volume = {23},
       number = {8},
          eid = {085018},
        pages = {085018},
          doi = {10.1088/1674-4527/acdabe},
archivePrefix = {arXiv},
       eprint = {2302.10533},
 primaryClass = {astro-ph.HE},
       adsurl = {https://ui.adsabs.harvard.edu/abs/2023RAA....23h5018C}
}

@ARTICLE{tan26,
       author = {{Tan}, W.-J. and {Wang}, Y. and {Wang}, C.-W. and {Xiong}, S.-L. and {Li}, X.-B. and {Zhang}, S.-N. and {Cai}, C. and {Xue}, W.-C. and {Zhang}, P. and {Wu}, B.-B. and {An}, Z.-H. and {Gao}, M. and {Ge}, M.-Y. and {Gong}, K. and {Guo}, D.-Y. and {Guo}, H.-X. and {Hao}, L.-F. and {Huang}, Y. and {Huang}, Y.-X. and {Lee}, K.-J. and {Li}, B. and {Li}, K.-C. and {Li}, X.-Q. and {Liu}, J.-C. and {Liu}, X.-J. and {Liu}, Y.-Q. and {Ma}, X. and {Peng}, W.-X. and {Qiao}, R. and {Ren}, Y.-Z. and {Song}, L.-M. and {Sun}, X.-L. and {Wang}, J. and {Wang}, J.-Z. and {Wang}, P. and {Wen}, X.-Y. and {Xiao}, S. and {Xie}, L.-S. and {Xu}, H. and {Yang}, S. and {Yi}, S.-X. and {Yi}, Q.-B. and {Yu}, Z.-H. and {Zhang}, L.-D. and {Zhang}, F. and {Zhang}, H.-M. and {Zhang}, J.-P. and {Zhang}, Y.-Q. and {Zhang}, Z. and {Zhao}, X.-Y. and {Zhao}, Y. and {Zheng}, C. and {Zheng}, S.-J.},
        title = "{GECAM discovery of a peculiar magnetar X-ray burst (MXB 221120) from SGR J1935+2154 associated with a fast radio burst}",
      journal = {\aap},
         year = 2026,
        month = mar,
       volume = {707},
          eid = {A289},
        pages = {A289},
          doi = {10.1051/0004-6361/202658841},
archivePrefix = {arXiv},
       eprint = {2604.02261},
 primaryClass = {astro-ph.HE},
       adsurl = {https://ui.adsabs.harvard.edu/abs/2026A&A...707A.289T}
}

@ARTICLE{dem25,
       author = {{Demirer}, M. and {G{\"o}{\u{g}}{\"u}{\textcommabelow s}}, E. and {Kaneko}, Y. and {Keskin}, {\"O}. and {{\c{S}}a{\c{s}}maz}, S. and {Yamasaki}, S.},
        title = "{Detailed Time-resolved Spectral and Temporal Investigations of SGR J1550-5418 Bursts Detected with Fermi Gamma-Ray Burst Monitor}",
      journal = {\apj},
         year = 2025,
        month = aug,
       volume = {988},
       number = {2},
          eid = {282},
        pages = {282},
          doi = {10.3847/1538-4357/ade237},
archivePrefix = {arXiv},
       eprint = {2506.04414},
 primaryClass = {astro-ph.HE},
       adsurl = {https://ui.adsabs.harvard.edu/abs/2025ApJ...988..282D}
}

@ARTICLE{rob23,
       author = {{Roberts}, O.~J. and {Baring}, M.~G. and {Huppenkothen}, D. and {Kouveliotou}, C. and {G{\"o}{\u{g}}{\"u}{\textcommabelow s}}, E. and {Kaneko}, Y. and {Lin}, L. and {van der Horst}, A.~J. and {Younes}, G.},
        title = "{Quasiperiodic Peak Energy Oscillations in X-Ray Bursts from SGR J1935+2154}",
      journal = {\apjl},
         year = 2023,
        month = oct,
       volume = {956},
       number = {1},
          eid = {L27},
        pages = {L27},
          doi = {10.3847/2041-8213/acfcad},
archivePrefix = {arXiv},
       eprint = {2306.08130},
 primaryClass = {astro-ph.HE},
       adsurl = {https://ui.adsabs.harvard.edu/abs/2023ApJ...956L..27R}
}

@ARTICLE{hupp14b,
       author = {{Huppenkothen}, D. and {D'Angelo}, C. and {Watts}, A.~L. and {Heil}, L. and {van der Klis}, M. and {van der Horst}, A.~J. and {Kouveliotou}, C. and {Baring}, M.~G. and {G{\"o}{\u{g}}{\"u}{\textcommabelow s}}, E. and {Granot}, J. and {Kaneko}, Y. and {Lin}, L. and {von Kienlin}, A. and {Younes}, G.},
        title = "{Quasi-periodic Oscillations in Short Recurring Bursts of the Soft Gamma Repeater J1550-5418}",
      journal = {\apj},
         year = 2014,
        month = jun,
       volume = {787},
       number = {2},
          eid = {128},
        pages = {128},
          doi = {10.1088/0004-637X/787/2/128},
archivePrefix = {arXiv},
       eprint = {1404.2756},
 primaryClass = {astro-ph.HE},
       adsurl = {https://ui.adsabs.harvard.edu/abs/2014ApJ...787..128H}
}

@ARTICLE{hupp14,
       author = {{Huppenkothen}, Daniela and {Watts}, Anna L. and {Levin}, Yuri},
        title = "{Intermittency and Lifetime of the 625 Hz Quasi-periodic Oscillation in the 2004 Hyperflare from the Magnetar SGR 1806-20 as Evidence for Magnetic Coupling between the Crust and the Core}",
      journal = {\apj},
         year = 2014,
        month = oct,
       volume = {793},
       number = {2},
          eid = {129},
        pages = {129},
          doi = {10.1088/0004-637X/793/2/129},
archivePrefix = {arXiv},
       eprint = {1408.0734},
 primaryClass = {astro-ph.HE},
       adsurl = {https://ui.adsabs.harvard.edu/abs/2014ApJ...793..129H}
}

@ARTICLE{wad26,
       author = {{Wadiasingh}, Zorawar and {Dinh Thi}, Hoa and {Kalapotharakos}, Constantinos and {Hu}, Kun and {Baring}, Matthew G. and {Harding}, Alice K. and {Younes}, George and {Guillot}, Sebastien and {Sanna}, Andrea and {Negro}, Michela and {Schnittman}, Jeremy D. and {Roberts}, Oliver J. and {Burns}, Eric and {Hu}, Chin-Ping and {G{\"o}{\u{g}}{\"u}{\textcommabelow s}}, Ersin},
        title = "{Magnetar Fireballs and Short Bursts: Curved Spacetime Lensing, Quantum Electrodynamical Effects, Spectra, Polarization, and Impulse Responses}",
      journal = {\apjs},
         year = 2026,
        month = oct,
       volume = {286},
       number = {2},
          eid = {55},
        pages = {55},
          doi = {10.3847/1538-4365/ae9754},
archivePrefix = {arXiv},
       eprint = {2605.22323},
 primaryClass = {astro-ph.HE},
       adsurl = {https://ui.adsabs.harvard.edu/abs/2026ApJS..286...55W}
}

@ARTICLE{sotani2024,
       author = {{Sotani}, Hajime},
        title = "{Magnetar QPOs and Neutron Star Crust Elasticity}",
      journal = {Universe},
         year = 2024,
        month = may,
       volume = {10},
       number = {6},
          eid = {231},
        pages = {231},
          doi = {10.3390/universe10060231},
archivePrefix = {arXiv},
       eprint = {2405.11858},
 primaryClass = {astro-ph.HE},
       adsurl = {https://ui.adsabs.harvard.edu/abs/2024Univ...10..231S}
}

@ARTICLE{cutl87,
       author = {{Cutler}, Curt and {Lindblom}, Lee},
        title = "{The Effect of Viscosity on Neutron Star Oscillations}",
      journal = {\apj},
         year = 1987,
        month = mar,
       volume = {314},
        pages = {234},
          doi = {10.1086/165052},
       adsurl = {https://ui.adsabs.harvard.edu/abs/1987ApJ...314..234C}
}

@ARTICLE{sotani2008,
       author = {{Sotani}, H. and {Colaiuda}, A. and {Kokkotas}, K.~D.},
        title = "{Constraints on the magnetic field geometry of magnetars}",
      journal = {\mnras},
         year = 2008,
        month = apr,
       volume = {385},
       number = {4},
        pages = {2161-2165},
          doi = {10.1111/j.1365-2966.2008.12977.x},
archivePrefix = {arXiv},
       eprint = {0711.1518},
 primaryClass = {gr-qc},
       adsurl = {https://ui.adsabs.harvard.edu/abs/2008MNRAS.385.2161S}
}

@ARTICLE{pl25,
       author = {{Pnigouras}, P. and {Lander}, S.~K.},
        title = "{Nonlinear Coupling between Magnetar QPOs}",
      journal = {\apj},
         year = 2025,
        month = jul,
       volume = {987},
       number = {1},
          eid = {35},
        pages = {35},
          doi = {10.3847/1538-4357/adceee},
archivePrefix = {arXiv},
       eprint = {2501.04556},
 primaryClass = {astro-ph.HE},
       adsurl = {https://ui.adsabs.harvard.edu/abs/2025ApJ...987...35P}
}

@ARTICLE{fuji14,
       author = {{Fujisawa}, Kotaro and {Kisaka}, Shota},
        title = "{Magnetic field configurations of a magnetar throughout its interior and exterior - core, crust and magnetosphere}",
      journal = {\mnras},
         year = 2014,
        month = dec,
       volume = {445},
       number = {3},
        pages = {2777-2793},
          doi = {10.1093/mnras/stu1911},
archivePrefix = {arXiv},
       eprint = {1409.4547},
 primaryClass = {astro-ph.HE},
       adsurl = {https://ui.adsabs.harvard.edu/abs/2014MNRAS.445.2777F}
}

@ARTICLE{konno99,
       author = {{Konno}, K. and {Obata}, T. and {Kojima}, Y.},
        title = "{Deformation of relativistic magnetized stars}",
      journal = {\aap},
         year = 1999,
        month = dec,
       volume = {352},
        pages = {211-216},
          doi = {10.48550/arXiv.gr-qc/9910038},
archivePrefix = {arXiv},
       eprint = {gr-qc/9910038},
 primaryClass = {gr-qc},
       adsurl = {https://ui.adsabs.harvard.edu/abs/1999A&A...352..211K}
}

@ARTICLE{suvg23,
       author = {{Suvorov}, Arthur G. and {Glampedakis}, Kostas},
        title = "{Magnetic equilibria of relativistic axisymmetric stars: The impact of flow constants}",
      journal = {\prd},
         year = 2023,
        month = oct,
       volume = {108},
       number = {8},
          eid = {084006},
        pages = {084006},
          doi = {10.1103/PhysRevD.108.084006},
archivePrefix = {arXiv},
       eprint = {2309.08071},
 primaryClass = {gr-qc},
       adsurl = {https://ui.adsabs.harvard.edu/abs/2023PhRvD.108h4006S}
}

@ARTICLE{hk09,
       author = {{Horowitz}, C.~J. and {Kadau}, Kai},
        title = "{Breaking Strain of Neutron Star Crust and Gravitational Waves}",
      journal = {\prl},
         year = 2009,
        month = may,
       volume = {102},
       number = {19},
          eid = {191102},
        pages = {191102},
          doi = {10.1103/PhysRevLett.102.191102},
archivePrefix = {arXiv},
       eprint = {0904.1986},
 primaryClass = {astro-ph.SR},
       adsurl = {https://ui.adsabs.harvard.edu/abs/2009PhRvL.102s1102H}
}

@ARTICLE{cap2026b,
       author = {{Caplan}, Matthew E and {Smith}, Nevin T and {Bransgrove}, Ashley J and {Horowitz}, Charles J},
        title = "{Plasticity of Neutron Star Crusts}",
      journal = {arXiv e-prints},
         year = 2026,
        month = jun,
          eid = {arXiv:2606.06706},
        pages = {arXiv:2606.06706},
archivePrefix = {arXiv},
       eprint = {2606.06706},
 primaryClass = {astro-ph.HE},
       adsurl = {https://ui.adsabs.harvard.edu/abs/2026arXiv260606706C}
}

@ARTICLE{cap2026,
       author = {{Caplan}, Matthew E. and {Bransgrove}, Ashley},
        title = "{Revisiting the Rheology of Neutron Star Crusts with Molecular Dynamics}",
      journal = {Research Notes of the American Astronomical Society},
         year = 2026,
        month = may,
       volume = {10},
       number = {5},
          eid = {107},
        pages = {107},
          doi = {10.3847/2515-5172/ae67f1},
archivePrefix = {arXiv},
       eprint = {2605.02101},
 primaryClass = {astro-ph.HE},
       adsurl = {https://ui.adsabs.harvard.edu/abs/2026RNAAS..10..107C}
}

@ARTICLE{stroh91,
       author = {{Strohmayer}, T. and {Ogata}, S. and {Iyetomi}, H. and {Ichimaru}, S. and {van Horn}, H.~M.},
        title = "{The Shear Modulus of the Neutron Star Crust and Nonradial Oscillations of Neutron Stars}",
      journal = {\apj},
         year = 1991,
        month = jul,
       volume = {375},
        pages = {679},
          doi = {10.1086/170231},
       adsurl = {https://ui.adsabs.harvard.edu/abs/1991ApJ...375..679S}
}

@ARTICLE{carter72,
       author = {{Carter}, B. and {Quintana}, H.},
        title = "{Foundations of General Relativistic High-Pressure Elasticity Theory}",
      journal = {Proceedings of the Royal Society of London Series A},
         year = 1972,
        month = nov,
       volume = {331},
       number = {1584},
        pages = {57-83},
          doi = {10.1098/rspa.1972.0164},
       adsurl = {https://ui.adsabs.harvard.edu/abs/1972RSPSA.331...57C}
}

@ARTICLE{pot18,
       author = {{Potekhin}, A.~Y. and {Chabrier}, G.},
        title = "{Magnetic neutron star cooling and microphysics}",
      journal = {\aap},
         year = 2018,
        month = jan,
       volume = {609},
          eid = {A74},
        pages = {A74},
          doi = {10.1051/0004-6361/201731866},
archivePrefix = {arXiv},
       eprint = {1711.07662},
 primaryClass = {astro-ph.HE},
       adsurl = {https://ui.adsabs.harvard.edu/abs/2018A&A...609A..74P}
}

@ARTICLE{zem23,
       author = {{Zemlyakov}, Nikita A. and {Chugunov}, Andrey I.},
        title = "{Neutron star inner crust: reduction of shear modulus by nuclei finite size effect}",
      journal = {\mnras},
         year = 2023,
        month = jan,
       volume = {518},
       number = {3},
        pages = {3813-3819},
          doi = {10.1093/mnras/stac3377},
archivePrefix = {arXiv},
       eprint = {2209.05821},
 primaryClass = {astro-ph.HE},
       adsurl = {https://ui.adsabs.harvard.edu/abs/2023MNRAS.518.3813Z}
}

@ARTICLE{sotani2007,
       author = {{Sotani}, H. and {Kokkotas}, K.~D. and {Stergioulas}, N.},
        title = "{Torsional oscillations of relativistic stars with dipole magnetic fields}",
      journal = {\mnras},
         year = 2007,
        month = feb,
       volume = {375},
       number = {1},
        pages = {261-277},
          doi = {10.1111/j.1365-2966.2006.11304.x},
archivePrefix = {arXiv},
       eprint = {astro-ph/0608626},
 primaryClass = {astro-ph},
       adsurl = {https://ui.adsabs.harvard.edu/abs/2007MNRAS.375..261S}
}

@ARTICLE{bransgrove2018,
       author = {{Bransgrove}, Ashley and {Levin}, Yuri and {Beloborodov}, Andrei},
        title = "{Magnetic field evolution of neutron stars - I. Basic formalism, numerical techniques and first results}",
      journal = {\mnras},
         year = 2018,
        month = jan,
       volume = {473},
       number = {2},
        pages = {2771-2790},
          doi = {10.1093/mnras/stx2508},
archivePrefix = {arXiv},
       eprint = {1709.09167},
 primaryClass = {astro-ph.HE},
       adsurl = {https://ui.adsabs.harvard.edu/abs/2018MNRAS.473.2771B}
}

@ARTICLE{ponsv19,
       author = {{Pons}, Jos{\'e} A. and {Vigan{\`o}}, Daniele},
        title = "{Magnetic, thermal and rotational evolution of isolated neutron stars}",
      journal = {Living Reviews in Computational Astrophysics},
         year = 2019,
        month = dec,
       volume = {5},
       number = {1},
          eid = {3},
        pages = {3},
          doi = {10.1007/s41115-019-0006-7},
archivePrefix = {arXiv},
       eprint = {1911.03095},
 primaryClass = {astro-ph.HE},
       adsurl = {https://ui.adsabs.harvard.edu/abs/2019LRCA....5....3P}
}

@ARTICLE{suvpons25,
       author = {{Suvorov}, Arthur G. and {Pons}, Jos{\'e} A.},
        title = "{Magnetar structure in non-linear electrodynamics with mixed poloidal─toroidal fields}",
      journal = {\mnras},
         year = 2025,
        month = jun,
       volume = {539},
       number = {4},
        pages = {3655-3668},
          doi = {10.1093/mnras/staf704},
archivePrefix = {arXiv},
       eprint = {2503.01409},
 primaryClass = {astro-ph.HE},
       adsurl = {https://ui.adsabs.harvard.edu/abs/2025MNRAS.539.3655S}
}

@ARTICLE{brans25,
       author = {{Bransgrove}, Ashley and {Levin}, Yuri and {Beloborodov}, Andrei M.},
        title = "{Giant Hall Waves Launched by Superconducting Phase Transition in Pulsars}",
      journal = {\apj},
         year = 2025,
        month = feb,
       volume = {979},
       number = {2},
          eid = {144},
        pages = {144},
          doi = {10.3847/1538-4357/ad90a3},
archivePrefix = {arXiv},
       eprint = {2408.10888},
 primaryClass = {astro-ph.HE},
       adsurl = {https://ui.adsabs.harvard.edu/abs/2025ApJ...979..144B}
}

@ARTICLE{lan25,
       author = {{Lander}, S.~K.},
        title = "{The Meissner effect in neutron stars}",
      journal = {\mnras},
         year = 2024,
        month = dec,
       volume = {535},
       number = {3},
        pages = {2449-2468},
          doi = {10.1093/mnras/stae2453},
archivePrefix = {arXiv},
       eprint = {2411.08021},
 primaryClass = {astro-ph.HE},
       adsurl = {https://ui.adsabs.harvard.edu/abs/2024MNRAS.535.2449L}
}

@ARTICLE{chugunov10,
       author = {{Chugunov}, A.~I. and {Horowitz}, C.~J.},
        title = "{Breaking stress of neutron star crust}",
      journal = {\mnras},
         year = 2010,
        month = sep,
       volume = {407},
       number = {1},
        pages = {L54-L58},
          doi = {10.1111/j.1745-3933.2010.00903.x},
archivePrefix = {arXiv},
       eprint = {1006.2279},
 primaryClass = {astro-ph.SR},
       adsurl = {https://ui.adsabs.harvard.edu/abs/2010MNRAS.407L..54C}
}

@ARTICLE{pearson2018,
       author = {{Pearson}, J.~M. and {Chamel}, N. and {Potekhin}, A.~Y. and {Fantina}, A.~F. and {Ducoin}, C. and {Dutta}, A.~K. and {Goriely}, S.},
        title = "{Unified equations of state for cold non-accreting neutron stars with Brussels-Montreal functionals - I. Role of symmetry energy}",
      journal = {\mnras},
         year = 2018,
        month = dec,
       volume = {481},
       number = {3},
        pages = {2994-3026},
          doi = {10.1093/mnras/sty2413},
archivePrefix = {arXiv},
       eprint = {1903.04981},
 primaryClass = {astro-ph.HE},
       adsurl = {https://ui.adsabs.harvard.edu/abs/2018MNRAS.481.2994P}
}

@ARTICLE{koj24,
       author = {{Kojima}, Yasufumi},
        title = "{Correct Criterion of Crustal Failure Driven by Intense Magnetic Stress in Neutron Stars}",
      journal = {\apj},
         year = 2024,
        month = oct,
       volume = {974},
       number = {1},
          eid = {125},
        pages = {125},
          doi = {10.3847/1538-4357/ad7382},
archivePrefix = {arXiv},
       eprint = {2408.14100},
 primaryClass = {astro-ph.HE},
       adsurl = {https://ui.adsabs.harvard.edu/abs/2024ApJ...974..125K}
}

@ARTICLE{stef25,
       author = {{Stefanou}, Petros and {Suvorov}, Arthur G. and {Pons}, Jos{\'e} A.},
        title = "{General-relativistic magnetar magnetospheres in 3D with physics-informed neural networks}",
      journal = {\mnras},
         year = 2025,
        month = oct,
       volume = {543},
       number = {1},
        pages = {273-284},
          doi = {10.1093/mnras/staf1438},
archivePrefix = {arXiv},
       eprint = {2506.13519},
 primaryClass = {astro-ph.HE},
       adsurl = {https://ui.adsabs.harvard.edu/abs/2025MNRAS.543..273S}
}

@ARTICLE{brans20,
       author = {{Bransgrove}, Ashley and {Beloborodov}, Andrei M. and {Levin}, Yuri},
        title = "{A Quake Quenching the Vela Pulsar}",
      journal = {\apj},
         year = 2020,
        month = jul,
       volume = {897},
       number = {2},
          eid = {173},
        pages = {173},
          doi = {10.3847/1538-4357/ab93b7},
archivePrefix = {arXiv},
       eprint = {2001.08658},
 primaryClass = {astro-ph.HE},
       adsurl = {https://ui.adsabs.harvard.edu/abs/2020ApJ...897..173B}
}

@ARTICLE{coop24,
       author = {{Cooper}, A.~J. and {Wadiasingh}, Z.},
        title = "{Beyond the Rotational Deathline: Radio Emission from Ultra-long Period Magnetars}",
      journal = {\mnras},
         year = 2024,
        month = sep,
       volume = {533},
       number = {2},
        pages = {2133-2155},
          doi = {10.1093/mnras/stae1813},
archivePrefix = {arXiv},
       eprint = {2406.04135},
 primaryClass = {astro-ph.HE},
       adsurl = {https://ui.adsabs.harvard.edu/abs/2024MNRAS.533.2133C}
}

@ARTICLE{lyu15,
       author = {{Lyutikov}, Maxim},
        title = "{Magnetar activity mediated by plastic deformations of neutron star crust}",
      journal = {\mnras},
         year = 2015,
        month = feb,
       volume = {447},
       number = {2},
        pages = {1407-1417},
          doi = {10.1093/mnras/stu2413},
archivePrefix = {arXiv},
       eprint = {1407.5881},
 primaryClass = {astro-ph.HE},
       adsurl = {https://ui.adsabs.harvard.edu/abs/2015MNRAS.447.1407L}
}

@ARTICLE{passl13,
       author = {{Passamonti}, A. and {Lander}, S.~K.},
        title = "{Stratification, superfluidity and magnetar QPOs}",
      journal = {\mnras},
         year = 2013,
        month = feb,
       volume = {429},
       number = {1},
        pages = {767-774},
          doi = {10.1093/mnras/sts372},
archivePrefix = {arXiv},
       eprint = {1210.2969},
 primaryClass = {astro-ph.SR},
       adsurl = {https://ui.adsabs.harvard.edu/abs/2013MNRAS.429..767P}
}

@BOOK{stacey08,
  author    = {{Stacey}, Frank D. and {Davis}, Paul M.},
  title     = "{Physics of the Earth}",
  publisher = {Cambridge University Press},
  address   = {Cambridge},
  edition   = {4},
  year      = 2008,
  adsurl    = {https://ui.adsabs.harvard.edu/abs/2008phea.book.....S}
}

@ARTICLE{2007MNRAS.377..159L,
       author = {{Levin}, Yuri},
        title = "{On the theory of magnetar QPOs}",
      journal = {\mnras},
         year = 2007,
        month = may,
       volume = {377},
       number = {1},
        pages = {159-167},
          doi = {10.1111/j.1365-2966.2007.11582.x},
archivePrefix = {arXiv},
       eprint = {astro-ph/0612725},
 primaryClass = {astro-ph},
       adsurl = {https://ui.adsabs.harvard.edu/abs/2007MNRAS.377..159L}
}

@ARTICLE{lev15,
       author = {{Levin}, Yuri and {Lyutikov}, Maxim},
        title = "{On the dynamics of mechanical failures in magnetized neutron star crusts}",
      journal = {\mnras},
         year = 2012,
        month = dec,
       volume = {427},
       number = {2},
        pages = {1574-1579},
          doi = {10.1111/j.1365-2966.2012.22016.x},
archivePrefix = {arXiv},
       eprint = {1204.2605},
 primaryClass = {astro-ph.HE},
       adsurl = {https://ui.adsabs.harvard.edu/abs/2012MNRAS.427.1574L}
}

@ARTICLE{burn25,
       author = {{Burnaz}, Louis and {Most}, Elias R. and {Bransgrove}, Ashley},
        title = "{Crustal Quakes Spark Magnetospheric Blasts: Imprints of Realistic Magnetar Crust Oscillations on the Fast Radio Burst Signal}",
      journal = {\apjl},
         year = 2025,
        month = dec,
       volume = {995},
       number = {2},
          eid = {L57},
        pages = {L57},
          doi = {10.3847/2041-8213/ae2466},
archivePrefix = {arXiv},
       eprint = {2508.18033},
 primaryClass = {astro-ph.HE},
       adsurl = {https://ui.adsabs.harvard.edu/abs/2025ApJ...995L..57B}
}

@ARTICLE{kj15,
       author = {{Keer}, L. and {Jones}, D.~I.},
        title = "{Developing a model for neutron star oscillations following starquakes}",
      journal = {\mnras},
         year = 2015,
        month = jan,
       volume = {446},
       number = {1},
        pages = {865-891},
          doi = {10.1093/mnras/stu2123},
archivePrefix = {arXiv},
       eprint = {1408.1249},
 primaryClass = {astro-ph.SR},
       adsurl = {https://ui.adsabs.harvard.edu/abs/2015MNRAS.446..865K}
}

@ARTICLE{pv19,
       author = {{Pons}, J.~A. and {Miralles}, J.~A. and {Geppert}, U.},
        title = "{Magneto-thermal evolution of neutron stars}",
      journal = {\aap},
         year = 2009,
        month = mar,
       volume = {496},
       number = {1},
        pages = {207-216},
          doi = {10.1051/0004-6361:200811229},
archivePrefix = {arXiv},
       eprint = {0812.3018},
 primaryClass = {astro-ph},
       adsurl = {https://ui.adsabs.harvard.edu/abs/2009A&A...496..207P}
}

@ARTICLE{palm05,
       author = {{Palmer}, D.~M. and {Barthelmy}, S. and {Gehrels}, N. and {Kippen}, R.~M. and {Cayton}, T. and {Kouveliotou}, C. and {Eichler}, D. and {Wijers}, R.~A.~M.~J. and {Woods}, P.~M. and {Granot}, J. and {Lyubarsky}, Y.~E. and {Ramirez-Ruiz}, E. and {Barbier}, L. and {Chester}, M. and {Cummings}, J. and {Fenimore}, E.~E. and {Finger}, M.~H. and {Gaensler}, B.~M. and {Hullinger}, D. and {Krimm}, H. and {Markwardt}, C.~B. and {Nousek}, J.~A. and {Parsons}, A. and {Patel}, S. and {Sakamoto}, T. and {Sato}, G. and {Suzuki}, M. and {Tueller}, J.},
        title = "{A giant {\ensuremath{\gamma}}-ray flare from the magnetar SGR 1806 - 20}",
      journal = {\nat},
         year = 2005,
        month = apr,
       volume = {434},
       number = {7037},
        pages = {1107-1109},
          doi = {10.1038/nature03525},
archivePrefix = {arXiv},
       eprint = {astro-ph/0503030},
 primaryClass = {astro-ph},
       adsurl = {https://ui.adsabs.harvard.edu/abs/2005Natur.434.1107P}
}

@ARTICLE{lev16,
       author = {{Li}, Xinyu and {Levin}, Yuri and {Beloborodov}, Andrei M.},
        title = "{Magnetar Outbursts from Avalanches of Hall Waves and Crustal Failures}",
      journal = {\apj},
         year = 2016,
        month = dec,
       volume = {833},
       number = {2},
          eid = {189},
        pages = {189},
          doi = {10.3847/1538-4357/833/2/189},
archivePrefix = {arXiv},
       eprint = {1606.04895},
 primaryClass = {astro-ph.HE},
       adsurl = {https://ui.adsabs.harvard.edu/abs/2016ApJ...833..189L}
}

@ARTICLE{vh11,
       author = {{van Hoven}, Maarten and {Levin}, Yuri},
        title = "{Magnetar oscillations - I. Strongly coupled dynamics of the crust and the core}",
      journal = {\mnras},
         year = 2011,
        month = jan,
       volume = {410},
       number = {2},
        pages = {1036-1051},
          doi = {10.1111/j.1365-2966.2010.17499.x},
archivePrefix = {arXiv},
       eprint = {1006.0348},
 primaryClass = {astro-ph.HE},
       adsurl = {https://ui.adsabs.harvard.edu/abs/2011MNRAS.410.1036V}
}

@ARTICLE{mast15b,
       author = {{Mastrano}, A. and {Suvorov}, A.~G. and {Melatos}, A.},
        title = "{Interpreting the AXP 1E 2259+586 antiglitch as a change in internal magnetization}",
      journal = {\mnras},
         year = 2015,
        month = oct,
       volume = {453},
       number = {1},
        pages = {522-530},
          doi = {10.1093/mnras/stv1658},
archivePrefix = {arXiv},
       eprint = {1507.07308},
 primaryClass = {astro-ph.HE},
       adsurl = {https://ui.adsabs.harvard.edu/abs/2015MNRAS.453..522M}
}

@ARTICLE{col12,
       author = {{Colaiuda}, A. and {Kokkotas}, K.~D.},
        title = "{Magnetar oscillations in the presence of a crust}",
      journal = {\mnras},
         year = 2011,
        month = jul,
       volume = {414},
       number = {4},
        pages = {3014-3022},
          doi = {10.1111/j.1365-2966.2011.18602.x},
archivePrefix = {arXiv},
       eprint = {1012.3103},
 primaryClass = {gr-qc},
       adsurl = {https://ui.adsabs.harvard.edu/abs/2011MNRAS.414.3014C}
}

@ARTICLE{perna11,
       author = {{Perna}, Rosalba and {Pons}, Jose A.},
        title = "{A Unified Model of the Magnetar and Radio Pulsar Bursting Phenomenology}",
      journal = {\apjl},
         year = 2011,
        month = feb,
       volume = {727},
       number = {2},
          eid = {L51},
        pages = {L51},
          doi = {10.1088/2041-8205/727/2/L51},
archivePrefix = {arXiv},
       eprint = {1101.1098},
 primaryClass = {astro-ph.HE},
       adsurl = {https://ui.adsabs.harvard.edu/abs/2011ApJ...727L..51P}
}

@ARTICLE{pons11,
       author = {{Pons}, Jose A. and {Perna}, Rosalba},
        title = "{Magnetars versus High Magnetic Field Pulsars: A Theoretical Interpretation of the Apparent Dichotomy}",
      journal = {\apj},
         year = 2011,
        month = nov,
       volume = {741},
       number = {2},
          eid = {123},
        pages = {123},
          doi = {10.1088/0004-637X/741/2/123},
archivePrefix = {arXiv},
       eprint = {1109.5184},
 primaryClass = {astro-ph.HE},
       adsurl = {https://ui.adsabs.harvard.edu/abs/2011ApJ...741..123P}
}

@ARTICLE{suvm23,
       author = {{Suvorov}, Arthur G. and {Melatos}, Andrew},
        title = "{Evolutionary implications of a magnetar interpretation for GLEAM-X J162759.5-523504.3}",
      journal = {\mnras},
         year = 2023,
        month = mar,
       volume = {520},
       number = {1},
        pages = {1590-1600},
          doi = {10.1093/mnras/stad274},
archivePrefix = {arXiv},
       eprint = {2301.08541},
 primaryClass = {astro-ph.HE},
       adsurl = {https://ui.adsabs.harvard.edu/abs/2023MNRAS.520.1590S}
}

@ARTICLE{qu25,
       author = {{Qu}, Yuanhong and {Zhang}, Bing},
        title = "{Magnetic Interactions in White Dwarf Binaries as Mechanism for Long-period Radio Transients}",
      journal = {\apj},
         year = 2025,
        month = mar,
       volume = {981},
       number = {1},
          eid = {34},
        pages = {34},
          doi = {10.3847/1538-4357/adb1b5},
archivePrefix = {arXiv},
       eprint = {2409.05978},
 primaryClass = {astro-ph.HE},
       adsurl = {https://ui.adsabs.harvard.edu/abs/2025ApJ...981...34Q}
}

@ARTICLE{stroh06,
       author = {{Strohmayer}, Tod E. and {Watts}, Anna L.},
        title = "{The 2004 Hyperflare from SGR 1806-20: Further Evidence for Global Torsional Vibrations}",
      journal = {\apj},
         year = 2006,
        month = dec,
       volume = {653},
       number = {1},
        pages = {593-601},
          doi = {10.1086/508703},
archivePrefix = {arXiv},
       eprint = {astro-ph/0608463},
 primaryClass = {astro-ph},
       adsurl = {https://ui.adsabs.harvard.edu/abs/2006ApJ...653..593S}
}

@ARTICLE{lg19,
       author = {{Lander}, S.~K. and {Gourgouliatos}, K.~N.},
        title = "{Magnetic-field evolution in a plastically failing neutron-star crust}",
      journal = {\mnras},
         year = 2019,
        month = jul,
       volume = {486},
       number = {3},
        pages = {4130-4143},
          doi = {10.1093/mnras/stz1042},
archivePrefix = {arXiv},
       eprint = {1902.02121},
 primaryClass = {astro-ph.HE},
       adsurl = {https://ui.adsabs.harvard.edu/abs/2019MNRAS.486.4130L}
}

@ARTICLE{gabler2013,
       author = {{Gabler}, Michael and {Cerd{\'a}-Dur{\'a}n}, Pablo and {Font}, Jos{\'e} A. and {M{\"u}ller}, Ewald and {Stergioulas}, Nikolaos},
        title = "{Magneto-elastic oscillations of neutron stars: exploring different magnetic field configurations}",
      journal = {\mnras},
         year = 2013,
        month = apr,
       volume = {430},
       number = {3},
        pages = {1811-1831},
          doi = {10.1093/mnras/sts721},
archivePrefix = {arXiv},
       eprint = {1208.6443},
 primaryClass = {astro-ph.SR},
       adsurl = {https://ui.adsabs.harvard.edu/abs/2013MNRAS.430.1811G}
}

@ARTICLE{belo09,
       author = {{Beloborodov}, Andrei M.},
        title = "{Untwisting Magnetospheres of Neutron Stars}",
      journal = {\apj},
         year = 2009,
        month = sep,
       volume = {703},
       number = {1},
        pages = {1044-1060},
          doi = {10.1088/0004-637X/703/1/1044},
archivePrefix = {arXiv},
       eprint = {0812.4873},
 primaryClass = {astro-ph},
       adsurl = {https://ui.adsabs.harvard.edu/abs/2009ApJ...703.1044B}
}

@BOOK{lor04,
  author    = {{Lorimer}, D.~R. and {Kramer}, M.},
  title     = "{Handbook of Pulsar Astronomy}",
  publisher = {Cambridge University Press},
  address   = {Cambridge},
  year      = 2005,
  volume    = {4},
  series    = {Cambridge Observing Handbooks for Research Astronomers},
  adsurl    = {https://ui.adsabs.harvard.edu/abs/2004hpa..book.....L}
}

@article{MS,
    author = "Mueller, Horst and Serot, Brian D.",
    title = "{Relativistic mean field theory and the high density nuclear equation of state}",
    eprint = "nucl-th/9603037",
    archivePrefix = "arXiv",
    reportNumber = "IU-NTC-96-03",
    doi = "10.1016/0375-9474(96)00187-X",
    journal = "Nucl. Phys. A",
    volume = "606",
    pages = "508--537",
    year = "1996"
}

@ARTICLE{miller19qpos,
       author = {{Miller}, M. Coleman and {Chirenti}, Cecilia and {Strohmayer}, Tod E.},
        title = "{On the Persistence of QPOs during the SGR 1806-20 Giant Flare}",
      journal = {\apj},
         year = 2019,
        month = jan,
       volume = {871},
       number = {1},
          eid = {95},
        pages = {95},
          doi = {10.3847/1538-4357/aaf5ce},
archivePrefix = {arXiv},
       eprint = {1808.09483},
 primaryClass = {astro-ph.HE},
       adsurl = {https://ui.adsabs.harvard.edu/abs/2019ApJ...871...95M}
}

@ARTICLE{Andersson19,
       author = {{Andersson}, N. and {Haskell}, B. and {Comer}, G.~L. and {Samuelsson}, L.},
        title = "{The dynamics of neutron star crusts: Lagrangian perturbation theory for a relativistic superfluid-elastic system}",
      journal = {Classical and Quantum Gravity},
         year = 2019,
        month = may,
       volume = {36},
       number = {10},
          eid = {105004},
        pages = {105004},
          doi = {10.1088/1361-6382/ab12a1},
archivePrefix = {arXiv},
       eprint = {1811.09419},
 primaryClass = {gr-qc},
       adsurl = {https://ui.adsabs.harvard.edu/abs/2019CQGra..36j5004A}
}

@article{hyman26,
  title={Discovery of a 62-min long-period transient with near-orthogonal rotator geometry},
  author={Hyman, Scott D and Hurley-Walker, Natasha and Frail, Dale A and Polisensky, Emil and Cotton, WM},
  journal={arXiv preprint arXiv:2609.34102},
  year={2026}
}

@ARTICLE{gabler16,
       author = {{Gabler}, Michael and {Cerd{\'a}-Dur{\'a}n}, Pablo and {Stergioulas}, Nikolaos and {Font}, Jos{\'e} A. and {M{\"u}ller}, Ewald},
        title = "{Coherent magneto-elastic oscillations in superfluid magnetars}",
      journal = {\mnras},
         year = 2016,
        month = aug,
       volume = {460},
       number = {4},
        pages = {4242-4257},
          doi = {10.1093/mnras/stw1272},
archivePrefix = {arXiv},
       eprint = {1605.07638},
 primaryClass = {astro-ph.HE},
       adsurl = {https://ui.adsabs.harvard.edu/abs/2016MNRAS.460.4242G}
}

@ARTICLE{gabler12,
       author = {{Gabler}, Michael and {Cerd{\'a}-Dur{\'a}n}, Pablo and {Stergioulas}, Nikolaos and {Font}, Jos{\'e} A. and {M{\"u}ller}, Ewald},
        title = "{Magnetoelastic oscillations of neutron stars with dipolar magnetic fields}",
      journal = {\mnras},
         year = 2012,
        month = apr,
       volume = {421},
       number = {3},
        pages = {2054-2078},
          doi = {10.1111/j.1365-2966.2012.20454.x},
archivePrefix = {arXiv},
       eprint = {1109.6233},
 primaryClass = {astro-ph.HE},
       adsurl = {https://ui.adsabs.harvard.edu/abs/2012MNRAS.421.2054G}
}

@ARTICLE{gabler18,
       author = {{Gabler}, Michael and {Cerd{\'a}-Dur{\'a}n}, Pablo and {Stergioulas}, Nikolaos and {Font}, Jos{\'e} A. and {M{\"u}ller}, Ewald},
        title = "{Constraining properties of high-density matter in neutron stars with magneto-elastic oscillations}",
      journal = {\mnras},
         year = 2018,
        month = may,
       volume = {476},
       number = {3},
        pages = {4199-4212},
          doi = {10.1093/mnras/sty445},
archivePrefix = {arXiv},
       eprint = {1710.02334},
 primaryClass = {astro-ph.HE},
       adsurl = {https://ui.adsabs.harvard.edu/abs/2018MNRAS.476.4199G}
}

@ARTICLE{HeyvaertsPriest83,
       author = {{Heyvaerts}, J. and {Priest}, E.~R.},
        title = "{Coronal heating by phase-mixed shear Alfven waves.}",
      journal = {\aap},
         year = 1983,
        month = jan,
       volume = {117},
        pages = {220-234},
       adsurl = {https://ui.adsabs.harvard.edu/abs/1983A&A...117..220H}
}

@ARTICLE{ofeng15,
       author = {{Ofengeim}, D.~D. and {Yakovlev}, D.~G.},
        title = "{Shear viscosity in magnetized neutron star crust}",
      journal = {Europhys. Lett.},
         year = 2015,
        month = dec,
       volume = {112},
       number = {5},
          eid = {59001},
        pages = {59001},
          doi = {10.1209/0295-5075/112/59001},
archivePrefix = {arXiv},
       eprint = {1512.03915},
 primaryClass = {astro-ph.SR},
       adsurl = {https://ui.adsabs.harvard.edu/abs/2015EL....11259001O}
}

@ARTICLE{schut83,
       author = {{Schumaker}, B.~L. and {Thorne}, K.~S.},
        title = "{Torsional oscillations of neutron stars}",
      journal = {\mnras},
         year = 1983,
        month = may,
       volume = {203},
        pages = {457-489},
          doi = {10.1093/mnras/203.2.457},
       adsurl = {https://ui.adsabs.harvard.edu/abs/1983MNRAS.203..457S}
}

@ARTICLE{pons2013,
       author = {{Pons}, Jos{\'e} A. and {Vigan{\`o}}, Daniele and {Rea}, Nanda},
        title = "{A highly resistive layer within the crust of X-ray pulsars limits their spin periods}",
      journal = {Nature Physics},
         year = 2013,
        month = jul,
       volume = {9},
       number = {7},
        pages = {431-434},
          doi = {10.1038/nphys2640},
archivePrefix = {arXiv},
       eprint = {1304.6546},
 primaryClass = {astro-ph.SR},
       adsurl = {https://ui.adsabs.harvard.edu/abs/2013NatPh...9..431P}
}

@ARTICLE{Kaneko2015,
       author = {{Kaneko}, T. and {Goossens}, M. and {Soler}, R. and {Terradas}, J. and {Van Doorsselaere}, T. and {Yokoyama}, T. and {Wright}, A.~N.},
        title = "{Apparent Cross-field Superslow Propagation of Magnetohydrodynamic Waves in Solar Plasmas}",
      journal = {\apj},
         year = 2015,
        month = oct,
       volume = {812},
       number = {2},
          eid = {121},
        pages = {121},
          doi = {10.1088/0004-637X/812/2/121},
archivePrefix = {arXiv},
       eprint = {1509.03042},
 primaryClass = {astro-ph.SR},
       adsurl = {https://ui.adsabs.harvard.edu/abs/2015ApJ...812..121K}
}

@ARTICLE{lee25,
       author = {{Lee}, Y.~W.~J. and {Caleb}, M. and {Murphy}, Tara and {Lenc}, E. and {Kaplan}, D.~L. and {Ferrario}, L. and {Wadiasingh}, Z. and {Anumarlapudi}, A. and {Hurley-Walker}, N. and {Karambelkar}, V. and {Ocker}, S.~K. and {McSweeney}, S. and {Qiu}, H. and {Rajwade}, K.~M. and {Zic}, A. and {Bannister}, K.~W. and {Bhat}, N.~D.~R. and {Deller}, A. and {Dobie}, D. and {Driessen}, L.~N. and {Gendreau}, K. and {Glowacki}, M. and {Gupta}, V. and {Jahns-Schindler}, J.~N. and {Jaini}, A. and {James}, C.~W. and {Kasliwal}, M.~M. and {Lower}, M.~E. and {Shannon}, R.~M. and {Uttarkar}, P.~A. and {Wang}, Y. and {Wang}, Z.},
        title = "{The emission of interpulses by a 6.45-h-period coherent radio transient}",
      journal = {Nature Astronomy},
         year = 2025,
        month = mar,
       volume = {9},
        pages = {393-405},
          doi = {10.1038/s41550-024-02452-z},
archivePrefix = {arXiv},
       eprint = {2501.09133},
 primaryClass = {astro-ph.HE},
       adsurl = {https://ui.adsabs.harvard.edu/abs/2025NatAs...9..393L}
}

\appendix

%%%%%%%%%%%%%%%%%%%%%%%%%%%%%%%%%
\section{The no island limit} \label{sec:noislands}
%%%%%%%%%%%%%%%%%%%%%%%%%%%%%%%%%

A basic validation of the two-dimensional time-domain solver is obtained by switching off failure completely, so that the elasticity remains intact throughout the simulation such that we can (hopefully) recover published results.
In this limit different angular harmonics decouple; setting
\begin{equation}
\xi^{\phi} \propto \frac{1}{\sin\theta}\frac{\partial P_\ell(\cos\theta)} {\partial\theta} =-P_\ell'(x),
\end{equation}
allows us to project equation \eqref{eq:torsional_evolution} onto harmonics.
Thus, for example, a generic $n,\ell$ injection is achieved by setting $Y(t=0)=0$ and
\begin{equation}
\dot{Y}(t=0,r,x) = Q_{n}(r) P_{\ell}'(x),
\end{equation}
where $Q_{n}$ is some radial function with $n$ nodes (e.g., a cosine profile whose normalisation is unimportant).
This excites the relevant mode without needing to know the radial eigenfunction beforehand.
Note that removing the island does not make angular harmonics decouple exactly when the magnetic field is present as there are $\ell \to \ell \pm 2$ mixings, but these are weak unless $B \gtrsim 10^{15}$~G \citep{sotani2007}.

\begin{figure}
\centering
 \includegraphics[width=0.48\textwidth]{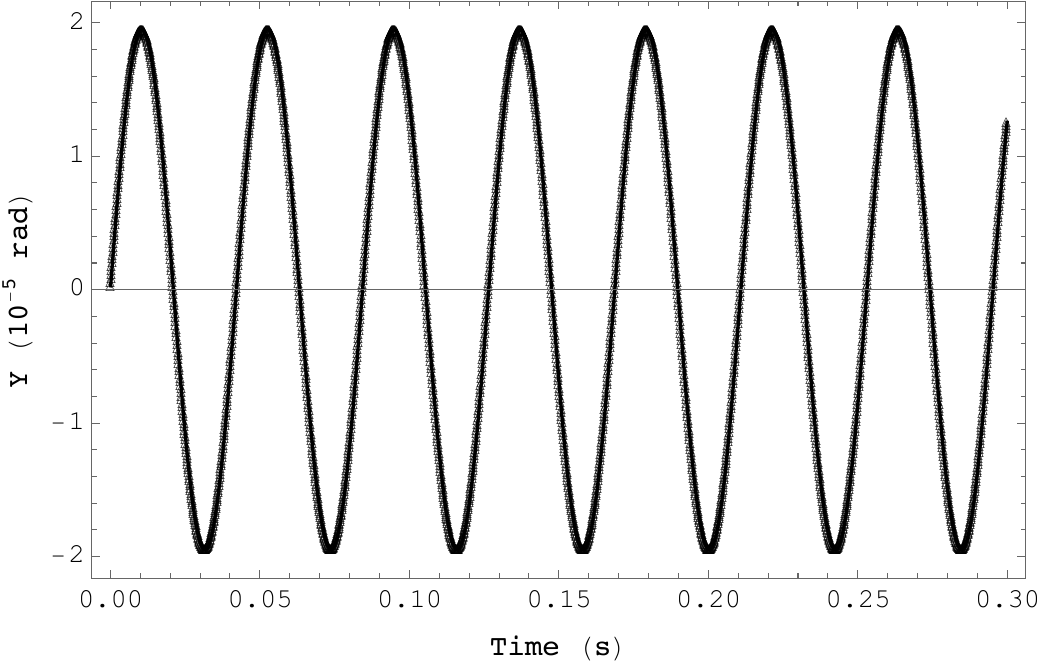}
 \caption{Displacement recorded by a single observer at $r = 11.69$~km and $\theta = 30.3^{\circ}$ for a fundamental ($\ell =2,n=0$) torsional excitation in the no-island, zero-field limit.}
  \label{fig:l2y}
\end{figure}

In the further simplification where we turn off the magnetic field though, equation \eqref{eq:torsional_evolution} simplifies to
\begin{equation}
\begin{aligned}
 0 =& \left( \mu r^4e^{\Phi-\Lambda}Y_{,r}\right)_{,r} -\mu(\ell-1)(\ell+2)r^2e^{\Phi+\Lambda} Y \\
 &-\left(\epsilon+{p}\right) e^{\Lambda-\Phi}r^4\ddot Y.
\end{aligned}
\end{equation}
Tests follow the same methodology of the main text in the sense that multiple observers record the value of $Y$ at their positions over time and then we Fourier transform their data via equation \eqref{eq:mean_observer_power}.
The recording of a single observer, where we switch off both the magnetic field and failure, is shown in Figure~\ref{fig:l2y} for an $\ell=2,n=0$ injection.
The profile is smooth and oscillatory and remains non-dissipative and non-drifting for the entire run (in this case up to $0.3$~s, or $\approx 7$ cycles). 

%%%%%% FIGURE %%%%%%
%%%%%%%%%%%%

Taking the Fourier transform we find, for $\ell =2,n=0$, a frequency of $23.8$~Hz. 
For $\ell =3,n=0$ we have $37.5$~Hz, for $\ell =2,n=1$ we find $828.7$~Hz, and for $\ell =3,n=1$ we get $829.3$~Hz.
Although we do not expect an exact match as our EOS and shear modulus are different, these are roughly consistent with the APR+DH runs of \cite{sotani2007} who found 24.6, 38.9, 761.3 and 762.0~Hz for these modes in a $1.4 M_{\odot}$ star with an eigenvalue code.
The first two of these frequencies agree to within 3.4 and 3.7 percent, respectively, which is the same level of difference between the radius of our nominal crust and theirs.
When using instead a polar field of $10^{14}$~G the magnetic terms start to contribute non-negligibly, and instead we get $23.74$~Hz for the fundamental mode (for instance), meaning a $\approx 0.3\%$ decrease. 
Though not shown, larger field strengths lead to a quadratic scaling in a manner similar to that found by \cite{sotani2007} and others.
The slopes are slightly different, however, since we use a crust-confined field while other works use core-threading ones.

%37.54

%%%%%%%%%%%%%%%%%%%%%%%%%%%%%%%%%%%%%%%%%%%%%%%%%%%%%%%%%%%%%%%%%%%%%%%%%%%%%%%%%
\section{Resolution tests} \label{sec:resolutiontests}
%%%%%%%%%%%%%%%%%%%%%%%%%%%%%%%%%%%%%%%%%%%%%%%%%%%%%%%%%%%%%%%%%%%%%%%%%%%%%%%%%

For all runs shown in the main text, a spatial resolution of $N_{r} \times N_{x} = 96 \times 448$ is used.
Such a value was determined as a tradeoff between computation time and convergence.
Initialising the global island model of Sec.~\ref{sec:global} except with $B_{p} = 5 \times 10^{13}$~G to enable the use of a larger timestep, Table~\ref{tab:resolution} reports the failure fraction \eqref{eq:plastic_fraction} obtained at $t= 0.1$~s for various resolutions.
The key feature is that, at the highest angular resolutions, changing $N_r$ has only a small effect.
For example, at $N_{x} = 448$ the final count is actually identical (within the displayed precision) between the $N_{r} = 64$ and $N_{r} = 128$ cases.

We caution the reader though that reporting only $f_{\rm fail}(0.1 \text{s})$ does not account for the late-time creep which could also, in principle, be resolution sensitive.
Owing to computational limitations we have not extended all of these test runs to $\gtrsim 1$~s, though the closeness of the $N_{x} = 448$ and $N_{x} = 544$ -- with $f_{\rm fail}$ differing only on the order of $\sim 1\%$ -- is at least indicative that avalanche phenomena are captured accurately.
The $96 \times 448$ runs reported in the main text take approximately $\sim 100$~hours of wall time on a local workstation (Apple M1 Max with 10 CPU cores and 64 GB of RAM).

\begin{table}
\centering
\caption{
Plastic volume fraction at $t=0.1$~s as a function of radial and angular resolution.  
All physical parameters and initial conditions are held fixed as in Sec.~\ref{sec:global}.
}
\label{tab:resolution}
\setlength{\tabcolsep}{4.5pt}
\begin{tabular}{c c c c c c}
\hline
\hline
$N_r \backslash N_x$
& 160 & 256 & 352 & 448 & 544 \\
\hline
64
& 0.2576 & 0.2883 & 0.2905 & 0.3006 & 0.3124 \\
96
& 0.2709 & 0.2799 & 0.2914 & 0.2995 & 0.3034 \\
128
& 0.2615 & 0.2748 & 0.2907 & 0.3006 & 0.3048 \\
\hline
\hline
\end{tabular}
\end{table}

%%%%%%%%%%%%%%%%%%%%%%%%%%%%%%%%%%%%%%%%%%%%%%%%%%%%%%%%%%%%%%%%%%%%%%%%
\section{Scaling with field strength and kick efficiency} \label{sec:fieldstrength}
%%%%%%%%%%%%%%%%%%%%%%%%%%%%%%%%%%%%%%%%%%%%%%%%%%%%%%%%%%%%%%%%%%%%%%%%

Here we consider another set of tests where all parameters are held fixed, again for the global island configuration of Sec.~\ref{sec:global}, except for the polar field strength, $B_p$, and kick efficiency, $\eta$.

%%%%%% FIGURE %%%%%%
\begin{figure}
\centering
 \includegraphics[width=0.48\textwidth]{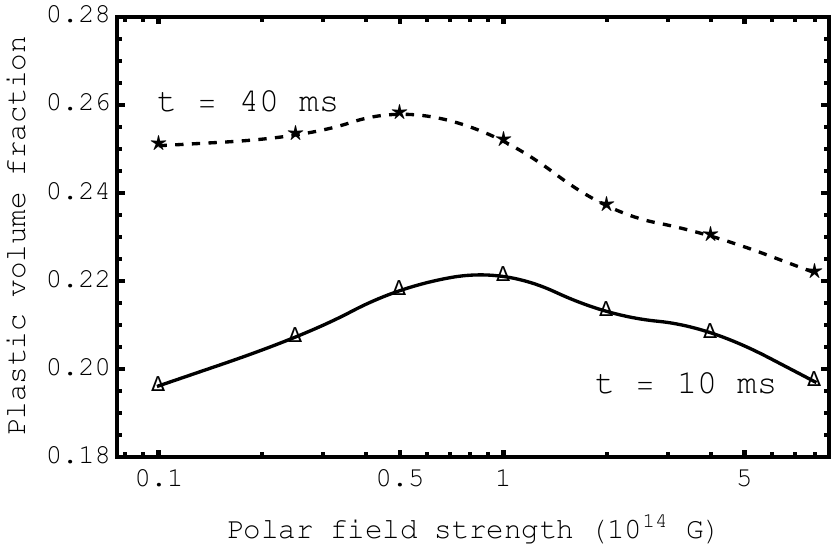}
 \caption{Plastic volume fraction (starting from $\approx 9\%$) as a function of $B_{p}$ as recorded at $t=10$~ms (solid, triangles) or $t=40$~ms (dashed, stars) with a linear interpolation.}
  \label{fig:bpscaling}
\end{figure}
%%%%%%%%%%%%

Figure~\ref{fig:bpscaling} reports the failure count by $10$~ms (solid, triangles) and by $40$~ms (dashed, stars) for field strengths in the range $0.1 \leq B_{p}/10^{14} \text{ G} \leq 8$.
By $10$~ms, we see that the failed cell count is highest for the $B_{p} = 10^{14}$~G model ($\approx 22$\%), though decreases on both sides as $B_{p}$ varies. 
The non-monotonic dependence can be interpreted as a competition between elastic and magnetic restoring stresses, characterised locally by the ratio $\mathcal{R} = B^2/4\pi\mu$ (a kind of magnetoelastic Mach number).
In cases where $\mathcal{R} \ll 1$ (e.g., for $B_{p} = 10^{13}$~G), elastic shear provides the dominant restoring force, so further reducing the field produces relatively little change in the response to the kick; this is why we observe a weak-field flattening at $40$~ms.
Conversely, when $\mathcal{R}\gg1$, magnetic tension strongly resists differential twisting \cite[see also][for a discussion on shear/Alfv{\'e}n-dominated regimes]{gabler12,lev15}.
For a fixed injection energy, the greater restoring stiffness can limit the elastic deformation (i.e., $E_{B} \gg E_{\rm kin,0}$).
This provides a plausible explanation for the reduced failure fraction at the strongest fields.
At intermediate strengths, magnetic coupling can transmit the disturbance through already failed material and redistribute stress
to additional near-threshold regions, while still allowing appreciable elastic deformation.
The maximum at $B_{p} \sim 10^{14}$~G can be interpreted as some kind of equality in the competition between the communication of the disturbance and magnetic stiffening (noting also the equatorial field is a factor $\approx 50$ stronger than the polar value in our model; Fig.~\ref{fig:fieldlines}).

\begin{figure}
\centering
 \includegraphics[width=0.48\textwidth]{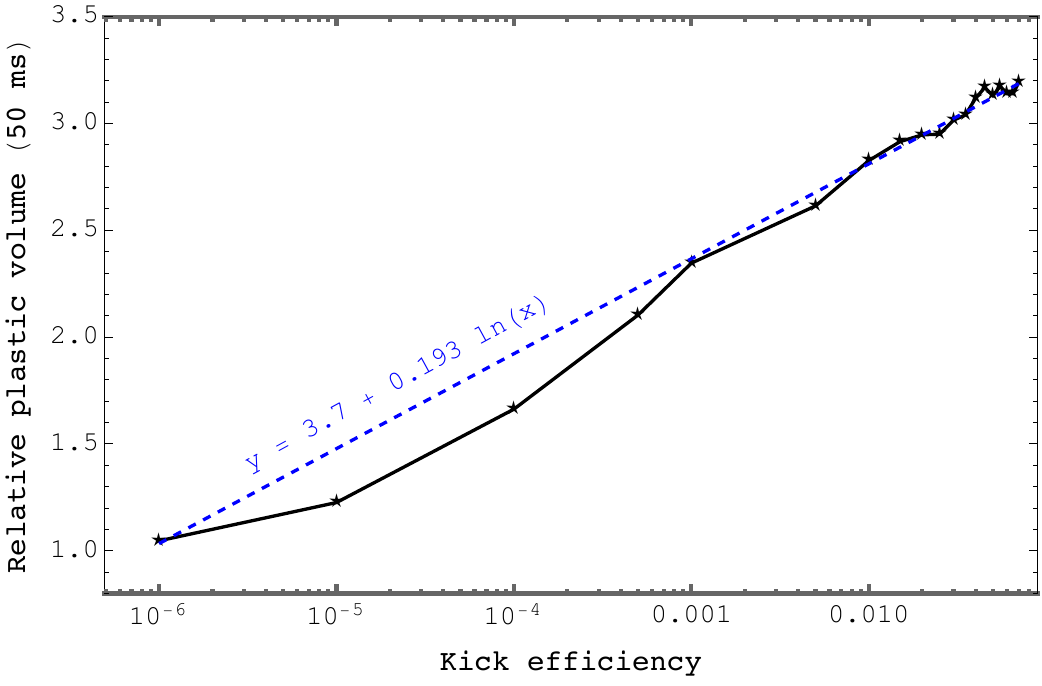}
 \caption{Relative failed-cell fraction as a function of $\eta$ as recorded at $t=50$~ms.
 Overlaid by the dashed, blue curve is the function $y=3.7+0.193 \ln x$ for reference (equation \ref{eq:fplfit}).}
  \label{fig:kickscaling}
\end{figure}

In Figure~\ref{fig:kickscaling} we instead show the \emph{relative} failed volume as recorded at $t=50$~ms -- as compared to the initial island -- as a function of the efficiency.
As expected, as the kick efficiency tends to zero no additional failures accumulate: there is simply not enough energy imparted to the oscillations for them to generate significant stress.
The increase in the fraction is monotonic until $\eta \approx 3\%$, where approximate saturation occurs.
This indicates there is a diminishing return, in the sense that kicking the star harder beyond a certain point does not accentuate the prompt avalanche, though could play a role still in the nature of creep.
Such a saturation may be sensitive to the island geometry however, in that providing energy to a fixed set of initial oscillations does not invite more mixing or constructive interference but changing the effective overlaps could.
On the figure we overlay the function 
\begin{equation} \label{eq:fplfit}
    \frac{f_{\rm fail}(50~\text{ms})}{f_{\rm fail}(0)}=3.7+0.193 \ln \eta,
\end{equation}
which matches both the low- and high-ends of the numerical data reasonably well and could be used as a rule of thumb for $10^{-6} \lesssim \eta \lesssim 10^{-2}$.
The results of the main text are unlikely to change therefore had we used a different efficiency in the range $3\% \lesssim \eta \lesssim 7\%$, but would for $\eta \lesssim 1\%$ (as in Fig.~\ref{fig:weakerkick}).

%%%%%% FIGURE %%%%%%
%%%%%%%%%%%%

\label{lastpage}

\end{document}